\documentclass[mnsc,nonblindrev]{informs3modified} 

\usepackage{amsmath,amsfonts,amssymb}
\usepackage{subfigure}
\usepackage[font=small,labelfont=bf]{caption}
\usepackage{algpseudocode}
\usepackage{algorithm}
\usepackage{epstopdf}

\OneAndAHalfSpacedXI

\usepackage{natbib}
 \bibpunct[, ]{(}{)}{,}{a}{}{,}%
 \def\bibfont{\small}%
 \def\bibsep{\smallskipamount}%
 \def\bibhang{24pt}%
\TheoremsNumberedThrough     
\ECRepeatTheorems

\EquationsNumberedThrough    

\MANUSCRIPTNO{MS-0001-1922.65}

\begin{document}


\RUNAUTHOR{Henry Lam}

\RUNTITLE{Sensitivity to Serial Dependency of Input Processes: A Robust Approach}

\TITLE{Sensitivity to Serial Dependency of Input Processes: A Robust Approach}

\ARTICLEAUTHORS{%
\AUTHOR{Henry Lam}
\AFF{Department of Industrial and Operations Engineering, University of Michigan, Ann Arbor, MI 48109, \EMAIL{khlam@umich.edu}} 
} 

\ABSTRACT{%
Procedures in assessing the impact of serial dependency on performance analysis are usually built on parametrically specified models. In this paper, we propose a robust, nonparametric approach to carry out this assessment, by computing the worst-case deviation of the performance measure due to arbitrary dependence. The approach is based on optimizations, posited on the model space, that have constraints specifying the level of dependency measured by a nonparametric distance to some nominal i.i.d.~input model. We study approximation methods for these optimizations via simulation and analysis-of-variance (ANOVA). Numerical experiments demonstrate how the proposed approach can discover the hidden impacts of dependency beyond those revealed by conventional parametric modeling and correlation studies.


}%



\maketitle

%


While stylized models in operations and service systems often assume independently generated inputs, studies have pointed to the ubiquity of serial dependency in many real-life applications. Examples include arrival streams in file transfers (\cite{ware1998automatic}), video frames (\cite{melamed1992tes}) and call centers (\cite{ibrahim2012modeling}). For decision makers, ignoring serial dependency in the input model for performance analysis potentially leads to substantial errors (e.g., \cite{livny1993impact}). Being able to assess the impact of dependency on a given performance analysis is, therefore, crucial for reliable decision making.

The common way of assessing the effect of dependency is model-based. Namely, one builds a parametric dependent model for the input, and carries out the performance analysis at different values of the parameter, which typically corresponds to correlation. While this line of study is no doubt convenient, it relies heavily on the correctness of the model: If the truth does not lie in the parametric model family, there is no guarantee that the estimate on the effect of dependency is accurate. As an illustration, consider the following simple example:
\vspace{2mm}
\begin{example}
\emph{
Consider two uniform random variables $X$ and $Y$ on the interval $[0,1]$. Suppose the task is to assess how dependency between $X$ and $Y$ affects a quantity such as $E[X^2Y^2]$. One way is to construct a Gaussian copula on $(X,Y)$ with a correlation parameter, vary this parameter starting from 0 and observe how $E[X^2Y^2]$ behaves. Despite being straightforward to apply, this approach may not provide valid information if the true dependency structure between $X$ and $Y$ is non-Gaussian (which might not even be expressible in closed-form).}\label{example:bivariate}
\end{example}
\vspace{4mm}

This issue arises more generally for stochastic inputs on a process-level. For instance:
\vspace{2mm}

\begin{example}
\emph{Consider a first-come-first-serve $M/M/1$ queue. Suppose the task is to assess what happens to a performance measure, say the probability of overload, if the interarrival times are not i.i.d. but are serially dependent. 
One can build a first-order autoregressive model with Gaussian noise, i.e. AR(1), and embed it into the interarrival sequence with exponential marginal distribution (the so-called autoregressive-to-anything (ARTA) procedure; see \cite{cario1996autoregressive}). Simulating the overload probability at different values of the first-order lag parameter away from zero in the AR(1) model gives an assessment of the effect of serial dependency. Again, the AR(1) model is specific. Even within the class of all processes that are dependent only at lag one (or equivalently, Markovian processes), there can be many choices that are nonlinear or non-Gaussian. If the true dependency structure is any of these other possibilities, the present assessment scheme can underestimate the actual impact of dependency.}\label{example:queue}
\end{example}
\vspace{4mm}

This paper investigates a new scheme for assessing the implication of dependency on performance measures like those above for \emph{any} form of dependency. Rather than tuning a correlation parameter in a pre-specified parametric family of dependent models, the proposed approach relies on tuning a unit of measurement for the level of dependency that applies universally to all models. This unit of measurement is the so-called Pearson's $\phi^2$-coefficient (\cite{cramer1999mathematical}), which can be defined on any bivariate joint probability distribution regardless of its parametric form. With this unit, the central ingredient of our methodology is a worst-case analysis: We aim to find the most extreme values of the performance measure of interest among all dependent models that are within a tunable threshold of the $\phi^2$-coefficient.
This can be formulated as optimization programs over the space of models, with constraints on the type of input processes and the dependency level. 
We will illustrate our methodology on Examples \ref{example:bivariate} and \ref{example:queue} in the sequel.

Our worst-case framework is based on the lag-dependent class of stationary processes, which is a generalization of autoregressive processes without restricting to parametric form, and has been used in nonparametric hypothesis tests in econometrics (e.g., \cite{robinson1991consistent,granger2004dependence,granger1994using,chan1992nonparametric}). In general, worst-case optimizations over such a nonparametric class of dependent processes are non-convex and intractable. To handle this issue, we develop approximations on the optimizations under the asymptotic regime that the $\phi^2$-coefficient shrinks to zero. Our main contributions lie in the extraction of conceptual insights and the computational methods for these approximations. First, we show that the worst-case dependency effects, measured by the deviations of the optimal objective values from a baseline independent model, can generally be approximated via the ``interaction" variance of a suitable analysis-of-variance (ANOVA) decomposition on the performance measure. The magnitude of this interaction variance depends on the degree to which the performance measure can be \emph{separated additively} among its input variates. Second, based on these insights, we build simulation-based algorithms that combine with ANOVA procedures to numerically compute the interaction variance, and in turn approximate the worst-case values of the optimizations. 
We provide detailed analyses of the design, sampling efficiencies and statistical properties of our algorithms. 




Asymptotic methods in approximating worst-case optimizations have been recently proposed in \cite{lam2013robust}, which studies uncertainty on the marginal distributions among i.i.d. variates. 
Regarding the motivation for quantifying model misspecification, our results to handle dependency structures beyond \cite{lam2013robust} are important for the following reason: Whereas marginal distributions can often be inferred consistently from data via the empirical distribution, or via goodness-of-fit selections among plenty of available parametric models, the counterparts for simulable dependent models are much less developed. Empirical methods for building transition or conditional densities in dependent processes involve kernel estimators (e.g. \cite{fan2003nonlinear}) that require bandwidth tuning whose statistical errors can be difficult to quantify, and the resulting model can also be difficult to simulate. Moreover, the choices of parametric models for serially dependent input processes are limited, mostly to linear dependence. Thus, unlike the estimation of marginal distributions, the errors in even the best statistical fits for serial dependency structures may not go away with more abundant input data. 
One major view suggested by this paper is to therefore take the pragmatic approach of running a simple baseline model, and \emph{robustifying} it by looking at the worst-case performance among models within a calibrated distance, such as the $\phi^2$-coefficient, that measures the amount of dependency not captured by the baseline. 
Our developed results set a foundation for more extended investigations along this line.



We close this introduction with a brief review of other related work. The literature of dependency modeling in stochastic simulation mostly surrounds the fitting of versatile, parametrically based models, e.g., ARTA (\cite{cario1996autoregressive,biller2005fitting}) and TES (\cite{melamed1992tes,melamed1991tes}) in the context of serial dependence, and NORTA (a sister scheme of ARTA; \cite{cario1997modeling}), chessboard distribution (\cite{ghosh2002chessboard}) and tree-dependent variables (\cite{bedford2002vines}) for cross-sectional and high-dimensional dependence. Simulation-based studies on the effect of dependency in single-server queues have been carried out by \cite{livny1993impact} and \cite{mitchell1977effect} among others. There are also some analytical results on the correlation effects in other stylized queueing models (e.g., \cite{pang2012impact,pang2012infinite}). 

Our formulation is inspired by the literature on distributionally robust optimization and model uncertainty, which considers stochastic optimization problems where the underlying probabilistic model is itself not fully specified and instead believed to lie in some feasible set (e.g. \cite{ben2013robust,goh2010distributionally}). In particular, \cite{delage2010distributionally} consider moment constraints, including cross-moments to represent correlation structure. \cite{glasserman2012robust} consider the scenario of bivariate variables with conditional marginal constraints. 
Our formulation specifically utilizes statistical distance to measure the level of uncertainty centered around a nominal model, which has been used in economics (\cite{hansen2008robustness}), finance (\cite{glasserman2012robust,glasserman2013robust}) and stochastic control (\cite{petersen2000minimax,nilim2005robust,iyengar2005robust}). 

The remainder of this paper is organized as follows. Section \ref{sec:notation} introduces the notion of $\phi^2$-coefficient and some notation. Section \ref{sec:bivariate} discusses the worst-case optimization programs for bivariate settings. Section \ref{sec:1-dependence} discusses more general input processes within the 1-dependence class. Section \ref{sec:numerics} gives the details of numerical implementations and illustrations. Section \ref{sec:general dependence} extends the results to the 2-dependence class. Lastly, Section \ref{extensions} discusses future extensions and concludes. All proofs are presented in the Appendix and the Supplementary Materials.


\section{$\chi^2$-Distance and $\phi^2$-Coefficient}\label{sec:notation}
The $\chi^2$-distance between two probability distributions, say $P_1$ and $P_2$, is defined as
$$\chi^2(P_1,P_2)=E_2\left(\frac{dP_1}{dP_2}-1\right)^2$$
where $dP_1/dP_2$ is the Radon-Nikodym derivative between $P_1$ and $P_2$, and $E_2[\cdot]$ denotes the expectation under $P_2$. As a canonical example, if $P_1$ and $P_2$ are absolutely continuous with respect to the Lebesgue measure on $\mathbb R$ with densities $g_1(x)$ and $g_2(x)$, then $dP_1/dP_2$ is represented by $g_1(x)/g_2(x)$, and $E_2(dP_1/dP_2-1)^2=\int_{\mathbb R}(g_1(x)/g_2(x)-1)^2g_2(x)dx$.

The $\phi^2$-coefficient, defined on a bivariate probability distribution, is the $\chi^2$-distance between this distribution and its independent counterpart. To be clear, for a bivariate random vector $(X,Y)$, let us denote $P(x,y)$ as its probability distribution, and $P_X(x)\times P_Y(y)$ as the distribution having the same marginal distributions as $P(x,y)$ but with $X$ and $Y$ being independent, e.g., if $P(x,y)$ is absolutely continuous with respect to the Lebesgue measure on $\mathbb R^2$ with density $g(x,y)$, then $P_X(x)\times P_Y(y)$ has density $g_X(x)g_Y(y)$, where $g_X(x)=\int_{\mathbb R}g(x,y)dy$ and $g_Y(y)=\int_{\mathbb R}g(x,y)dx$. The $\phi^2$-coefficient of $P(x,y)$ is defined as\footnote{It is also common to consider the square root of $\phi^2(P(x,y))$, which becomes the $\phi$-coefficient.}
$$\phi^2(P(x,y))=\chi^2(P(x,y),P_X(x)\times P_Y(y)).$$

For example, suppose $P(x,y)$ is a bivariate normal distribution with zero means, unit variances and correlation $\rho$. Then $P_X(x)\times P_Y(y)$ is a bivariate normal distribution with zero means, unit variances and zero correlation, and
$$\phi^2(P(x,y))=\int\int\left(\frac{\mathcal N(x,y;\rho)}{\mathcal N(x)\mathcal N(y)}-1\right)^2\mathcal N(x)\mathcal N(y)dxdy$$
where $\mathcal N(\cdot)$ denotes the standard normal density and $\mathcal N(\cdot,\cdot;\rho)$ denotes the bivariate normal density with standard marginals and correlation $\rho$.

Note that the $\phi^2$-coefficient of any independent joint distribution is zero. By definition, it is calculated by fixing the marginal distributions, thus isolating only the amount of dependency. Importantly, its definition is not restricted to a particular parametric model.

\section{Assessment of Bivariate Dependency}\label{sec:bivariate}
To illustrate the basic idea of our method, we first focus on the bivariate setting. Consider a bivariate random vector $(X,Y)\in\mathcal X^2$ for some space $\mathcal X$, and a performance measure denoted $E[h(X,Y)]$ where $h:\mathcal X^2\to\mathbb R$. Our premise is that a \emph{baseline} probability distribution $P_0(x,y)$ has been chosen for $(X,Y)$, where $X$ and $Y$ are independent under $P_0$. We denote $P_0(x)$ and $P_0(y)$ as the (potentially different) marginal distributions of $X$ and $Y$ respectively. In other words, $P_0(x,y)=P_0(x)\times P_0(y)$. We denote $E_0[\cdot]$ and $Var_0(\cdot)$ as the expectation and variance respectively under $P_0$. We also denote $\mathcal P_0$ as the space of all distributions that are absolutely continuous with respect to $P_0(x,y)$.

Our scheme relies on the worst-case optimization programs
\begin{equation}
\renewcommand\arraystretch{1.4}
\begin{array}{ll}
\max&E_f[h(X,Y)]\\
\text{subject to}&\phi^2(P_f(x,y))\leq\eta\\
&P_f(x)=P_0(x),\ P_f(y)=P_0(y)\\
&P_f\in\mathcal P_0
\end{array}
\text{\ \ \ \ and\ \ \ \ }
\begin{array}{ll}
\min&E_f[h(X,Y)]\\
\text{subject to}&\phi^2(P_f(x,y))\leq\eta\\
&P_f(x)=P_0(x),\ P_f(y)=P_0(y)\\
&P_f\in\mathcal P_0.
\end{array} \label{max P}
\end{equation}
The decision variable is $P_f(x,y)$, and we use $E_f[\cdot]$ to denote the expectation under $P_f(x,y)$. 
The constraints $P_f(x)=P_0(x)$ and $P_f(y)=P_0(y)$ state that $P_f$ has the same marginal distributions as $P_0$. The constraint $\phi^2(P_f(x,y))\leq\eta$ controls the level of dependency of $P_f(x,y)$ to be within $\eta$ units of $\phi^2$-coefficient, or equivalently $\eta$ units of $\chi^2$-distance away from $P_0(x,y)$. Overall, \eqref{max P} gives the worst-case values (on both upper and lower ends) of the performance measure among all models with the same marginals as the baseline and $\phi^2$-coefficient within $\eta$. 

The optimality of \eqref{max P} is characterized by the following proposition:
\begin{proposition}
Define $r(x,y)=h(x,y)-E_0[h(X,Y)|X=x]-E_0[h(X,Y)|Y=y]+E_0[h(X,Y)]$. Suppose $h$ is bounded and $Var_0(r(X,Y))>0$. Then, for sufficiently small $\eta>0$, the optimal values of the max and min formulations in \eqref{max P} satisfy
\begin{equation}
\max E_f[h(X,Y)]=E_0[h(X,Y)]+\sqrt{Var_0(r(X,Y))\eta}\label{expansion bivariate max}
\end{equation}
and
\begin{equation}
\min E_f[h(X,Y)]=E_0[h(X,Y)]-\sqrt{Var_0(r(X,Y))\eta}\label{expansion bivariate min}
\end{equation}
respectively.\label{prop:max solution}
\end{proposition}

Therefore, to obtain the worst-case performance measure only requires calculating the standard deviation of $r(X,Y)$. The intuition on $r(X,Y)$ boils down to the question: With fixed marginal distributions of $X$ and $Y$, how does dependency between $X$ and $Y$ affect $E[h(X,Y)]$? Obviously, when $h(x,y)$ is additively separable, i.e. $h(x,y)=h_1(x)+h_2(y)$ for some functions $h_1$ and $h_2$, dependency does not exert any effect on the value of $E[h(X,Y)]$. In other words, any effect of dependency on $E[h(X,Y)]$ must come from the structure of $h$ beyond additive separability, and this is exactly captured by the quantity $r(x,y)$. To see this, by the definition of $r$, we can decompose $h$ as
\begin{eqnarray}
h(x,y)&=&E_0[h(X,Y)]+(E_0[h(X,Y)|X=x]-E_0[h(X,Y)])+(E_0[h(X,Y)|Y=y]-E_0[h(X,Y)]){}\notag\\
&&{}+r(x,y).\label{ANOVA simple}
\end{eqnarray}
Using the terminology from ANOVA (\cite{cox2002theory}), $E_0[h(X,Y)|X]-E_0[h(X,Y)]$ and $E_0[h(X,Y)|Y]-E_0[h(X,Y)]$ can be regarded as the ``main effects" of the ``factors" $X$ and $Y$, and $r(X,Y)$ can be regarded as the ``interaction effect" that captures any nonlinear interaction beyond the main effects.
In particular, when $h(x,y)$ is additively separable, one can easily see that $r(x,y)$ is exactly zero. The standard deviation of $r(X,Y)$ represents the magnitude of interaction. The larger it is, the faster is the rate of deviation from the baseline with respect to the level of dependency measured by $\eta$.
\vspace{2mm}

\emph{Example \ref{example:bivariate} Revisited. }
We illustrate the use of Proposition \ref{prop:max solution} with Example \ref{example:bivariate}. Consider the objective $E[X^2Y^2]$ with uniform marginal distributions for $X$ and $Y$ over $[0,1]$. One standard way to assess the effect of dependency is to construct a Gaussian copula (\cite{mai2012simulating}), i.e. $F_{\mathcal{MN},\rho}(F_{\mathcal N}^{-1}(x),F_{\mathcal N}^{-1}(y))$, as the bivariate distribution function of $(X,Y)$, where $F_{\mathcal N}^{-1}(\cdot)$ is the quantile function of a standard normal variable and $F_{\mathcal{MN},\rho}(\cdot,\cdot)$ is the distribution function of a bivariate normal distribution with standard marginals and correlation $\rho$. Instead of this choice, a modeler could also use other copulas, such as:
\begin{equation*}
\renewcommand\arraystretch{1.4}
\begin{array}{ll}
\text{Gumbel:}&\exp(-((-\log x)^\theta+(-\log y)^\theta)^{1/\theta})\text{\ \ for\ \ }\theta\geq1\\
\text{Clayton:}&(\max\{x^{-\theta}+y^{-\theta}-1,0\})^{-1/\theta}\text{\ \ for\ \ }\theta>0\\
\text{Ali-Mikhail-Haq (AMH):}&\frac{xy}{1-\theta(1-x)(1-y)}\text{\ \ for\ \ }-1\leq\theta\leq1.
\end{array}
\end{equation*}
The effect of dependency in these models can each be assessed by varying the parameter $\rho$ or $\theta$ 
(with $\rho=0$ corresponding to independence for Gaussian copula, $\theta=1$ for Gumbel, and $\theta=0$ for Clayton and AMH). Each model operates on its own specific unit and form of dependency.

\begin{figure}[h]
\centering
\includegraphics[scale=.35]{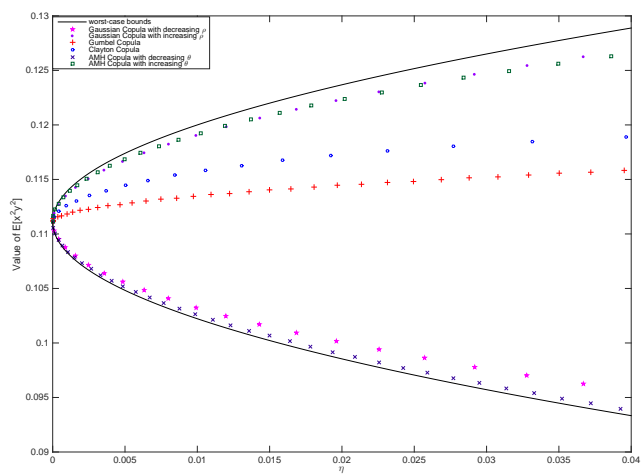}
\caption{Worst-case bounds on the value of $E[X^2Y^2]$ compared against different dependent models}
\label{pic:copulas}
\end{figure}

Alternately, without any presumption on the true model, we can use the $\phi^2$-coefficient as the universal measurement of dependency. Figure \ref{pic:copulas} shows the upper and lower worst-case bounds, as functions of $\eta$, obtained by simply calculating $\sqrt{Var_0(r(X,Y))}=4/45$ in Proposition \ref{prop:max solution}. For comparison, we also plot the values of $E[X^2Y^2]$ when evaluated at the specific copulas above with increasing or decreasing parameters. The worst-case bounds can be seen to dominate all these copulas. It appears that among them, Gaussian and AMH are the closest to the worst-case bounds, whereas Clayton is farther away and Gumbel has the least effect. To give some perspective, if a modeler uses the Gaussian model to assess dependency by only varying the correlation parameter, he/she is restricted to the Gaussian dots in Figure \ref{pic:copulas}, which has a gap with the worst-case scenario and hence may underestimate the actual effect of dependency if the true model is not Gaussian. On the other hand, if the modeler uses the $\phi^2$-coefficient, then he/she can output a range of all possible performance values that cover any model form including Gaussian, hence avoiding the issue of potential model misspecification.


As a further detail, the worst-case joint densities that achieve the depicted bounds in Figure \ref{pic:copulas} are
$$1+\frac{45\sqrt\eta}{4}\left(x^2y^2-\frac{1}{3}x^2-\frac{1}{3}y^2+\frac{1}{9}\right),\ \ 0\leq x,y\leq1\text{\ \ for the upper bound, and}$$
$$1-\frac{45\sqrt\eta}{4}\left(x^2y^2-\frac{1}{3}x^2-\frac{1}{3}y^2+\frac{1}{9}\right),\ \ 0\leq x,y\leq1\text{\ \ for the lower bound.}$$
This can be seen easily from the proof of Proposition \ref{prop:max solution} (in Appendix \ref{appendix:proof bivariate}).

In practice, using our method requires interpreting and choosing $\eta$. Some preliminary sense of its scale can be drawn by comparing with specific models. For instance, in Example \ref{example:bivariate} above, a $\phi^2$-coefficient of $0.04$ approximately corresponds to a correlation of $\pm0.2$ if the model is known to be a bivariate normal distribution. In general, we suggest calibrating $\eta$ from historical data (see Section \ref{sec:calibration} for a discussion and related references).


%

\section{Serial Dependency Assessment within the One-Dependence Class}\label{sec:1-dependence}
We now turn our focus to serial dependency in more general stochastic input sequences. To fix some notation, we denote an input process as $\mathbf X_T=(X_1,\ldots,X_T)\in\mathcal X^T$ and $T$ is the time horizon. Let the performance measure of interest be $E[h(\mathbf X_T)]$, where $h:\mathcal X^T\to\mathbb R$ is a known cost function (i.e.~that can be evaluated by computer, but not necessarily has closed form). Our premise is that the modeler adopts a baseline model for the input process that describes $X_1,\ldots,X_T$ as i.i.d. random variables. As in Section \ref{sec:bivariate}, we denote the probability distribution of the baseline model as $P_0$, and similar definitions for $E_0[\cdot]$ and $Var_0(\cdot)$.

For instance, in the queueing setting in Example \ref{example:queue}, $\mathbf X_T$ is the sequence of interarrival times, $h(\mathbf X_T)$ can be the indicator of whether the $30$-th customer has waiting time longer than a threshold, so that $T=30$, and the baseline model is described by the joint distribution $P_0$ of the i.i.d.~sequence of exponential $X_t$'s.

To define our scope of assessment, we introduce the concept of $p$-dependence, commonly used in nonparametric time series modeling (see, e.g., \cite{robinson1991consistent}). A $p$-dependent process is a process where, conditioned on the last $p$ steps, the current state is independent of all states that are more than $p$ steps back in the past, i.e.~for any $t$, conditioning on $X_t,X_{t+1},\ldots,X_{t+p-1}$, the sequence $\{\ldots,X_{t-2},X_{t-1}\}$ and $\{X_{t+p},X_{t+p+1},\ldots\}$ are independent. This definition in particular contains AR($p$) models and is the natural nonlinear generalization of autoregression. Our assessment scheme for serial dependency operates on the space of all $p$-dependent processes that are stationary, where the lag parameter $p$ is pre-specified to be 1 or 2. The most basic and important case, $p=1$, is discussed in this section. 

On a high level, the worst-case optimization is written as
\begin{equation}
\renewcommand\arraystretch{1.4}
\begin{array}{ll}
\max/\min&E_f[h(\mathbf X_T)]\\
\text{subject to}&P_f\text{\ is within an $\eta$-neighborhood of\ }P_0\\
&\{X_t\}_{t=1,\ldots,T}\text{\ is a $1$-dependent stationary process under $P_f$}\\
&P_f\text{\ generates the same marginal distribution as\ }P_0\\
&P_f\in\mathcal P_0
\end{array}\label{main formulation}
\end{equation}
where $\max/\min$ denotes a pair of max and min problems, and $\mathcal P_0$ denotes the set of all distributions on $\mathbf X_T$ that are absolutely continuous to the i.i.d. baseline distribution $P_0$. The second and the third constraints restrict attention to $1$-dependent processes with the same marginal structure, thus isolating the effect of dependency. Note that the decision variable is the probability measure of the whole sequence $\mathbf X_T$.

The question remains how to define the $\eta$-neighborhood in the first constraint in terms of the $\phi^2$-coefficient. By definition, for a $1$-dependent process under stationarity, the distribution of any two consecutive states completely characterizes the process. Hence one can define an $\eta$-neighborhood of $P_0$ to be any $P_f$ that satisfies $\chi^2(P_f(x_{t-1},x_t),P_0(x_{t-1},x_t))\leq\eta$, where $P_f(x_{t-1},x_t)$ and $P_0(x_{t-1},x_t)$ are the joint distributions of $(X_{t-1},X_t)$ under $P_f$ and $P_0$, for any $t$, under stationarity. Since the baseline model $P_0$ entails i.i.d. $X_t$'s, this neighborhood is equivalent to $\phi^2(P_f(X_{t-1},X_t))\leq\eta$.

Therefore, formulation \eqref{main formulation} can be written mathematically as
\begin{equation}
\renewcommand\arraystretch{1.4}
\begin{array}{ll}
\max/\min&E_f[h(\mathbf{X}_T)]\\
\text{subject to}&\phi^2(P_f(x_{t-1},x_t))\leq\eta\\
&\{X_t:t=1,\ldots,T\}\text{\ is a 1-dependent stationary process under $P_f$}\\
&P_f(x_t)=P_0(x_t)\\
&P_f\in\mathcal P_0
\end{array} \label{max P many}
\end{equation}
where $P_f(x_t)$ and $P_0(x_t)$ are short-hand for the marginal distributions of $X_t$ under $P_f$ and $P_0$ respectively (which clearly do not rely on $t$ under stationarity).

Unlike \eqref{max P}, the optimization \eqref{max P many} is generally non-convex. This challenge is placed on top of the fact that its decision variable lies on the huge space of probability measures on the sample paths with time horizon $T$. Therefore, we consider an approximation of \eqref{max P many} via an asymptotic analysis by shrinking $\eta$ to zero. Given $h$, we define the function $H(x,y):\mathcal X^2\to\mathbb R$ as
\begin{equation}
H(x,y)=\sum_{t=2}^TE_0[h(\mathbf X_T)|X_{t-1}=x,X_t=y].\label{symmetrization}
\end{equation}
We further define $R:\mathcal X^2\to\mathbb R$ as
\begin{equation}
R(x,y)=H(x,y)-E_0[H(X,Y)|X=x]-E_0[H(X,Y)|Y=y]+E_0[H(X,Y)]\label{residual symmetrization}
\end{equation}
where $X$ and $Y$ are i.i.d. random variables each under the marginal distribution of $P_0$.

With the above definitions, our main result is that:
\begin{theorem}
Assume $h$ is bounded and $Var_0(R(X,Y))>0$. The optimal values of the max and min formulations in \eqref{max P many} possess the expansions
\begin{equation}
\max E_f[h(\mathbf X_T)]=E_0[h(\mathbf X_T)]+\Xi_1\sqrt{\eta}+\Xi_2\eta+O(\eta^{3/2}) \label{expansion upper}
\end{equation}
and
\begin{equation}
\min E_f[h(\mathbf X_T)]=E_0[h(\mathbf X_T)]-\Xi_1\sqrt{\eta}+\Xi_2\eta+O(\eta^{3/2}) \label{expansion lower}
\end{equation}
respectively as $\eta\to0$, where
\begin{align*}
\Xi_1&=\sqrt{Var_0(R(X,Y))},\\
\Xi_2&=\frac{\sum_{\substack{s,t=2,\ldots,T\\s<t}}E_0[R(X_{s-1},X_s)R(X_{t-1},X_t)h(\mathbf X_T)]}{Var_0(R(X,Y))}
\end{align*}
and $X$ and $Y$ are i.i.d. random variables each distributed under the marginal distribution of $P_0$.
\label{thm:expansion}
\end{theorem}
The function $R(x,y)$ captures the nonlinear interaction effect analogous to $r(x,y)$ in Section \ref{sec:bivariate}, but now it is applied to the function $H$ instead of the cost function $h$ itself. The function $H$ can be interpreted as a summary of the infinitesimal effect when there is a change in the bivariate distribution of any pair of consecutive elements in the input sequence. 

Note that the first-order coefficients in the asymptotic expansions for the max and min formulations in Theorem \ref{thm:expansion} differ by the sign, whereas the second-order coefficients (i.e.~the ``curvature") are the same for both formulations. Thus, in terms of first- or second-order approximation, obtaining either the upper or lower bound will give another one ``for free".

\section{Numerical Implementation}\label{sec:numerics}
\vspace{2mm}

\subsection{Computing the First-order Expansion Coefficient}\label{sec:implementation}
We discuss how to compute the expansions in Theorem \ref{thm:expansion}, focusing on the first-order coefficient. Our first observation is that we can write the function $H$ as
\begin{equation}
H(x,y)=E_0\left[\sum_{t=2}^Th(\mathbf X_T^{(X_{t-1}=x,X_t=y)})\right].\label{simulation1}
\end{equation}
Here $h(\mathbf X_T^{(X_{t-1}=x,X_t=y)})$ denotes the cost function with $X_{t-1}$ fixed at $x$ and $X_t$ fixed at $y$, and all other $X_s$'s are independently randomly generated under the marginal distribution of $P_0$. Algorithm \ref{ANOVA procedure} generates an unbiased sample for $Var_0(R(X,Y))$.

\begin{algorithm}
  \caption{Generating an unbiased sample for $Var_0(R(X,Y))$}
  \begin{algorithmic}
  \State Given the cost function $h$ and the marginal distribution of $P_0$:
  \State \textbf{1.} Generate $2K$ \emph{outer} copies of $X_t$'s under $P_0$ and divide into two groups of size $K$. Call these realizations $\{x_i\}_{i=1,\ldots,K}$ and $\{y_j\}_{j=1,\ldots,K}$.
  \State\textbf{2. } For each $i,j=1,\ldots,K$, given $x_i$ and $y_j$, simulate $n$ \emph{inner} copies of $\sum_{t=2}^Th(\mathbf X_T^{(X_{t-1}=x_i,X_t=y_j)})$ (where the involved $X_s$'s are simulated using new replications each time they show up, with the exception of the ones fixed as $x_i$ or $y_j$). Call these copies $Z_{ijl},l=1,\ldots,n$.
  \State \textbf{3.} Calculate
  \begin{itemize}
  \item $\bar Z_{ij}=\frac{1}{n}\sum_{l=1}^{n}Z_{ijl}$ for each $i,j=1,\ldots,K$
  \item $\bar Z_{i\cdot}=\frac{1}{K}\sum_{j=1}^{K}\bar Z_{ij}$ for each $i=1,\ldots,K$
  \item $\bar Z_{\cdot j}=\frac{1}{K}\sum_{i=1}^{K}\bar Z_{ij}$ for each $j=1,\ldots,K$
  \item $\bar Z=\frac{1}{K^2}\sum_{i,j=1}^{K}\bar Z_{ij}$
  \end{itemize}
\State \textbf{4.} Compute
$$\frac{1}{n}(s_I^2-s_\epsilon^2)$$
where
$$s_I^2=\frac{n}{(K-1)^2}\sum_{i,j=1}^K(\bar Z_{ij}-\bar Z_{i\cdot}-\bar Z_{\cdot j}+\bar Z)^2$$
and
$$s_\epsilon^2=\frac{1}{K^2(n-1)}\sum_{i,j=1}^K\sum_{l=1}^n(Z_{ijl}-\bar Z_{ij})^2.$$
  \end{algorithmic}\label{ANOVA procedure}
\end{algorithm}

Algorithm \ref{ANOVA procedure} is essentially a manifestation of the ``interaction effect" interpretation of $R(x,y)$ as discussed in Section \ref{sec:1-dependence}. Casting in the setting of a two-way random effect model (\cite{cox2002theory}), we can regard each group of outer samples of $X_t$'s as the realizations of a factor. The inner samples are the responses for each realized pair of factors. Then, $s_I^2$ and $s_\epsilon^2$ are the interaction mean square and the residual mean square, respectively, in the corresponding two-way ANOVA table. The final output $(1/n)(s_I^2-s_\epsilon^2)$ is an unbiased estimator of the interaction effect in the random effect model, which leads to:
\begin{theorem}
Algorithm \ref{ANOVA procedure} gives an unbiased sample for $Var_0(R(X,Y))$.\label{unbiased}
\end{theorem}
Moreover, this estimator possesses the following sampling error bound:
\begin{theorem}
Let $M=E_0[H(X,Y)^4]$ where $X$ and $Y$ are i.i.d. each under the marginal distribution of $P_0$. The variance of the estimator from Algorithm \ref{ANOVA procedure} is of order
$$O\left(\frac{1}{K^2}\left(\frac{T^2}{n^2}+M\right)\right).$$\label{sampling error}
\end{theorem}

To construct a confidence interval for $\sqrt{Var_0(R(X,Y))}$, one can generate $N$ (for example, $N=20$) replications of the output in Algorithm \ref{ANOVA procedure}. Say they are $W_i,i=1,\ldots,N$. Using the delta method (\cite{asmussen2007stochastic}), an asymptotically valid $1-\alpha$ confidence interval for $\sqrt{Var_0(R(X,Y))}$ is given by
$$\left(\sqrt{\bar W}\pm\frac{\nu}{2\sqrt{\bar W}}\cdot\frac{t_{1-\alpha/2,N-1}}{\sqrt{N}}\right)$$
where $\bar W=(1/N)\sum_{i=1}^NW_i$ is the sample mean, and $\nu^2=(1/(N-1))\sum_{i=1}^N(W_i-\bar W)^2$ is the sample variance of $W_i$'s. The quantity $t_{1-\alpha/2,N-1}$ is the $(1-\alpha/2)$-quantile of the $t$-distribution with $N-1$ degrees of freedom.

\subsection{Example \ref{example:queue} Revisited}
Consider the example of $M/M/1$ queue. The baseline model is described by i.i.d.~interarrival times and service times. We are interested in assessing the impact on performance measures when the interarrival times are not i.i.d. but instead serially dependent.


First, we check the qualitative behavior of the first-order coefficient $\Xi_1$ in Theorem \ref{thm:expansion}. 
We consider the performance measure $P(W_T>b)$, where $W_T$ is the waiting time of the $T$-th customer, and $b$ is a threshold. We set the arrival rate for the baseline model to be $0.8$ and the service rate is 1. We compute $\Xi_1$ for different values of $T$ and $b$. For the first set of experiments, we fix $b=2$ and vary $T$ from $10$ to $50$; for the second set, we fix $T=30$ and vary $b$ from 1 to $10$. For each setting we run $50$ replications of Algorithm \ref{ANOVA procedure} to obtain point estimates and $95\%$ confidence intervals (CIs), and use an outer sample size $K=20$ and an inner sample size $n=100$ in Algorithm \ref{ANOVA procedure}.

Figure \ref{pic:T} and Table \ref{table:T} show the point estimates and CIs of $\Xi_1$ against $T$. Note the steady increase in the magnitude of $\Xi_1$: The point estimate of $\Xi_1$ increases from around $0.13$ when $T=10$ to $0.20$ when $T=50$. 
Intuitively, $\Xi_1$ should increase with the number $T$, since $T$ represents the number of variates from the time-series input process consumed in each replication and hence should positively relate to the impact of serial dependency. The depicted trend coincides with this intuition. On the other hand, Figure \ref{pic:S} and Table \ref{table:S} show an increasing-decreasing trend of $\Xi_1$ against $b$. $\Xi_1$ increases first from $0.13$ for $b=1$ to around $0.20$ for $b=4$, before dropping to $0.08$ for $b=10$, showing that the impact of dependency is biggest for the tail probabilities that are at the central region, i.e.~not too big or too small.

To put things into context, the tail probability of the $30$-th customer's waiting time above $b=2$ is around $0.48$, and the expected waiting time is around 3. 
Furthermore, the point estimate of $\Xi_1$ for the expected waiting time is $1.8066$ with 95\% CI $[1.6016,2.0116]$.

\hspace{-.5cm}
\begin{minipage}{\textwidth}
  \begin{minipage}[b]{0.47\textwidth}
    \centering
    \includegraphics[scale=.23]{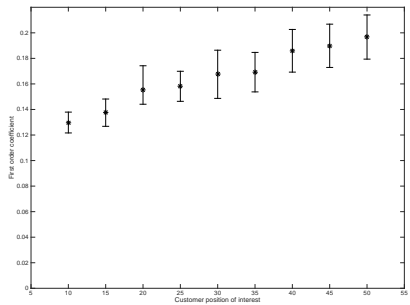}
    \captionof{figure}{Plot of first-order coefficient in Theorem \ref{thm:expansion} against the customer position of interest\\}
    \label{pic:T}
  \end{minipage}
  \hspace{.3cm}
  \begin{minipage}[b]{0.47\textwidth}
    \centering
    {\footnotesize\begin{tabular}{c|c|c}
    Customer Position&Point&$95\%$ C.I.\\
    of Interest&Estimate&\\
    \hline
10&0.130&(0.122, 0.138)\\
15&0.138&(0.127, 0.148)\\
20&0.155&(0.144, 0.174)\\
25&0.158&(0.146, 0.170)\\
30&0.168&(0.149, 0.186)\\
35&0.169&(0.154, 0.185)\\
40&0.186&(0.169, 0.203)\\
45&0.190&(0.173, 0.207)\\
50&0.197&(0.179, 0.214)
    \end{tabular}}
      \captionof{table}{Point estimates and CIs for the first-order coefficients in Theorem \ref{thm:expansion} in relation to the customer position of interest}
      \label{table:T}
    \end{minipage}
  \end{minipage}

\hspace{-.5cm}
\begin{minipage}{\textwidth}
  \begin{minipage}[b]{0.47\textwidth}
    \centering
    \includegraphics[scale=.23]{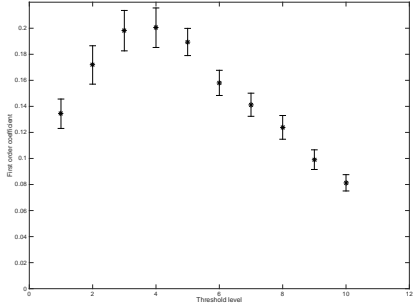}
    \captionof{figure}{Plot of first-order coefficient in Theorem \ref{thm:expansion} against the threshold level in the tail probability\\}
    \label{pic:S}
  \end{minipage}
  \hspace{.3cm}
  \begin{minipage}[b]{0.47\textwidth}
    \centering
    {\footnotesize\begin{tabular}{c|c|c}
    Threshold&Point&$95\%$ C.I.\\
    Level&Estimate&\\
    \hline
1&0.134&(0.123, 0.146)\\
2&0.172&(0.157, 0.187)\\
3&0.198&(0.183, 0.214)\\
4&0.200&(0.185, 0.216)\\
5&0.190&(0.179, 0.200)\\
6&0.158&(0.148, 0.168)\\
7&0.141&(0.132, 0.150)\\
8&0.124&(0.115, 0.133)\\
9&0.099&(0.092, 0.107)\\
10&0.081&(0.075, 0.088)
    \end{tabular}}
      \captionof{table}{Point estimates and CIs for the first-order coefficients in Theorem \ref{thm:expansion} in relation to the threshold level in the tail probability}
      \label{table:S}
    \end{minipage}
  \end{minipage}
\bigskip

%

Next, we assess the performance of our worst-case bounds against some parametric alternatives. Let $U_t,t=1,2,\ldots$ be the interarrival times. We consider:
\vspace{2mm}

 \begin{enumerate}
 \item\emph{Embedded AR(1) process: }$U_t=F_{\mathcal E,0.8}^{-1}(F_{\mathcal N,\beta_0/(1-\beta_1),\sigma^2/(1-\beta_1^2)}(\zeta_t))$, where $F_{\mathcal E,0.8}^{-1}$ is the quantile function of Exp($0.8$) and $F_{\mathcal N,a,b}(\cdot)$ is the distribution function of a normal random variable with mean $a$ and variance $b$. The process $\zeta_t,\ t=1,2,\ldots$ follows a stationary AR(1) model, i.e. $\zeta_t=\beta_0+\beta_1\zeta_{t-1}+\epsilon_t$, with $\epsilon_t\sim N(0,\sigma^2)$, and $\zeta_0\sim N(\beta_0/(1-\beta_1),\sigma^2/(1-\beta_1^2))$ (which is the stationary distribution of this AR(1) model). Note that by construction $U_t$ has marginal distribution Exp($0.8$). This is because under stationarity all $\zeta_t$'s have distribution $N(\beta_0/(1-\beta_1),\sigma^2/(1-\beta_1^2))$, and so $F_{\mathcal N,\beta_0/(1-\beta_1),\sigma^2/(1-\beta_1^2)}(\zeta_t)\sim\text{Unif}[0,1]$, which in turn implies $U_t=F_{\mathcal E,0.8}^{-1}(F_{\mathcal N,\beta_0/(1-\beta_1),\sigma^2/(1-\beta_1^2)}(\zeta_t))\sim\text{Exp}(0.8)$. $U_t$'s are correlated due to the embedded AR(1) process $\{\zeta_t\}$.
     \vspace{2mm}

     \item\emph{Embedded MC process: }$U_t=F_{\mathcal E,0.8}^{-1}(I_t)$ where
     $$I_t=\left\{\begin{array}{ll}
     \text{Unif}\left[0,\frac{a}{1-\theta}\right]&\text{\ \ if\ \ }J_t=0\\
     \text{Unif}\left[\frac{a}{1-\theta},1\right]&\text{\ \ if\ \ }J_t=1
     \end{array}\right.$$
     and $J_t$ is a discrete-time Markov chain with two states $\{0,1\}$ and transition matrix
     $$\left[\begin{array}{cc}a+\theta&1-a-\theta\\a&1-a\end{array}\right]$$
     for some parameter $-a<\theta<1-a$ and $0<a<1$. The distribution of $J_0$ is set to be $a/(1-\theta)$ for state 0 and $1-a/(1-\theta)$ for state 1.

     Note that $a/(1-\theta)$ is the stationary probability of state 0 for the Markov chain $\{J_t\}$. Thus, $I_t$ by construction has a uniform distribution over $[0,1]$, and hence $U_t\sim\text{Exp}(0.8)$. The correlation of $U_t$ is induced by the Markov chain $\{J_t\}$.
     \vspace{2mm}

\end{enumerate}

Figures \ref{pic:queue AR} to \ref{pic:queue all} compare our worst-case bounds using only the first-order terms in Theorem \ref{thm:expansion} to the two parametric dependent models of interarrival times above. We use the mean waiting time of the $30$-th customer as our performance measure. In the figures, the curves are the lower and upper worst-case bounds (the solid ones are calculated using the point estimates of $\Xi_1$, and the dashed ones using the 95\% CIs), and the dots are the estimated performance measures of each parametric model (each dot is estimated using 1 million samples, and the white dots above and below are the $95\%$ CIs). In Figure \ref{pic:queue AR}, we use the embedded AR(1) model, fix $\beta_0=1$, $\sigma^2=0.5$ and vary $\beta_1$ from $-0.2$ to $0.2$. In Figure \ref{pic:queue MC5}, we use the embedded MC model, fix $a=0.5$, and vary $\theta$ from $-0.2$ to $0.2$, while in Figure \ref{pic:queue MC3} we fix $a=0.3$ and vary $\theta$ from $-0.2$ to $0.2$. Finally, Figure \ref{pic:queue all} plots all the parametric models on the same scale of $\eta$. As can be seen, the parametric models are all dominated by the worst-case bounds. The embedded AR(1) model is closest to the worst-case, followed by the embedded MC with $a=0.5$ and then $a=0.3$. Similar to Example \ref{example:bivariate} in Section \ref{sec:bivariate}, the key message here is that using $\phi^2$-coefficient and Theorem \ref{thm:expansion} alleviate the possibility of underestimating the impact of dependency due to model misspecification.

\begin{minipage}{\textwidth}
  \begin{minipage}[b]{0.47\textwidth}
    \centering
    \includegraphics[scale=.34]{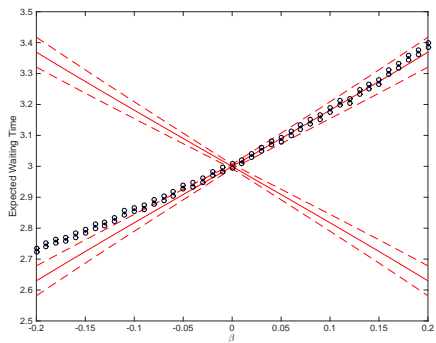}
    \captionof{figure}{Comparison of worst-case bounds with embedded AR($1$) process}
    \label{pic:queue AR}
  \end{minipage}
  \hspace{.3cm}
  \begin{minipage}[b]{0.47\textwidth}
    \centering
    \centering
    \includegraphics[scale=.34]{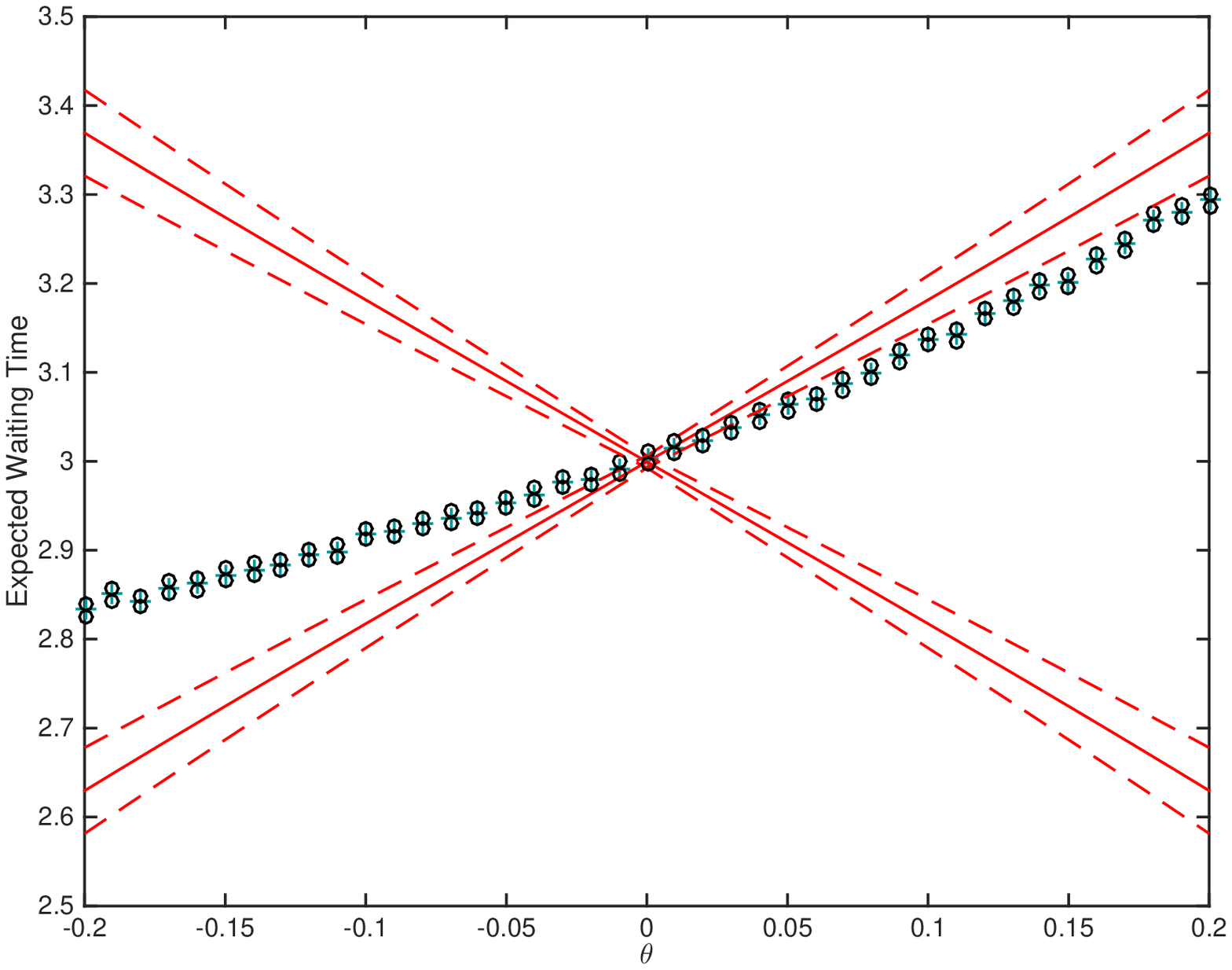}
    \captionof{figure}{Comparison of worst-case bounds with embedded MC($0.5$) process}
    \label{pic:queue MC5}
    \end{minipage}
  \end{minipage}

\hspace{-.5cm}
\begin{minipage}{\textwidth}
  \begin{minipage}[b]{0.47\textwidth}
    \centering
    \centering
    \includegraphics[scale=.34]{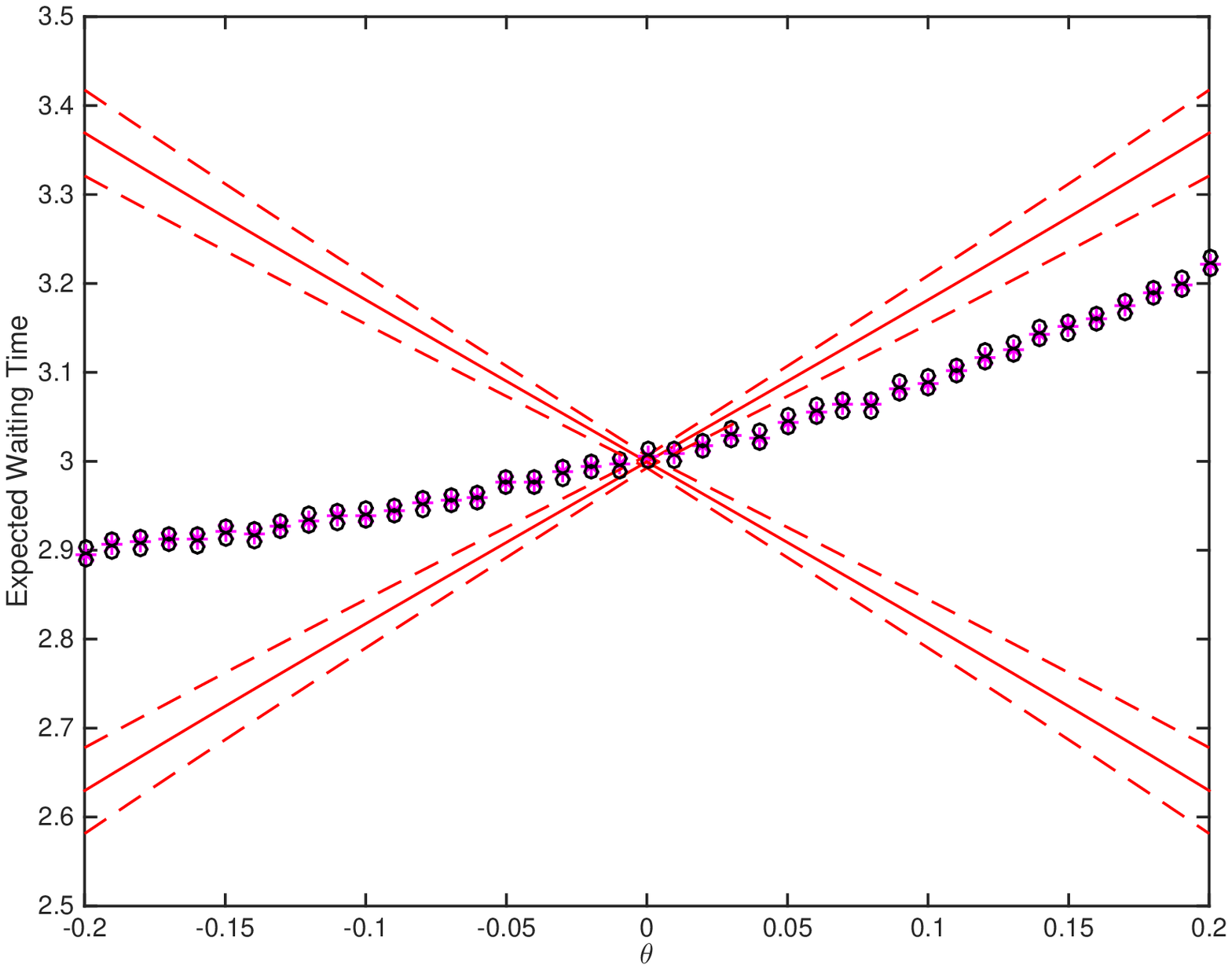}
    \captionof{figure}{Comparison of worst-case bounds with embedded MC($0.3$) process}
    \label{pic:queue MC3}
  \end{minipage}
  \hspace{.3cm}
  \begin{minipage}[b]{0.47\textwidth}
    \centering
    \includegraphics[scale=.25]{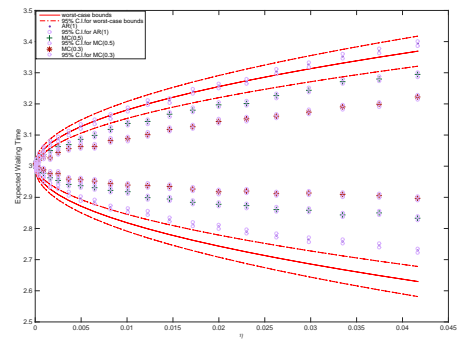}
    \captionof{figure}{Comparison of worst-case bounds with various dependent processes}
    \label{pic:queue all}
    \end{minipage}
  \end{minipage}

\subsection{A Delta Hedging Example}\label{delta hedging}
The following example demonstrates another application of our proposed method. Consider the calculation of delta hedging error for a European option, studied in \cite{glasserman2012robust}. Our focus here is the case where the stock price of an underlying asset deviates from the geometric Brownian motion assumption to serially dependent processes. 

We adopt some notation from \cite{glasserman2012robust}. Consider a European call option, i.e.~the payoff of the option is $(X_T-K)^+$ where $T$ is the maturity, $\{X_t\}_{t\geq0}$ is the stock price, and $K$ is the strike price. We assume a Black-Scholes baseline model for the stock, namely that the stock price follows a geometric Brownian motion
$$dX_t=\mu X_tdt+\sigma X_tdB_t$$
where $B_t$ is a standard Brownian motion, $\mu$ is the growth rate of the stock, and $\sigma$ is the volatility. Assuming trading can be executed continuously over time without transaction fees, it is well known that a seller of the option at time 0 can perfectly hedge its payoff by investing $\delta(t,X_t)=\Phi((\log(X_t/K)+(r-(1/2)\sigma^2)(T-t))/(\sigma\sqrt{T-t}))$, i.e.~the ``delta",  units of stock at any point of time, with the remainder of the portfolio held as cash. Here $r$ is the risk-free rate. This self-balancing portfolio will exactly offset the payoff of the option at maturity. In practice, however, trading cannot be done continuously, and implementing the Black-Scholes strategy will incur a discretization error.

Specifically, suppose that trading is allowed at times $k\Delta t$ for $k=0,1,2,\ldots,T/(\Delta t)$, and that the trader holds the amounts of cash $C(0)$ and stock $S(0)$:
\begin{align*}
C(0)&=BS(0,T,X_0)-X_0\delta(0,X_0)\\
S(0)&=X_0\delta(0,X_0)
\end{align*}
at time 0, and the amounts of cash $C(k\Delta t)$ and stock $S(k\Delta t)$:
\begin{align*}
C(k\Delta t)&=e^{r\Delta t}C((k-1)\Delta t)-X_{k\Delta t}[\delta(k\Delta,X_{k\Delta t})-\delta((k-1)\Delta t,X_{(k-1)\Delta t})]\\
S(k\Delta t)&=X_{k\Delta t}\delta(k\Delta t,X_{k\Delta t})
\end{align*}
at time $k\Delta t$. $BS(0,T,X_0)$ denotes the Black-Scholes price for a European call option with maturity $T$ and initial stock price $X_0$, and $\delta(k\Delta t,X_{k\Delta t})$ is the ``delta" of the option at time $k\Delta t$ with current stock price $X_{k\Delta t}$. The delta hedging error is given by
$$H_e=(X_T-K)^+-C(T)-S(T)$$
and the performance measure is set to be $E|H_e|$.

Our interest is to assess the impact on $E|H_e|$ if $\{X_{k\Delta t}\}$ deviates from i.i.d.~lognormal distribution to a 1-dependent process.
 We set $T=1$ and $\Delta t=0.01$, so that the number of periods in consideration is 100. The initial stock price and the strike price are both set to be 100, i.e.~the call option is at-the-money. We set the baseline geometric Brownian motion to have a growth rate $\mu=0.1$ and a volatility $\sigma=0.2$, and the risk-free rate as $r=0.05$.

\begin{figure}[h]
\centering
\includegraphics[scale=.5]{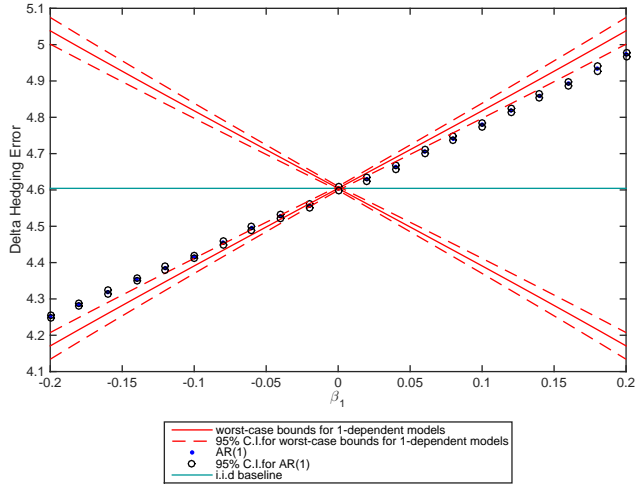}
\caption{Performance of worst-case bounds against the AR(1) model for the hedging error}
\label{picAR}
\end{figure}

 The first-order coefficient $\Xi_1$ in Theorem \ref{thm:expansion} is estimated to be $2.1247$, with 95\% CI $[1.9709,2.2786]$. Figure \ref{picAR} depicts the performance of the worst-case bounds in Theorem \ref{thm:expansion} against an AR(1) model. The delta hedging error for the baseline model, i.e. geometric Brownian motion, is estimated to be about $4.81$ using 1 million sample paths, as depicted by the horizontal line. The curves are the upper and lower bounds using only the first-order term in Theorem \ref{thm:expansion}. As a comparison to our bounds, we compute $E|H_e|$ for $\{X_{k\Delta t}\}$'s that have logarithmic increments satisfying the AR(1) model, i.e.~$\Delta\log X_t=\beta_0+\beta_1\Delta\log X_{t-1}+\epsilon_t$ with $\beta_1$ ranging from $-0.2$ to $0.2$, and keeping the stationary distribution at $\text{lognormal}((\mu-(1/2)\sigma^2)\Delta t,\sigma^2\Delta t)$. The dots show the point estimates of $E|H_e|$ at different $\beta_1$, each using 1 million sample paths. The fact that the dots are quite close to the worst-case bounds (although still bearing a gap) demonstrates that the AR(1) model is a conservative model choice here.


\section{Assessment for Higher-Lag Serial Dependency}\label{sec:general dependence}
So far we have focused on bivariate and one-lag serial dependencies. In practice, the type of dependency may exceed these scopes. This section provides some discussion on how to generalize our results to higher-lag dependency, especially focusing on 2-dependent processes. Further generalizations will be left to future work.

\subsection{Extensions of the $\phi^2$-Coefficient, Formulations and Main Results}

Consider a baseline $P_0$ that generates i.i.d.~$X_t$'s, and we are interested in assessing the impact of 2-dependency. Our formulation for the assessment is:
\begin{equation}
\renewcommand\arraystretch{1.4}
\begin{array}{ll}
\max/\min&E_f[h(\mathbf X_T)]\\
\text{subject to}&\phi^2(P_f(x_{t-1},x_t))\leq\eta_1\\
&\phi^2_2(P_f(x_{t-2},x_{t-1},x_t))\leq\eta_2\\
&\{X_t:t=1,2,\ldots,T\}\text{\ is a 2-dependent stationary process under $P_f$}\\
&P_f(x_t)=P_0(x_t)\\
&P_f\in\mathcal P_0
\end{array} \label{max multi-lag}
\end{equation}
where $\phi_2^2$ is a suitable extension of the definition of $\phi^2$-coefficient to trivariate probability distributions. The constraints $\phi^2(P_f(x_{t-1},x_t))\leq\eta_1$ and $\phi^2_2(P_f(x_{t-2},x_{t-1},x_t))\leq\eta_2$ together describe the level of dependency among all the 2-dependent stationary processes that have the same marginal distribution as $P_0$. The first constraint restricts the amount on $1$-lag dependence, much like formulation \eqref{main formulation}, whereas the second constraint restricts the amount of any additional $2$-lag dependence. 

More precisely, note that for any stationary 2-dependent $P_f$, the joint distribution of any three consecutive states, i.e. $P_f(x_{t-2},x_{t-2},x_t)$, completely determines $P_f$. 
Next, given any $P_f$ that generates a stationary $2$-dependent sequence of $X_t$'s, one can find a measure, $P_{\tilde f_1}$, that generates a $1$-dependent sequence such that the joint distribution of any pair of consecutive states is unchanged, i.e. $P_{\tilde f_1}(x_{t-1},x_t)=P_f(x_{t-1},x_t)$.
Now, in the first constraint of \eqref{max multi-lag}, we have the Pearson's $\phi^2$-coefficient
$$\phi^2(P_f(x_{t-1},x_t))=E_0\left(\frac{dP_f(X_{t-1},X_t)}{dP_0(X_{t-1},X_t)}-1\right)^2.$$
In the second constraint, we shall define
$$\phi^2_2(P_f(x_{t-2},x_{t-1},x_t))=E_{\tilde f_1}\left(\frac{dP_f(X_{t-2},X_{t-1},X_t)}{dP_{\tilde{f}_1}(X_{t-2},X_{t-1},X_t)}-1\right)^2.$$
This definition is illustrated with the following example:
\vspace{2mm}

\begin{example}
\emph{Suppose $(X_{t-2},X_{t-1},X_t)$ is multivariate normal with mean 0 and covariance matrix $\Sigma$ under $P_f$, where
$$\Sigma=\left[\begin{array}{ccc}1&\rho_1&\rho_2\\\rho_1&1&\rho_1\\\rho_2&\rho_1&1\end{array}\right].$$
Then under $P_{\tilde f_1}$, $(X_{t-2},X_{t-1},X_t)$ will still remain as a zero-mean multivariate normal distribution, but with a covariance matrix replaced by
$$\tilde\Sigma=\left[\begin{array}{ccc}1&\rho_1&\rho_1^2\\\rho_1&1&\rho_1\\\rho_1^2&\rho_1&1\end{array}\right]$$
which is essentially the covariance matrix of an AR(1) model with lag parameter $\rho_1$. Details of derivation is left to the Supplmentary Materials.}
\label{Gaussian example}
\end{example}
\vspace{4mm}

With the above definitions and formulation \eqref{max multi-lag}, we have the following approximation:
\begin{theorem}
For $T>2$, the optimal value of the max formulation in \eqref{max multi-lag} satisfies
\begin{equation}
\max E_f[h(\mathbf X_T)]\leq E_0[h(\mathbf X_T)]+\sqrt{Var_0(R(X,Y))\eta_1}+\sqrt{Var_0(S(X,Y,Z))\eta_2}+o(\sqrt{\eta_1}+\sqrt{\eta_2})\label{2-lag upper}
\end{equation}
and the min formulation satisfies
\begin{equation}
\min E_f[h(\mathbf X_T)]\geq E_0[h(\mathbf X_T)]-\sqrt{Var_0(R(X,Y))\eta_1}-\sqrt{Var_0(S(X,Y,Z))\eta_2}+o(\sqrt{\eta_1}+\sqrt{\eta_2})\label{2-lag lower}
\end{equation}
where $X$, $Y$ and $Z$ are i.i.d. each distributed under the marginal distribution of $P_0$, $R$ is defined as in \eqref{residual symmetrization}, i.e.
\begin{eqnarray*}
R(x,y)&=&\sum_{t=2}^TE_0[h(\mathbf X_T)|X_{t-1}=x,X_t=y]-\sum_{t=2}^TE_0[h(\mathbf X_T)|X_{t-1}=x]-\sum_{t=2}^TE_0[h(\mathbf X_T)|X_t=y]{}\\
&&{}+(T-1)E_0[h(\mathbf X_T)]
\end{eqnarray*}
and $S$ is defined as
\begin{eqnarray}
S(x,y,z)&=&\sum_{t=3}^TE_0[h(\mathbf X_T)|X_{t-2}=x,X_{t-1}=y,X_t=z]-\sum_{t=3}^TE_0[h(\mathbf X_T)|X_{t-2}=x,X_{t-1}=y]{}\notag\\
&&{}-\sum_{t=3}^TE_0[h(\mathbf X_T)|X_{t-1}=y,X_t=z]+\sum_{t=3}^TE_0[h(\mathbf X_T)|X_{t-1}=y].\label{H2}
\end{eqnarray}
\label{thm:2-lag}
\end{theorem}

Compared to Theorem \ref{thm:expansion}, the bounds \eqref{2-lag upper} and \eqref{2-lag lower} in Theorem \ref{thm:2-lag} are weaker as they are expressed as inequalities instead of tight asymptotic equalities, and also their remainder terms are $o(\sqrt{\eta_1}+\sqrt{\eta_2})$ instead of $O(\eta^{3/2})$. These relaxations come from the decomposition analysis of \eqref{max multi-lag} into a one-lag optimization and its increment to the two-lag formulation \eqref{max multi-lag} (\eqref{triangle} in Appendix \ref{proof:2-lag}). The latter involves two layers of asymptotic approximation as both the one-lag and two-lag threshold parameters go to zero, which curtails the accuracy.

\subsection{Implementation}
Similar to the 1-dependence case, we present here the computation procedure for the expansion coefficient for the 2-dependent term, namely $\sqrt{Var_0(S(X,Y,Z))}$. The notation $h(\mathbf X_T^{(X_{t-2}=x,X_{t-1}=y,X_t=z)})$ denotes the cost function with $X_{t-2}$ fixed at $x$, $X_{t-1}$ fixed at $y$, and $X_t$ fixed at $z$, and all other $X_s$'s are independently randomly generated under the marginal distribution of $P_0$. Algorithm \ref{ANOVA procedure 2-dependence} generates an unbiased sample for $Var_0(S(X,Y,Z))$.

\begin{algorithm}
  \caption{Generating an unbiased sample for $Var_0(S(X,Y,Z))$}
  \begin{algorithmic}
  \State Given the cost function $h$ and the marginal distribution of $P_0$:
  \State \textbf{1.} Generate one copy of $X_t$, say the realization is $y$.
  \State \textbf{2.} Generate $2K$ \emph{outer} copies of $X_t$'s and divide into two groups of size $K$. Call these realizations $\{x_i\}_{i=1,\ldots,K}$ and $\{z_j\}_{j=1,\ldots,K}$.
  \State\textbf{3. } For each $i,j=1,\ldots,K$, given $y$, $x_i$ and $z_j$, simulate $n$ \emph{inner} copies of $\sum_{t=2}^Th(\mathbf X_T^{(X_{t-2}=x_i,X_{t-1}=y,X_t=z_j)})$ (where the involved $X_s$'s are simulated using new replications each time they show up, with the exception of the ones fixed as $x_i$, $y$ or $z_j$). Call these copies $Z_{ijl},l=1,\ldots,n$.
  \State \textbf{4.} Calculate
  \begin{itemize}
  \item $\bar Z_{ij}=\frac{1}{n}\sum_{l=1}^{n}Z_{ijl}$ for each $i,j=1,\ldots,K$
  \item $\bar Z_{i\cdot}=\frac{1}{K}\sum_{j=1}^{K}\bar Z_{ij}$ for each $i=1,\ldots,K$
  \item $\bar Z_{\cdot j}=\frac{1}{K}\sum_{i=1}^{K}\bar Z_{ij}$ for each $j=1,\ldots,K$
  \item $\bar Z=\frac{1}{K^2}\sum_{i,j=1}^{K}\bar Z_{ij}$
  \end{itemize}
\State \textbf{5.} Compute
$$\frac{1}{n}(s_I^2-s_\epsilon^2)$$
where
$$s_I^2=\frac{n}{(K-1)^2}\sum_{i,j=1}^K(\bar Z_{ij}-\bar Z_{i\cdot}-\bar Z_{\cdot j}+\bar Z)^2$$
and
$$s_\epsilon^2=\frac{1}{K^2(n-1)}\sum_{i,j=1}^K\sum_{l=1}^n(Z_{ijl}-\bar Z_{ij})^2.$$
  \end{algorithmic}\label{ANOVA procedure 2-dependence}
\end{algorithm}

The only differences between Algorithm \ref{ANOVA procedure 2-dependence} and Algorithm \ref{ANOVA procedure} are that a sample $y$ must be generated at the very beginning, and the conditioning in the inner copies $\sum_{t=3}^Th(\mathbf X_T^{(X_{t-2}=x_i,X_{t-1}=y,X_t=z_j)})$ include $X_{t-1}=y$ as well as $X_{t-2}=x_i,X_t=z_j$. Consequently, there is essentially no increase in sampling complexity for generating one unbiased sample in Algorithm \ref{ANOVA procedure 2-dependence} compared to Algorithm \ref{ANOVA procedure}. 
We have the following statistical guarantees:
\begin{theorem}
Algorithm \ref{ANOVA procedure 2-dependence} gives an unbiased sample for $Var_0(S(X,Y,Z))$.\label{unbiased 2-dependence}
\end{theorem}
\begin{theorem}
Let $G(x,y,z)=\sum_{t=2}^TE_0[h(\mathbf X_T)|X_{t-2}=x,X_{t-1}=y,X_t=z]$. Suppose $E_0[G(X,Y,Z)^4|Y]\leq M_1$ uniformly in $Y$ a.s. under $P_0$, and $Var_0(Var_0(S(X,Y,Z)|Y))\leq M_2$, where $X$, $Y$ and $Z$ are i.i.d. each under the marginal distribution of $P_0$. Then the variance of the estimator from Algorithm \ref{ANOVA procedure 2-dependence} is of order
$$O\left(M_2+\frac{1}{K^2}\left(\frac{T^2}{n^2}+M_1\right)\right).$$\label{sampling error 2-dependence}
\end{theorem}

The same procedure in Section \ref{sec:implementation} using $N$ replications of Algorithm \ref{ANOVA procedure 2-dependence} can be employed to construct confidence intervals.

\subsection{Delta Hedging Example Continued}
We use our example in Section \ref{delta hedging} further to illustrate Theorem \ref{thm:2-lag}, now using a parametric AR(2) model as a comparison. Consider a stock price with logarithmic increment $\Delta\log X_t=\beta_0+\beta_1\Delta\log X_{t-1}+\beta_2\Delta\log X_{t-2}+\epsilon_t$. We allow $\beta_1$ and $\beta_2$ to vary, with $\beta_0$ and the variance of $\epsilon_t$ chosen such that the stationary distribution is kept at $\text{lognormal}((\mu-(1/2)\sigma^2)\Delta t,\sigma^2\Delta t)$.

Figures \ref{pic:fix1} and \ref{pic:fix2} compare our bounds with the AR(2) model. Figure \ref{pic:fix1} considers $\beta_1$ fixed at $0.1$, and $\beta_2$ varies from $-0.2$ to $0.2$. Figure \ref{pic:fix2} considers $\beta_2$ fixed at $0.1$, and $\beta_1$ varies from $-0.2$ to $0.2$. The dots depict the point estimates and the $95\%$ CIs of the hedging errors of the AR(2) model each using 1 million sample paths. The horizontal lines represent the hedging error of the i.i.d. baseline model. The curves are the upper and lower bounds from Theorem \ref{thm:2-lag}. We use $20$ replications of Algorithm \ref{ANOVA procedure 2-dependence} to generate point estimate ($2.0602$) and $95\%$ CI ($[1.9026, 2.2179]$) of the coefficient $\sqrt{Var_0(S(X,Y,Z))}$ with outer sample size $K=20$ and inner sample size $n=100$. Clearly, the curves provide valid bounds for both AR(1) and AR(2) (note that $\beta_2=0$ reduces to the AR(1) model). If the modeler is uncertain about whether the process elicits a one or two-lag dependence, the bounds in Theorem \ref{thm:2-lag} is a more conservative choice to use.

%

\begin{minipage}{\textwidth}
  \begin{minipage}[b]{0.47\textwidth}
    \centering
    \includegraphics[scale=.37]{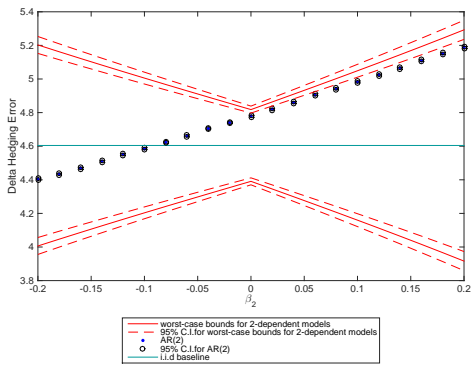}
    \captionof{figure}{Comparison of 2-dependence worst-case bounds against AR(2) processes with $\beta_1$ fixed}
    \label{pic:fix1}
  \end{minipage}
  \hspace{.3cm}
  \begin{minipage}[b]{0.47\textwidth}
    \centering
    \centering
    \includegraphics[scale=.37]{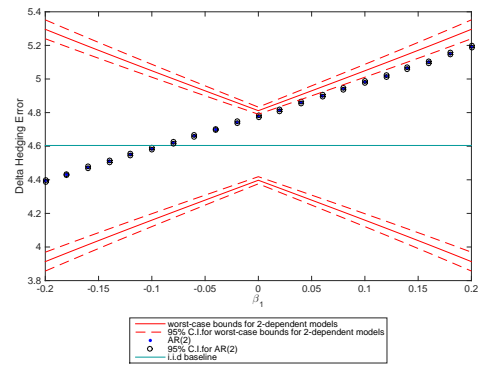}
    \captionof{figure}{Comparison of 2-dependence worst-case bounds against AR(2) processes with $\beta_2$ fixed}
    \label{pic:fix2}
    \end{minipage}
  \end{minipage}


\section{Discussion}\label{extensions}
In this paper, we developed optimization formulations, derived conceptual insights and constructed approximate solution schemes for quantifying the worst-case impacts of dependency within the class of serially dependent processes of dependence lags $1$ and $2$. Our proposed approach has the capability of assessing impacts of model misspecification beyond the conventional parametric regime used in dependency modeling. We view our results here as a first step towards a general robust, nonparametric framework to input model uncertainty in the presence of dependency. Five key issues, however, have to be resolved to make our framework practical.

\subsection{Calibration of $\phi^2$-coefficient}\label{sec:calibration}
First, we need to interpret and choose the value of $\eta$. Although a preliminary sense of the scale of $\phi^2$-coefficient can be drawn by using specific parametric models (e.g. see discussion at the end of Example \ref{example:bivariate} in Section \ref{sec:bivariate}), 
its scale (or that of other similar divergences) can be quite abstruse. One approach is to carry out statistical estimation when historical data are available. 
For discrete state-space, $\phi^2$-coefficients can be approximated by substituting the empirical frequencies at each state 
(this yields exactly the $\chi^2$-statistic used in the classical hypothesis test of independence for contingency table (\cite{cramer1999mathematical})). For continuous state-space, estimation is much more involved and requires some variants of binning or density estimation on the data. This problem falls into the classical and still growing literature of divergence estimation. 
For instance, \cite{liu2012exponential} study finite-sample bounds for histogram-based estimators, \cite{moon2014multivariate,pal2010estimation,poczos2011estimation,poczos2012nonparametric} study asymptotic convergence of nearest-neighbor approaches, and \cite{krishnamurthy2014nonparametric} derive minimax bounds for further corrections of these estimators. Other estimation techniques include the use of so-called sample-spacings (e.g. \cite{van1992estimating}) and the variational representation for divergences (e.g. \cite{nguyen2007estimating}). \cite{beirlant1997nonparametric} provide an excellent review of older references.

Note that when using the estimate of $\eta$, it should be the confidence upper bound that is placed onto the $\phi^2$-constraint. The confidence guarantee for $\eta$ then translates to the guarantee that the min and max values of the optimizations like \eqref{max P} and \eqref{max P many} contain the true value of the performance measure with the same confidence. Such translation of statistical guarantees is commonly used in data-driven robust optimizations (e.g., \cite{delage2010distributionally}).

Some flexibility can be employed in calibrating $\eta$ in addition to statistical inference. One situation of interest is when data exhibit non-stationary behavior over different segments of time. In that case, one can perform a separate estimation of $\eta$ for each period where stationarity is more sensibly assumed, and obtain a robust estimate of $\eta$ by taking the maximum among the estimates for all periods. This exemplifies a situation where subjective judgement can be injected to calibrate a more sensible $\eta$.

Another perspective in calibrating $\eta$ is to use output data if available. \cite{glasserman2012robust}, for example, suggest finding the least $\eta$ such that the worst-case bounds are consistent, i.e. overlap, with the historical outputs. The $\eta$ obtained in this way is an optimistic choice since the resulting optimizations give a lower estimate on the actual model risks, thus helping to identify situations where the involved model risks are unacceptably large.

Future work will investigate the above approaches to the important problem of calibrating $\eta$.

\subsection{Non-i.i.d.~Baseline}\label{sec:non-iid}
This paper has focused on i.i.d.~baseline model, but when an input process is observed to possess strong serial dependency, it is more useful to adopt a baseline model that is not i.i.d.. In this case, optimizations like \eqref{max P many} can be used to assess the impact on the performance measure when the model deviates from a dependent baseline within the 1-dependence class. The formulation in \eqref{max P many} will remain the same except that the first constraint will be $\chi^2(P_f(x_{t-1},x_t),P_0(x_{t-1},x_t))\leq\eta$, where $P_0$ is no longer an i.i.d.~model.

Under this setting, Theorem \ref{thm:expansion} still holds, except for a technical change in the definition of $R(x,y)$. Specifically, $R(x,y)$ will be the $L_2$-projection of the function $H(x,y)$, defined in \eqref{symmetrization}, onto the subspace $\mathcal A_X\cap A_Y$, where $\mathcal A_X$ and $\mathcal A_Y$ are the set of measurable functions, say $V(X,Y)$, satisfying $E_0[V(X,Y)|X]=0$ and $E_0[V(X,Y)|Y]=0$ a.s. respectively. This can be seen by following the proof of Theorem \ref{thm:expansion} (presented in Appendix \ref{appendix:proof expansion} and the Supplementary Materials).

In the case where $P_0$ describes an i.i.d.~process, this definition reduces to that of $R(x,y)$ in \eqref{residual symmetrization}. For a non-i.i.d.~baseline, $R(x,y)$ is represented by the Anderson-Duffin formula (see, for example, \cite{fillmore1971operator}):
\begin{equation}
R(x,y)=2\Pi^X(\Pi^X+\Pi^Y)^\dagger\Pi^YH(x,y)\label{general R}
\end{equation}
where $\Pi^X$ is the projection onto $\mathcal A_X$, $\Pi^Y$ is the projection onto $\mathcal A_Y$, and $\dagger$ denotes the Moore-Penrose pseudoinverse. Future work will investigate the computational procedures for this more general projection operation.
\vspace{2mm}


\subsection{Enhancement of Computational Efficiency}
Algorithms \ref{ANOVA procedure} and \ref{ANOVA procedure 2-dependence} involve simulating $\sum_{t=2}^Th(\mathbf X_T^{(X_{t-1}=x,X_t=y)})$ or $\sum_{t=3}^Th(\mathbf X_T^{(X_{t-2}=x,X_{t-1}=y,X_t=z)})$, which consists of running an order $T$ number of replications of the system. This can be computationally intensive for performance analysis over a long time horizon. One approach to reduce the amount of computation is to explore the connection between the function $H(x,y)$ in \eqref{simulation1} (also the correspondence for 2-dependence) and the so-called influence function, an object used commonly in robust statistics to quantify the infinitesimal effect on a statistical functional due to contamination of data (\cite{hampel2011robust,hampel1974influence}). Essentially, the influence function is a functional derivative taken with respect to the input probability distribution. It is conjectured that $H(x,y)$ bears such interpretation by restricting the perturbation direction to 1-dependent probability measures. The key is that the variance of influence function can be approximated by suitably resampling empirical distributions from the baseline (an observation also used to develop central limit theorems for symmetric statistics (\cite{serfling2009approximation}). This points to an alternate bootstrap-based scheme for computing the coefficients in this paper by resampling empirical distributions from the baseline, propagating them to the simulation outputs, followed by suitably defined projection operations like the ANOVA decomposition. This method eliminates the need to run an order $T$ number of replications on the system, but it introduces extra estimation bias. Its full algorithmic developments and bias-variance tradeoffs will be investigated in future work.


\subsection{Lag Dependence of Full Generality}
The natural extension of formulation \eqref{max multi-lag} to lag $p>2$ is to place additional constraints $\phi_k^2(P_f(x_{t-k},x_{t-k+1},\ldots,x_t))\leq\eta_k$ for $k=2,\ldots,p$, where
$$\phi_k^2(P_f(x_{t-k},x_{t-k+1},\ldots,x_t))=E_{\tilde f_{k-1}}\left(\frac{dP_f(X_{t-k},\ldots,X_t)}{dP_{\tilde f_{k-1}}(X_{t-k},\ldots,X_t)}-1\right)^2$$
with $P_{\tilde f_{k-1}}$ recursively defined as the $(k-1)$-dependent counterparts of $P_{\tilde f_k}$ that has the same $(k-1)$-dimensional marginals, and $P_{\tilde f_p}=P_f$. By generalizing the decomposition \eqref{triangle} in the proof of Theorem \ref{thm:2-lag} (see Appendix \ref{proof:2-lag}), each additional lag will lead to an additional term $\sqrt{Var_0(S_k(X_{t-k},X_{t-k+1},\ldots,X_t))\eta_k}$ in the worst-case bounds, where
\begin{eqnarray*}
&&S_k(x_1,x_2,\ldots,x_{k+1})\\
&=&\sum_{t=k+1}^TE_0[h(\mathbf X_T)|X_{t-k}=x_1,X_{t-k+1}=x_2,\ldots,X_t=x_{k+1}]-\sum_{t=k+1}^TE_0[h(\mathbf X_T)|X_{t-k}=x_1,\ldots,X_{t-1}=x_k]{}\\
&&{}-\sum_{t=k+1}^TE_0[h(\mathbf X_T)|X_{t-k+1}=x_2,\ldots,X_t=x_{k+1}]+\sum_{t=k+1}^TE_0[h(\mathbf X_T)|X_{t-k+1}=x_2,\ldots,X_{t-1}=x_k]
\end{eqnarray*}
Consequently, Algorithm \ref{ANOVA procedure 2-dependence} generalizes to the $k$-lag counterparts by replacing $y$ in the algorithm with $x_2,\ldots,x_k$. Moreover, its computational complexity favorably remains fixed with respect to the lag $k$.


\subsection{Other Extensions: Uncertainty on multiple Input Models, Joint Uncertainty with Marginal Distribution, and Large Amount of Model Uncertainty}
In this paper we have assumed known marginal distributions and have focused on a single input process. We can place additional constraints, e.g., a separate $\phi^2$-coefficient on each independent input process, a $\chi^2$-distance from a baseline marginal distribution etc., to extend the methodology to multiple inputs and joint uncertainty on the marginal distribution. A future direction is the generalization of the approximation and computation schemes to handle these additional constraints.
To deal with model uncertainty beyond a neighborhood, one approach is to build general iterative algorithms for approximating the more global worst-case optimizations. Our future work will use the insights into the local worst-case behaviors gained in this paper to design gradient-based optimization procedures for these problems. 



\begin{APPENDICES}
\section{Proof of Proposition \ref{prop:max solution}}\label{appendix:proof bivariate}
To avoid redundancy, let us focus on the maximization in \eqref{max P}; the minimization counterpart can be tackled by merely replacing $h$ by $-h$. First, we rewrite the maximization formulation \eqref{max P} in terms of the likelihood ratio $L=L(X,Y)=dP_f(X,Y)/dP_0(X,Y)$:
\begin{equation}
\begin{array}{ll}
\max&E_0[h(X,Y)L(X,Y)]\\
\text{subject to}&E_0(L-1)^2\leq\eta\\
&E_0[L|X]=1,\ E_0[L|Y]=1\text{\ a.s.}\\
&L\geq0\text{\ a.s.}
\end{array} \label{max}
\end{equation}
where the decision variable is $L$, over the set of measurable functions under $P_0$. The constraints $E_0[L|X]=1$ and $E_0[L|Y]=1$ come from the marginal constraints in \eqref{max}. To see this, note that $P_f(X\in A)=E_0[L(X,Y);X\in A]=E_0[E_0[L|X]I(X\in A)]$ for any measurable set $A$. Thus $P_f(X\in A)=P_0(X\in A)$ for any set $A$, depicted in the original formulation \eqref{max P}, is equivalent to $E_0[L|X]=1$ a.s.. Similarly for $E_0[L|Y]=1$. Moreover, observe also that $E_0[L|X]=E_0[L|Y]=1$ implies $E_0[L]=1$, and so $L$ in the formulation is automatically a valid likelihood ratio (and hence, there is no need to add this extra constraint). Furthermore, since $L=dP_f(X,Y)/dP_0(X,Y)=dP_f(X,Y)/(dP_0(X)dP_0(Y))=dP_f(X,Y)/(dP_f(X)dP_f(Y))$, we have $\phi^2(P_f(X,Y))=E_0(L(X,Y)-1)^2$, and hence the $\phi^2$-coefficient constraint in formulation \eqref{max P} is equivalent to $E_0(L-1)^2\leq\eta$.

We consider the Lagrangian of \eqref{max}:
\begin{equation}
\max_{L\in\mathcal{L}}E_0[h(X,Y)L]-\alpha(E_0(L-1)^2-\eta) \label{dual}
\end{equation}
where $\mathcal L=\{L\geq0\text{\ a.s.}:E_0[L|X]=E_0[L|Y]=1\text{\ a.s.}\}$. The optimal solution to \eqref{dual} can be characterized as follows:
\begin{proposition}
Under the condition that $h$ is bounded, for any large enough $\alpha>0$,
\begin{equation}
L^*(x,y)=1+\frac{r(x,y)}{2\alpha} \label{optimal L}
\end{equation}
maximizes $E_0[h(X,Y)L]-\alpha E_0(L-1)^2$ over $L\in\mathcal L$. \label{prop:characterization bivariate}
\end{proposition}

We explain how to obtain Proposition \ref{prop:characterization bivariate}. For convenience, we use shorthand $h$ to denote $h(X,Y)$, $r$ to denote $r(X,Y)$, $E_0[\cdot|x]$ to denote $E_0[\cdot|X=x]$ and $E_0[\cdot|y]$ to denote $E_0[\cdot|Y=y]$. To solve \eqref{dual}, we relax the constraints $E_0[L|X]=E_0[L|Y]=1$ a.s., and consider
\begin{equation}
E_0[h(X,Y)L]-\alpha(E_0(L-1)^2-\eta)+\int(E_0[L|x]-1)\beta(x)dP_0(x)+\int(E_0[L|y]-1)\gamma(y)dP_0(y) \label{Lagrangian}
\end{equation}
where $\beta$ and $\gamma$ are in the $L_2$-space under $P_0$, i.e.~$\int\beta(x)^2dP_0(x),\int\gamma(y)^2dP_0(y)<\infty$. We first look for a candidate optimal solution, which we will verify later. To this end, differentiate \eqref{Lagrangian} heuristically and set it to zero to get
\begin{equation}
h(x,y)-2\alpha L(x,y)+2\alpha+\beta(x)+\gamma(y)=0.\label{heuristic}
\end{equation}
This differentiation can be seen especially clearly in the case when $P_0$ has discrete support, as in the following example:
\vspace{2mm}

\begin{example}
\emph{Consider $X$ and $Y$ lying on $\mathcal X=\{1,\ldots,n\}$. Let $P_0(x),P_0(y)>0$, $x,y=1,\ldots,n$, denote the marginal probability masses on $X=x$ and $Y=y$ respectively. Under this setting, \eqref{Lagrangian} can be written as
\begin{eqnarray}
&&\sum_{x=1}^n\sum_{y=1}^nh(x,y)L(x,y)P_0(x)P_0(y)-\alpha\left(\sum_{x=1}^n\sum_{y=1}^n(L(x,y)-1)^2P_0(x)P_0(y)-\eta\right){}\notag\\
&&{}+\sum_{x=1}^n\left(\sum_{y=1}^nL(x,y)P_0(y)-1\right)\beta(x)P_0(x)+\sum_{y=1}^n\left(\sum_{x=1}^nL(x,y)P_0(x)-1\right)\gamma(y)P_0(y).\label{discrete Lagrangian}
\end{eqnarray}
Note that \eqref{discrete Lagrangian} is a function of $n^2$ variables, namely $L(x,y),x,y=1,\ldots,n$. The derivative of \eqref{discrete Lagrangian} with respect to $L(x,y)$ is
$$h(x,y)P_0(x)P_0(y)-2\alpha(L(x,y)-1)P_0(x)P_0(y)+\beta(x)P_0(x)P_0(y)+\gamma(y)P_0(x)P_0(y)$$
Setting it to zero, we have
$$h(x,y)-2\alpha L(x,y)+2\alpha+\beta(x)+\gamma(y)=0$$
which coincides with \eqref{heuristic}.}
\end{example}
\vspace{4mm}

Now, \eqref{heuristic} leads to
\begin{equation}
L=1+\frac{1}{2\alpha}(h(x,y)+\beta(x)+\gamma(y)). \label{interim L}
\end{equation}
Since $E_0[L]=1$, we must have $E_0[L]=1+\frac{1}{2\alpha}(E_0[h(X,Y)]+E_0\beta(X)+E_0\gamma(Y))=1$ or that
\begin{equation}
E_0\beta(X)+E_0\gamma(Y)=-E_0[h(X,Y)]. \label{eq1}
\end{equation}
Moreover, $E_0[L(X,Y)|x]=1$ and $E_0[L(X,Y)|y]=1$ give
$$1+\frac{1}{2\alpha}(E_0[h(X,Y)|x]+\beta(x)+E_0\gamma(Y))=1+\frac{1}{2\alpha}(E_0[h(X,Y)|y]+E_0\beta(X)+\gamma(y))=1$$
so that
\begin{align}
\beta(x)&=-E_0[h(X,Y)|x]-E_0\gamma(Y) \label{eq2}\\
\gamma(y)&=-E_0[h(X,Y)|y]-E_0\beta(Y). \label{eq3}
\end{align}
Therefore, substituting \eqref{eq1}, \eqref{eq2} and \eqref{eq3} into \eqref{interim L}, we get
\begin{align}
L(x,y)&=1+\frac{1}{2\alpha}(h(x,y)-E_0[h(X,Y)|x]-E_0\gamma(Y)-E_0[h(X,Y)|y]-E_0\beta(Y)) \notag\\
&=1+\frac{1}{2\alpha}(h(x,y)-E_0[h(X,Y)|x]-E_0[h(X,Y)|y]+E_0[h(X,Y)])\label{optimal L1}
\end{align}
which coincides with \eqref{optimal L}. The expression \eqref{optimal L1} is our candidate optimal solution, and our next task is to verify that it is indeed an optimal solution:

\proof{Proof of Proposition \ref{prop:characterization bivariate}.}
We need to prove that
\begin{equation}
E_0[hL]-\alpha E_0(L-1)^2\leq E_0[hL^*]-\alpha E_0(L^*-1)^2 \label{verification}
\end{equation}
for any $L\in\mathcal L$, where $L^*$ denotes the candidate optimal solution \eqref{optimal L}. To this end, we denote $\bar h=h-r$. It is straightforward to verify that $h$ possesses the orthogonal decomposition $h=\bar h+r$ on $\mathcal M^\perp\oplus\mathcal M=\mathcal L_2$, where $\mathcal M=\{V(X,Y)\in\mathcal L_2:E_0[V|X]=0,\ E_0[V|Y]=0\text{\ a.s.}\}\subset\mathcal L_2$ and $\mathcal L_2$ is the natural $L_2$-space under $P_0$. 
Then
\begin{align*}
E_0[hL^*]-\alpha E_0(L^*-1)^2&=E_0h+\frac{1}{2\alpha}E_0[hr]-\frac{1}{4\alpha}E_0r^2\\
&=E_0h+\frac{1}{2\alpha}Var_0(r)-\frac{1}{4\alpha}Var_0(r)\text{\ \ since $\bar h\perp r$ and $E_0r=0$}\\
&=E_0h+\frac{1}{4\alpha}Var_0(r).
\end{align*}
To prove \eqref{verification}, consider any $L\in\mathcal L$. For convenience, we write $U=L-1$ so that $E_0[U|X]=E_0[U|Y]=E_0U=0$ a.s.. Then we need to prove, for any such $U$,
\begin{eqnarray}
&&E_0[h(U+1)]-\alpha E_0U^2\leq E_0h+\frac{1}{4\alpha}Var_0(r)\notag\\
&\Leftrightarrow&E_0[hU]-\alpha E_0U^2\leq\frac{1}{4\alpha}Var_0(r)\notag\\
&\Leftrightarrow&E_0r^2-4\alpha E_0[hU]+4\alpha^2E_0U^2\geq0\notag\\
&\Leftrightarrow&E_0r^2-4\alpha E_0[rU]-4\alpha E_0[\bar hU]+4\alpha^2E_0U^2\geq0\notag\\
&\Leftrightarrow&E_0(r-2\alpha U)^2-4\alpha E_0[\bar hU]\geq0.\label{inequality verification}
\end{eqnarray}
But $E_0[\bar hU]=0$ since $U\in\mathcal M$. So \eqref{inequality verification} clearly holds.
\endproof


\proof{Proof of Proposition \ref{prop:max solution}.}
We use the notation in the proof of Proposition \ref{prop:characterization bivariate}. 
By the duality theorem in \cite{luenberger69}, Chapter 8 Theorem 1 (rewritten in Theorem \ref{thm:duality} in the Supplementary Materials), if one can find an $\alpha=\alpha^*\geq0$ such that there is an $L^*$ that maximizes \eqref{dual} and that $E_0(L^*-1)^2=\eta$, then $L^*$ must be the optimal solution for \eqref{max}. To find such $\alpha^*$ and $L^*$, note that Proposition \ref{prop:characterization bivariate} has already shown that $L^*$ defined in \eqref{optimal L} maximizes \eqref{dual} when $\alpha$ is large enough. With this $L^*$, $E_0(L^*-1)^2=\eta$ is equivalent to $\frac{1}{4{\alpha^*}^2}E_0r^2=\eta$
or
\begin{equation}
\frac{1}{4{\alpha^*}^2}Var_0(r)=\eta \label{complementary slackness1}
\end{equation}
since $E_0r=0$. Thus, when $\eta$ is small enough, we can find a large $\alpha^*$ and $L^*$ defined in \eqref{optimal L} such that \eqref{complementary slackness1} holds, or equivalently that $E_0(L^*-1)^2=\eta$ holds, and also that $L^*$ maximizes \eqref{dual}, implying that $L^*$ is an optimal solution for \eqref{max}. The optimal objective value is
\begin{align}
E_0[hL^*]&=E_0\left[h\left(1+\frac{r}{2\alpha^*}\right)\right]=E_0h+\frac{E_0[hr]}{2\alpha^*}=E_0h+\frac{E_0[r^2]}{2\alpha^*}=E_0h+\frac{Var_0(r)}{2\alpha^*}\label{objective}
\end{align}
since $h=\bar h+r$ with $\bar h\perp r$, and $E_0r=0$.
Combining with \eqref{complementary slackness1}, we get $E_0[hL^*]=E_0h+\sqrt{Var_0(r)\eta}$.

\endproof

\section{Sketch of Proof for Theorem \ref{thm:expansion}}\label{appendix:proof expansion}
%
%

As in the proof for the bivariate case, let us focus on the maximization formulation of \eqref{max P many}. First, we recast it in terms of likelihood ratio in the following form:
\begin{proposition}
The max formulation in \eqref{max P many} is equivalent to:
\begin{equation}
\begin{array}{ll}
\max&E_0\left[h(\mathbf{X}_T)\prod_{t=2}^TL(X_{t-1},X_t)\right]\\
\text{subject to}&E_0(L(X_{t-1},X_t)-1)^2\leq\eta\\
&E_0[L|X_{t-1}]=1,\ E_0[L|X_t]=1\text{\ a.s.}\\
&L\geq0\text{\ a.s.}
\end{array} \label{max many}
\end{equation}
The decision variable is $L(x_{t-1},x_t)$ (same for all $t$). \label{prop:max many equivalence}
\end{proposition}
The new feature in \eqref{max many}, compared to \eqref{max}, is the product form of the likelihood ratios in the objective function. Proposition \ref{prop:max many equivalence} is a consequence of the Markov property of $P_f$ and the martingale property of the products of $L$'s (see Appendix \ref{sec:proofs 1-lag} in the Supplementary Materials for the proof).

In contrast to \eqref{max}, the product form in the objective function of \eqref{max many} makes the posited optimization non-convex in general, and we shall use asymptotic analysis as a resolution. As in the bivariate case, we consider the Lagrangian of \eqref{max many}:
\begin{equation}
\max_{L\in\mathcal L^c(M)}E_0[h(\mathbf X_T)\underline L]-\alpha(E_0(L-1)^2-\eta) \label{Lagrangian many1}
\end{equation}
where for convenience we denote $\mathcal L^c(M)=\{L(X,Y)\geq0\text{\ a.s.}:E_0[L(X,Y)|X]=E_0[L(X,Y)|Y]=1\text{\ a.s.},\ E_0[L(X,Y)^2]\leq M\}$ for some $M>0$ large enough (the extra constraint $E_0[L(X,Y)^2]\leq M$ is a consequence of $E_0(L-1)^2\leq\eta$ for small enough $\eta$, and is required for technicality reason). We also denote $\underline L=\prod_{t=2}^TL(X_{t-1},X_t)$. For convenience, we denote $X$ and $Y$ as two i.i.d. random variables under the marginal distribution of $P_0$, and write $E_0[L|x]=E_0[L(X,Y)|X=x]$ and $E_0[L|y]=E_0[L(X,Y)|Y=y]$. To find a candidate local optimal solution, we relax the constraints $E_0[L|X]=E_0[L|Y]=1$ a.s. to get
\begin{equation}
\max_{L\in\mathcal L^+(M)}E_0[h(\mathbf X_T)\underline L]-\alpha(E_0(L(X,Y)-1)^2-\eta)+\int(E_0[L|x]-1)\beta(x)dP_0(x)+\int(E_0[L|y]-1)\gamma(y)dP_0(y) \label{Lagrangian many}
\end{equation}
where $\mathcal L^+(M)=\{L\geq0\text{\ a.s.}:E_0L^2\leq M\}$, and $\beta(X)$ and $\gamma(Y)$ are in the $L_2$-space under $P_0$. 
A similar differentiation argument as in \eqref{heuristic} in the bivariate case (but with the use of an additional heuristic product rule) gives
$$L(x,y)=1+\frac{1}{2\alpha}(H^{L}(x,y)+\beta(x)+\gamma(y))$$
where $H^{L}(x,y)$ is defined as
\begin{equation}
H^{L}(x,y)=\sum_{t=2}^TE_0[h(\mathbf X_T)\underline L_{2:T}^t|X_{t-1}=x,X_t=y] \label{HL1}
\end{equation}
and $\underline L_{2:T}^t=\prod_{\substack{s=2,\ldots,T\\s\neq t}}L(X_{s-1},X_s)$ is the leave-one-out product of likelihood ratios. 
Much like the proof of Proposition \ref{prop:characterization bivariate}, 
it can be shown that under the condition $E_0[L|X]=E_0[L|Y]=E_0L=1$ a.s., we can rewrite $\beta(x)$ and $\gamma(y)$ to get
\begin{equation}
L(x,y)=1+\frac{R^L(x,y)}{2\alpha}\label{optimal solution many}
\end{equation}
where
\begin{equation}
R^L(x,y)=H^{L}(x,y)-E_0[H^L(X,Y)|X=x]-E_0[H^L(X,Y)|Y=y]+E_0[H^{L}(X,Y)].\label{optimal residual}
\end{equation}
Here $E_0[H^L(X,Y)|X=x]$ and $E_0[H^L(X,Y)|Y=y]$ are the expectations of $H^L(X,Y)$ conditioned on $X$ and $Y$ which are i.i.d. each described by the marginal distribution of $P_0$.

Note that unlike Proposition \ref{prop:characterization bivariate}, $H^L(x,y)$ and $R^L(x,y)$ in \eqref{HL1} and \eqref{optimal residual} now depend on $L$ and hence \eqref{optimal solution many} does not explicitly specify $L$, but instead it is in the form of a fixed point equation. The key is to show that the $L$ defined as such is indeed optimal for \eqref{Lagrangian many1} for large enough $\alpha$:
\begin{proposition}
For any large enough $\alpha>0$, the $L^*$ that satisfies \eqref{optimal solution many} is an optimal solution of
\begin{equation}
\max_{L\in\mathcal L^c(M)}E_0[h(\mathbf X_T)\underline L]-\alpha E_0(L-1)^2. \label{inner maximization}
\end{equation}
\label{prop:characterization}
\end{proposition}
The proof of Proposition \ref{prop:characterization} is in Appendix \ref{sec:proofs 1-lag} in the Supplementary Materials. By invoking a suitable duality theorem (\cite{luenberger69}, Chapter 8 Theorem 1; also rewritten in Theorem \ref{thm:duality} in the Supplementary Materials), we can asymptotically expand the characterization of $L^*$ in Proposition \ref{prop:characterization} to arrive at Theorem \ref{thm:expansion}. Details are left to Appendix \ref{sec:proofs 1-lag} in the Supplementary Materials.

\bibliographystyle{ormsv080}

\bibliography{bibliography_dependency}

\begin{thebibliography}{47}
\expandafter\ifx\csname natexlab\endcsname\relax\def\natexlab#1{#1}\fi
\expandafter\ifx\csname url\endcsname\relax
  \def\url#1{{\tt #1}}\fi
\expandafter\ifx\csname urlprefix\endcsname\relax\def\urlprefix{URL }\fi
\expandafter\ifx\csname urlstyle\endcsname\relax
  \expandafter\ifx\csname doi\endcsname\relax
  \def\doi#1{doi:\discretionary{}{}{}#1}\fi \else
  \expandafter\ifx\csname doi\endcsname\relax
  \def\doi{doi:\discretionary{}{}{}\begingroup \urlstyle{rm}\Url}\fi \fi

\bibitem[{Asmussen and Glynn(2007)}]{asmussen2007stochastic}
Asmussen, S{\o}ren, Peter~W Glynn. 2007.
\newblock {\it Stochastic Simulation: Algorithms and Analysis\/}, vol.~57.
\newblock Springer.

\bibitem[{Bedford and Cooke(2002)}]{bedford2002vines}
Bedford, Tim, Roger~M Cooke. 2002.
\newblock Vines: A new graphical model for dependent random variables.
\newblock {\it Annals of Statistics\/}  1031--1068.

\bibitem[{Beirlant et~al.(1997)Beirlant, Dudewicz, Gy{\"o}rfi, and Van~der
  Meulen}]{beirlant1997nonparametric}
Beirlant, Jan, Edward~J Dudewicz, L{\'a}szl{\'o} Gy{\"o}rfi, Edward~C Van~der
  Meulen. 1997.
\newblock Nonparametric entropy estimation: An overview.
\newblock {\it International Journal of Mathematical and Statistical
  Sciences\/} {\bf 6}(1) 17--39.

\bibitem[{Ben-Tal et~al.(2013)Ben-Tal, Den~Hertog, De~Waegenaere, Melenberg,
  and Rennen}]{ben2013robust}
Ben-Tal, Aharon, Dick Den~Hertog, Anja De~Waegenaere, Bertrand Melenberg, Gijs
  Rennen. 2013.
\newblock Robust solutions of optimization problems affected by uncertain
  probabilities.
\newblock {\it Management Science\/} {\bf 59}(2) 341--357.

\bibitem[{Biller and Nelson(2005)}]{biller2005fitting}
Biller, Bahar, Barry~L Nelson. 2005.
\newblock Fitting time-series input processes for simulation.
\newblock {\it Operations Research\/} {\bf 53}(3) 549--559.

\bibitem[{Cario and Nelson(1996)}]{cario1996autoregressive}
Cario, Marne~C, Barry~L Nelson. 1996.
\newblock Autoregressive to anything: Time-series input processes for
  simulation.
\newblock {\it Operations Research Letters\/} {\bf 19}(2) 51--58.

\bibitem[{Cario and Nelson(1997)}]{cario1997modeling}
Cario, Marne~C, Barry~L Nelson. 1997.
\newblock Modeling and generating random vectors with arbitrary marginal
  distributions and correlation matrix.
\newblock Tech. rep., Citeseer.

\bibitem[{Chan and Tran(1992)}]{chan1992nonparametric}
Chan, Ngai~Hang, Lanh~Tat Tran. 1992.
\newblock Nonparametric tests for serial dependence.
\newblock {\it Journal of Time Series Analysis\/} {\bf 13}(1) 19--28.

\bibitem[{Cox and Reid(2002)}]{cox2002theory}
Cox, David~Roxbee, Nancy Reid. 2002.
\newblock {\it The Theory of the Design of Experiments\/}.
\newblock CRC Press.

\bibitem[{Cram{\'e}r(1999)}]{cramer1999mathematical}
Cram{\'e}r, Harald. 1999.
\newblock {\it Mathematical Methods of Statistics\/}, vol.~9.
\newblock Princeton university press.

\bibitem[{Delage and Ye(2010)}]{delage2010distributionally}
Delage, Erick, Yinyu Ye. 2010.
\newblock Distributionally robust optimization under moment uncertainty with
  application to data-driven problems.
\newblock {\it Operations Research\/} {\bf 58}(3) 595--612.

\bibitem[{Fan and Yao(2003)}]{fan2003nonlinear}
Fan, Jianqing, Qiwei Yao. 2003.
\newblock {\it Nonlinear Time Series: Nonparametric and Parametric Methods\/}.
\newblock Springer Science \& Business Media.

\bibitem[{Fillmore and Williams(1971)}]{fillmore1971operator}
Fillmore, PA, JP~Williams. 1971.
\newblock On operator ranges.
\newblock {\it Advances in Mathematics\/} {\bf 7}(3) 254--281.

\bibitem[{Ghosh and Henderson(2002)}]{ghosh2002chessboard}
Ghosh, Soumyadip, Shane~G Henderson. 2002.
\newblock Chessboard distributions and random vectors with specified marginals
  and covariance matrix.
\newblock {\it Operations Research\/} {\bf 50}(5) 820--834.

\bibitem[{Glasserman and Xu(2013)}]{glasserman2013robust}
Glasserman, Paul, Xingbo Xu. 2013.
\newblock Robust portfolio control with stochastic factor dynamics.
\newblock {\it Operations Research\/} {\bf 61}(4) 874--893.

\bibitem[{Glasserman and Xu(2014)}]{glasserman2012robust}
Glasserman, Paul, Xingbo Xu. 2014.
\newblock Robust risk measurement and model risk.
\newblock {\it Quantitative Finance\/} {\bf 14}(1) 29--58.

\bibitem[{Goh and Sim(2010)}]{goh2010distributionally}
Goh, Joel, Melvyn Sim. 2010.
\newblock Distributionally robust optimization and its tractable
  approximations.
\newblock {\it Operations Research\/} {\bf 58}(4-Part-1) 902--917.

\bibitem[{Granger and Lin(1994)}]{granger1994using}
Granger, Clive, Jin-Lung Lin. 1994.
\newblock Using the mutual information coefficient to identify lags in
  nonlinear models.
\newblock {\it Journal of time series analysis\/} {\bf 15}(4) 371--384.

\bibitem[{Granger et~al.(2004)Granger, Maasoumi, and
  Racine}]{granger2004dependence}
Granger, CW, Esfandiar Maasoumi, Jeffrey Racine. 2004.
\newblock A dependence metric for possibly nonlinear processes.
\newblock {\it Journal of Time Series Analysis\/} {\bf 25}(5) 649--669.

\bibitem[{Hampel(1974)}]{hampel1974influence}
Hampel, Frank~R. 1974.
\newblock The influence curve and its role in robust estimation.
\newblock {\it Journal of the American Statistical Association\/} {\bf 69}(346)
  383--393.

\bibitem[{Hampel et~al.(2011)Hampel, Ronchetti, Rousseeuw, and
  Stahel}]{hampel2011robust}
Hampel, Frank~R, Elvezio~M Ronchetti, Peter~J Rousseeuw, Werner~A Stahel. 2011.
\newblock {\it Robust Statistics: The Approach Based on Influence Functions\/},
  vol. 114.
\newblock Wiley. com.

\bibitem[{Hansen and Sargent(2008)}]{hansen2008robustness}
Hansen, Lars~Peter, Thomas~J Sargent. 2008.
\newblock {\it Robustness\/}.
\newblock Princeton university press.

\bibitem[{Ibrahim et~al.(2012)Ibrahim, Regnard, L'Ecuyer, and
  Shen}]{ibrahim2012modeling}
Ibrahim, Rouba, Nazim Regnard, Pierre L'Ecuyer, Haipeng Shen. 2012.
\newblock On the modeling and forecasting of call center arrivals.
\newblock {\it Proceedings of the Winter Simulation Conference\/}. IEEE, 23.

\bibitem[{Iyengar(2005)}]{iyengar2005robust}
Iyengar, Garud~N. 2005.
\newblock Robust dynamic programming.
\newblock {\it Mathematics of Operations Research\/} {\bf 30}(2) 257--280.

\bibitem[{Krishnamurthy et~al.(2014)Krishnamurthy, Kandasamy, Poczos, and
  Wasserman}]{krishnamurthy2014nonparametric}
Krishnamurthy, Akshay, Kirthevasan Kandasamy, Barnabas Poczos, Larry Wasserman.
  2014.
\newblock Nonparametric estimation of renyi divergence and friends.
\newblock {\it arXiv preprint arXiv:1402.2966\/} .

\bibitem[{Lam(2013)}]{lam2013robust}
Lam, Henry. 2013.
\newblock Robust sensitivity analysis for stochastic systems.
\newblock {\it arXiv preprint arXiv:1303.0326\/} .

\bibitem[{Liu et~al.(2012)Liu, Wasserman, and Lafferty}]{liu2012exponential}
Liu, Han, Larry Wasserman, John~D Lafferty. 2012.
\newblock Exponential concentration for mutual information estimation with
  application to forests.
\newblock {\it Advances in Neural Information Processing Systems\/}.
  2537--2545.

\bibitem[{Livny et~al.(1993)Livny, Melamed, and Tsiolis}]{livny1993impact}
Livny, Miron, Benjamin Melamed, Athanassios~K Tsiolis. 1993.
\newblock The impact of autocorrelation on queuing systems.
\newblock {\it Management Science\/} {\bf 39}(3) 322--339.

\bibitem[{Luenberger(1969)}]{luenberger69}
Luenberger, D.~G. 1969.
\newblock {\it Optimization by Vector Space Methods\/}.
\newblock Wiley.

\bibitem[{Mai and Scherer(2012)}]{mai2012simulating}
Mai, Jan-Frederik, Matthias Scherer. 2012.
\newblock {\it Simulating Copulas: Stochastic Models, Sampling Algorithms, and
  Applications\/}, vol.~4.
\newblock World Scientific.

\bibitem[{Melamed(1991)}]{melamed1991tes}
Melamed, Benjamin. 1991.
\newblock {TES}: A class of methods for generating autocorrelated uniform
  variates.
\newblock {\it ORSA Journal on Computing\/} {\bf 3}(4) 317--329.

\bibitem[{Melamed et~al.(1992)Melamed, Hill, and Goldsman}]{melamed1992tes}
Melamed, Benjamin, Jon~R Hill, David Goldsman. 1992.
\newblock The {TES} methodology: Modeling empirical stationary time series.
\newblock {\it Proceedings of the 24th Winter Simulation Conference\/}. ACM,
  135--144.

\bibitem[{Mitchell et~al.(1977)Mitchell, Paulson, and
  Beswick}]{mitchell1977effect}
Mitchell, CR, AS~Paulson, CA~Beswick. 1977.
\newblock The effect of correlated exponential service times on single server
  tandem queues.
\newblock {\it Naval Research Logistics Quarterly\/} {\bf 24}(1) 95--112.

\bibitem[{Moon and Hero(2014)}]{moon2014multivariate}
Moon, Kevin, Alfred Hero. 2014.
\newblock Multivariate f-divergence estimation with confidence.
\newblock {\it Advances in Neural Information Processing Systems\/}.
  2420--2428.

\bibitem[{Nguyen et~al.(2007)Nguyen, Wainwright, and
  Jordan}]{nguyen2007estimating}
Nguyen, XuanLong, Martin~J Wainwright, Michael~I Jordan. 2007.
\newblock Estimating divergence functionals and the likelihood ratio by
  penalized convex risk minimization.
\newblock {\it Advances in Neural Information Processing Systems\/}.

\bibitem[{Nilim and El~Ghaoui(2005)}]{nilim2005robust}
Nilim, Arnab, Laurent El~Ghaoui. 2005.
\newblock Robust control of markov decision processes with uncertain transition
  matrices.
\newblock {\it Operations Research\/} {\bf 53}(5) 780--798.

\bibitem[{P{\'a}l et~al.(2010)P{\'a}l, P{\'o}czos, and
  Szepesv{\'a}ri}]{pal2010estimation}
P{\'a}l, D{\'a}vid, Barnab{\'a}s P{\'o}czos, Csaba Szepesv{\'a}ri. 2010.
\newblock Estimation of {R}{\'e}nyi entropy and mutual information based on
  generalized nearest-neighbor graphs.
\newblock {\it Advances in Neural Information Processing Systems\/}.
  1849--1857.

\bibitem[{Pang and Whitt(2012{\natexlab{a}})}]{pang2012impact}
Pang, Guodong, Ward Whitt. 2012{\natexlab{a}}.
\newblock The impact of dependent service times on large-scale service systems.
\newblock {\it Manufacturing \& Service Operations Management\/} {\bf 14}(2)
  262--278.

\bibitem[{Pang and Whitt(2012{\natexlab{b}})}]{pang2012infinite}
Pang, Guodong, Ward Whitt. 2012{\natexlab{b}}.
\newblock Infinite-server queues with batch arrivals and dependent service
  times.
\newblock {\it Probability in the Engineering and Informational Sciences\/}
  {\bf 26}(02) 197--220.

\bibitem[{Petersen et~al.(2000)Petersen, James, and
  Dupuis}]{petersen2000minimax}
Petersen, Ian~R, Matthew~R James, Paul Dupuis. 2000.
\newblock Minimax optimal control of stochastic uncertain systems with relative
  entropy constraints.
\newblock {\it IEEE Transactions on Automatic Control\/} {\bf 45}(3) 398--412.

\bibitem[{P{\'o}czos and Schneider(2011)}]{poczos2011estimation}
P{\'o}czos, Barnab{\'a}s, Jeff~G Schneider. 2011.
\newblock On the estimation of alpha-divergences.
\newblock {\it International Conference on Artificial Intelligence and
  Statistics\/}. 609--617.

\bibitem[{P{\'o}czos and Schneider(2012)}]{poczos2012nonparametric}
P{\'o}czos, Barnab{\'a}s, Jeff~G Schneider. 2012.
\newblock Nonparametric estimation of conditional information and divergences.
\newblock {\it International Conference on Artificial Intelligence and
  Statistics\/}. 914--923.

\bibitem[{Robinson(1991)}]{robinson1991consistent}
Robinson, Peter~M. 1991.
\newblock Consistent nonparametric entropy-based testing.
\newblock {\it The Review of Economic Studies\/} {\bf 58}(3) 437--453.

\bibitem[{Serfling(2009)}]{serfling2009approximation}
Serfling, Robert~J. 2009.
\newblock {\it Approximation Theorems of Mathematical Statistics\/}, vol. 162.
\newblock John Wiley \& Sons.

\bibitem[{Tsybakov(2008)}]{tsybakov2008introduction}
Tsybakov, Alexandre~B. 2008.
\newblock {\it Introduction to Nonparametric Estimation\/}.
\newblock Springer Science \& Business Media.

\bibitem[{Van~Es(1992)}]{van1992estimating}
Van~Es, Bert. 1992.
\newblock Estimating functionals related to a density by a class of statistics
  based on spacings.
\newblock {\it Scandinavian Journal of Statistics\/}  61--72.

\bibitem[{Ware et~al.(1998)Ware, Page~Jr, and Nelson}]{ware1998automatic}
Ware, Peter~P, Thomas~W Page~Jr, Barry~L Nelson. 1998.
\newblock Automatic modeling of file system workloads using two-level arrival
  processes.
\newblock {\it ACM Transactions on Modeling and Computer Simulation (TOMACS)\/}
  {\bf 8}(3) 305--330.

\end{thebibliography}

\newpage

{\large\noindent\textbf{Supplementary Materials}}



\section{Proofs for the One-Dependence Case}
\subsection{Proof of Proposition \ref{prop:max many equivalence}} \label{sec:proofs 1-lag}
We prove in both directions. First, we show that each feasible solution in \eqref{max P many} can be re-expressed as a feasible solution in \eqref{max many} that has the same objective value. Given $P_f$ that satisfies $P_f(x_t)=P_0(x_t)$, we let $L_t(x_{t-1},x_t)=dP_f(x_t|x_{t-1})/dP_0(x_t)=dP_f(x_{t-1},x_t)/(dP_0(x_{t-1})dP_f(x_t))$ be the likelihood ratio of a transition step under $P_f$ w.r.t.~the baseline. Since $P_f(x_t)=P_0(x_t)$, we have $L_t(x_{t-1},x_t)=dP_f(x_{t-1},x_t)/(dP_0(x_{t-1})dP_0(x_t))=dP_f(x_{t-1},x_t)/(dP_f(x_{t-1})dP_f(x_t))$. Also, since by stationarity $P_f(x_{t-1},x_t)=P_f(x_{s-1},x_s)$ for all $s,t$, $L_t(\cdot,\cdot)$ are all the same, so we can let $L$ be the identical likelihood ratio. Then
\begin{align*}
E_f[h(\mathbf{X}_T)]&=\int h(\mathbf{x}_T)dP_f(x_1)dP_f(x_2|x_1)\cdots dP_f(x_T|x_{T-1})\\
&=\int h(\mathbf{x}_T)\frac{dP_f(x_1)}{dP_0(x_1)}\frac{dP_f(x_2|x_1)}{dP_0(x_2)}\cdots\frac{dP_f(x_T|x_{T-1})}{dP_0(x_T)}dP_0(x_1)\cdots dP_0(x_T)\\
&=\int h(\mathbf{x}_T)L(x_1,x_2)\cdots L(x_{T-1},x_T)dP_0(x_1)\cdots dP_0(x_T)\\
&=E_0\left[h(\mathbf{X}_T)\prod_{t=2}^TL(X_{t-1},X_t)\right].
\end{align*}
Hence the objectives are the same. To show that $L$ is feasible, first note that $\phi^2(P_f(x,y))=E(L(X,Y)-1)^2$ by the definition of $L$, so $\phi^2(P_f(x,y))\leq\eta$ implies $E(L(X,Y)-1)^2\leq\eta$. Next, by the constraint that $P_f(X_{t-1})=P_0(X_{t-1})$, we have
$$P_f(X_{t-1}\in A)=E_0[L(X_{t-1},X_t)I(X_{t-1}\in A)]=E_0[E_0[L(X_{t-1},X_t)|X_{t-1}]I(X_{t-1}\in A)]=P_0(X_{t-1}\in A)$$
for any measurable $A$, and so we must have $E_0[L(X_{t-1},X_t)|X_{t-1}]=1$. Similarly, $E_0[L(X_{t-1},X_t)|X_t]=1$. Hence $L$ satisfies all the constraints in \eqref{max many}.

To prove the other direction, given any $L$ in the feasible set of \eqref{max many}, define the Markov transition kernel $dP_f(x_t|x_{t-1})=L(x_{t-1},x_t)dP_0(x_t)$ for any $t=2,\ldots,T$, and $P_f(x_1)=P_0(x_1)$. Note that $dP_f(x_t|x_{t-1})$ is well-defined since
$$\int dP_f(x_t|x_{t-1})=\int L(x_{t-1},x_t)dP_0(x_t)=E_0[L(X_{t-1},X_t)|X_{t-1}=x_{t-1}]=1$$
a.s., by the first marginal constraint in \eqref{max many}. Now
\begin{align*}
E_0\left[h(\mathbf{X}_T)\prod_{t=2}^TL(X_{t-1},X_t)\right]&=\int h(\mathbf{x}_T)dP_0(x_1)\prod_{t=2}^TL(x_{t-1},x_t)dP_0(x_t)\\
&=\int h(\mathbf{x}_T)dP_f(x_1)\prod_{t=2}^TdP_f(x_t|x_{t-1})\\
&=E_f[h(\mathbf{X}_T)].
\end{align*}
Hence the objective is matched. Next we prove $P_f(x_t)=P_0(x_t)$ by induction. Note that $P_f(X_1\in A)=P_0(X_1\in A)$ for any measurable $A$ by definition. Assume $P_f(x_{t-1})=P_0(x_{t-1})$. Then
\begin{align*}
P_f(X_t\in A)&=\int_{x_t\in A}dP_f(x_t|x_{t-1})dP_f(x_{t-1})\\
&=\int_{x_t\in A}L(x_{t-1},x_t)dP_0(x_t)dP_0(x_{t-1})\\
&=\int_{x_t\in A}E_0[L(X_{t-1},X_t)|X_t=x_t]dP_0(x_t)\\
&=P_0(X_t\in A)
\end{align*}
for any measurable $A$, by the second marginal constraint in \eqref{max many}. Hence $P_f(x_t)=P_0(x_t)$, and we conclude our induction.

Finally, note that $dP_f(X_t|X_{t-1})=L(X_{t-1},X_t)dP_0(X_t)$ implies $L=dP_f(X_t|X_{t-1})/dP_0(X_t)=dP_f(X_{t-1},X_t)/(dP_0(X_{t-1})dP_0(X_t))=dP_f(X_{t-1},X_t)/(dP_f(X_{t-1})dP_f(X_t))$. Hence $E_0(L(X,Y)-1)^2\leq\eta$ implies $\phi^2(P_f(X,Y))\leq\eta$.

\subsection{Proof of Proposition \ref{prop:characterization}}
To make things clear, we first explain how the space $\mathcal L^c(M)$ arises in the proposition. This stems from the fact that \eqref{max many} is equivalent to
\begin{equation}
\renewcommand\arraystretch{1.4}
\begin{array}{ll}
\max&E_0\left[h(\mathbf{X}_T)\prod_{t=2}^TL(X_{t-1},X_t)\right]\\
\mbox{subject to}&E_0(L(X_{t-1},X_t)-1)^2\leq\eta\\
&E_0[L|X_{t-1}]=1,\ E_0[L|X_t]=1\mbox{\ a.s.}\\
&L(X_{t-1},X_t)\geq0\mbox{\ a.s.}\\
&E_0[L(X_{t-1},X_t)^2]\leq M
\end{array} \label{max many M}
\end{equation}
for some constant $M>1$. This can be seen easily by noting that the first constraint in \eqref{max many} implies that $E_0L^2\leq1+\eta$, and so we can choose some large enough $M$ so that the last constraint in \eqref{max many M} holds for any small enough $\eta$.
Thus \eqref{inner maximization} in Proposition \ref{prop:characterization} is the Lagrangian relaxation of the first constraint in \eqref{max many M}.

Next, define $\underline L_{s:t}=\prod_{k=s}^tL(X_{k-1},X_k)$ for any $2\leq s\leq t\leq T$. Also, for any $1\leq s\leq t\leq T$, define $\mathcal F_{s:t}=\mathcal F(X_s,\ldots,X_t)$ as the $\sigma$-algebra generated by $\{X_s,\ldots,X_t\}$. We point out the following properties of $\underline L_{s:t}$ viewed as a process: 
\begin{lemma}
\begin{enumerate}
\item Under $P_0$, fix the starting point of time $2\leq s\leq T$ under stationarity. The process $\{\underline L_{s:t}:t=s-1,s,s+1,\ldots\}$ (where we define $\underline L_{s:(s-1)}=1$) is a martingale adapted to $\{\mathcal{F}_{(s-1):t}:t=s-1,s,s+1,\ldots\}$, with an initial value identical to 1. A similar property holds for the time-reverse setting. Namely, under $P_0$, and fixing the starting time $2\leq t\leq T$, the process $\{\underline L_{s:t}:s=t+1,t,t-1,\ldots\}$ (where we now define $\underline L_{(t+1):t}=1$) is a martingale adapted to $\{\mathcal F_{(s-1):t}:s=t+1,t,t-1,\ldots\}$ with initial value 1.

\item For any measurable $g(\mathbf X_{s:t}):=g(X_s,\ldots,X_t)$, for $s\leq t$, we have
$$E_f[g(\mathbf X_{s:t})]=E_0[g(\mathbf X_{s:t})\underline L_{2:t}]=E_0[g(\mathbf X_{s:t})\underline L_{(s+1):T}].$$
\end{enumerate}
Both properties can be generalized to the scenario when the factors $L$ in the likelihood ratio product $\underline L_{s:t}$ are non-identical. \label{lemma:properties}
\end{lemma}

\proof{Proof of Lemma \ref{lemma:properties}.}
To prove the first part, consider, for any $t\geq s$,
\begin{eqnarray*}
E_0[\underline L_{s:(t+1)}|\mathcal F_{(s-1):t}]&=&E_0[L(X_{s-1},X_s)\cdots L(X_t,X_{t+1})|\mathcal F_{(s-1):t}]\\
&=&L(X_{s-1},X_s)\cdots L(X_{t-1},X_t)E_0[L(X_t,X_{t+1})|\mathcal F_{(s-1):t}]\\
&=&L(X_{s-1},X_s)\cdots L(X_{t-1},X_t)\\
&&\mbox{\ \ since $E_0[L(X_t,X_{t+1})|\mathcal F_{(s-1):t}]=E_0[L(X_t,X_{t+1})|X_t]=1$}\\
&&\mbox{\ \ by the first marginal constraint in \eqref{max many}}\\
&=&\underline L_{s:t}.
\end{eqnarray*}
The backward direction follows analogously, by using the second marginal constraint in \eqref{max many}. The second part of the lemma follows immediately from the forward and backward martingale properties of $\underline L_{s:t}$. Also, the generalization to non-identical $L$'s follows trivially.
\endproof

The proof of Proposition \ref{prop:characterization} relies on a contraction operator that we denote $\mathcal K:\mathcal L^c(M)^{T-2}\to\mathcal L^c(M)^{T-2}$. First, let us define a generalized notion of $H^L$, defined in \eqref{HL1}, to cover the case where the factors in $\underline L$ are not necessarily the same. For convenience, we write $h=h(\mathbf X_T)$. For any $\mathbf L^{(1):(T-2)}:=(L^{(1)},L^{(2)},\ldots,L^{(T-2)})\in\mathcal L^c(M)^{T-2}$, let
\begin{equation}
G^{\mathbf L^{(1):(T-2)}}(x,y)=\sum_{t=2}^T\sum_{\mathbf r\in\mathcal S_{1:(T-2)}}E_0\left[h\prod_{\substack{s=2,\ldots,T\\s\neq t}}L^{(r_s^t)}(X_{s-1},X_s)\Bigg|X_{t-1}=x,X_t=y\right]\label{G}
\end{equation}
where $\mathbf r^t=(r_2^t,r_3^t,\ldots,r_{t-1}^t,r_{t+1}^t,\ldots,r_T^t)$ is a vector of length $T-2$ and $\mathcal S_{1:(T-2)}$ is the symmetric group of all permutations of $(1,\ldots,T-2)$. In other words, \eqref{G} is the sum over all conditional expectations given each consecutive pairs of states $(X_{t-1},X_t)$, and for each conditioning, the sum is taken of all permutations of $L^{(1)}(\cdot,\cdot),\cdots,L^{(T-2)}(\cdot,\cdot)$ applied to all other consecutive pairs of $(X_{s-1},X_s),s\neq t$.
Note that when $L^{(k)}$'s are all identical, then $G^{\mathbf L^{(1):(T-2)}}$ reduces to $(T-2)!H^L$.

With the definition above, we define a mapping $K:\mathcal L^c(M)^{T-2}\to\mathcal L^c(M)$ as
\begin{equation}
K(\mathbf L^{(1):(T-2)})(x,y)=1+\frac{R^{\mathbf L^{(1):(T-2)}}(x,y)}{2(T-2)!\alpha} \label{K}
\end{equation}
where $R^{\mathbf L^{(1):(T-2)}}=G^{\mathbf L^{(1):(T-2)}}(x,y)-E_0[G^{\mathbf L^{(1):(T-2)}}|x]-E_0[G^{\mathbf L^{(1):(T-2)}}|y]+E_0[G^{\mathbf L^{(1):(T-2)}}]$, 
with $E_0[G^{\mathbf L^{(1):(T-2)}}|x]=E_0[G^{\mathbf L^{(1):(T-2)}}(X,Y)|X=x]$ and $E_0[G^{\mathbf L^{(1):(T-2)}}|y]=E_0[G^{\mathbf L^{(1):(T-2)}}(X,Y)|Y=y]$ for $X$, $Y$ i.i.d. each under the marginal distribution of $P_0$.

The operator $\mathcal K$ is now defined as follows. Given $\mathbf L^{(1):(T-2)}=(L^{(1)},L^{(2)},\ldots,L^{(T-2)})\in\mathcal L^c(M)^{T-2}$, define
\begin{align}
\tilde L^{(1)}&=K(L^{(1)},L^{(2)},\ldots,L^{(T-2)})\notag\\
\tilde L^{(2)}&=K(\tilde L^{(1)},L^{(2)},\ldots,L^{(T-2)})\notag\\
\tilde L^{(3)}&=K(\tilde L^{(1)},\tilde L^{(2)},L^{(3)},\ldots,L^{(T-2)})\notag\\
&\vdots\notag\\
\tilde L^{(T-2)}&=K(\tilde L^{(1)},\tilde L^{(2)},\ldots,\tilde L^{(T-1)},L^{(T-2)}).\label{iteration}
\end{align}
Then $\mathcal K(\mathbf L^{(1:(T-2))})=(\tilde L^{(1)},\tilde L^{(2)},\ldots,\tilde L^{(T-2)})$.

The $\mathcal K$ constructed above is a well-defined contraction operator:
\begin{lemma}
For large enough $\alpha$, the operator $\mathcal K$ is a well-defined, closed contraction map on $\mathcal L^c(M)^{T-2}$, with the metric $d(\mathbf L,\mathbf L')=\max_{k=1,\ldots,T-2}\|L^{(k)}-L^{(k)'}\|_2$, where $\mathbf L=(L^{(k)})_{k=1,\ldots,T-2}$ and $\mathbf L'=(L^{(k)'})_{k=1,\ldots,T-2}$, and $\|\cdot\|_2$ is the $\mathcal L_2$-norm under $P_0$. As a result $\mathcal K$ possesses a unique fixed point $\mathbf L^*\in\mathcal L^c(M)^{T-2}$. Moreover, all $T-2$ components of $\mathbf L^*$ are identical and satisfy \eqref{optimal solution many}. \label{lemma:contraction}
\end{lemma}

\proof{Proof of Lemma \ref{lemma:contraction}.}
We divide the proofs into three components: first, we show that $\mathcal K$ is well-defined and closed; second, we prove that $\mathcal K$ is a contraction map; third, we show that the fixed point of $\mathcal K$ has all identical components.

We start by showing well-definedness and closedness. Note that each summand in the definition of $G^{\mathbf L^{(1):(T-2)}}(x,y)$ satisfies
\begin{eqnarray*}
&&|E_0[hL^{(r_2^t)}(X_1,X_2)L^{(r_3^t)}(X_2,X_3)\cdots L^{(r_{t-1}^t)}(X_{t-2},X_{t-1})L^{(r_{t+1}^t)}(X_t,X_{t+1})L^{(r_{t+2}^t)}(X_{t+1},X_{t+2}){}\\
&&{}\cdots L^{(r_T^t)}(X_{T-1},X_T)|X_{t-1}=x,X_t=y]|\\
&\leq&CE_0[L^{(r_2^t)}(X_1,X_2)L^{(r_3^t)}(X_2,X_3)\cdots L^{(r_{t-1}^t)}(X_{t-2},X_{t-1})L^{(r_{t+1}^t)}(X_t,X_{t+1})L^{(r_{t+2}^t)}(X_{t+1},X_{t+2}){}\\
&&{}\cdots L^{(r_T^t)}(X_{T-1},X_T)|X_{t-1}=x,X_t=y]\mbox{\ \ for some $C>0$}\\
&=&CE_0[L^{(r_2^t)}(X_1,X_2)L^{(r_3^t)}(X_2,X_3)\cdots L^{(r_{t-1}^t)}(X_{t-2},X_{t-1})|X_{t-1}=x]{}\\
&&{}E_0[L^{(r_{t+1}^t)}(X_t,X_{t+1})L^{(2)}(X_{t+1},X_{t+2})\cdots L^{(r_T^t)}(X_{T-1},X_T)|X_t=y]\\
&=&C\mbox{\ \ by the forward and backward martingale properties of Lemma \ref{lemma:properties}}.
\end{eqnarray*}
Hence
\begin{equation}
|G^{\mathbf L^{(1):(T-2)}}(x,y)|\leq(T-1)!C\label{interim G}
\end{equation}
and the mapping $K$ satisfies
$$|K(\mathbf L^{(1):(T-2)})-1|\leq\frac{4(T-1)C}{2\alpha}<\infty.$$
The same bound holds for each $K(\tilde L^{(1)},\tilde L^{(2)},\ldots,\tilde L^{(t)},L^{(t+1)},\ldots,L^{(T-2)})$ for $t=1,\ldots,T-3$. Hence each component of $\mathcal K(\mathbf L^{(1):(T-2)})$ is finite a.s..

Moreover, using \eqref{interim G}, for large enough $\alpha$, we have
$$|G^{\mathbf L^{(1):(T-2)}}(x,y)-E_0[G^{\mathbf L^{(1):(T-2)}}|x]-E_0[G^{\mathbf L^{(1):(T-2)}}|y]+E_0G^{\mathbf L^{(1):(T-2)}}|\leq4(T-1)!C\leq2(T-2)!\alpha$$
for any $\mathbf L^{(1):(T-2)}\in{\mathcal L^c(M)}^{T-2}$ and so each component of $\mathcal K(\mathbf L^{(1):(T-2)})$ is non-negative a.s.. Note that by construction $K(\mathbf L^{(1):(T-2)})$ satisfies $E_0[K(\mathbf L^{(1):(T-2)})|x]=E_0[K(\mathbf L^{(1):(T-2)})|y]=1$ a.s., and similarly along the iteration of $K$ in \eqref{iteration}.

To conclude that $\mathcal K$ is closed, we are left to show that $\mathcal K$ preserves the boundedness in the $\mathcal L_2$-norm. Note that when $\alpha$ is large enough,
\begin{eqnarray*}
&&E_0K(\mathbf L^{(1):(T-2)})^2-1=E_0(K(\mathbf L^{(1):(T-2)})-1)^2\\
&=&\frac{1}{4((T-2)!)^2\alpha^2}E_0\left(G^{\mathbf L^{(1):(T-2)}}(X,Y)-E_0[G^{\mathbf L^{(1):(T-2)}}|X]-E_0[G^{\mathbf L^{(1):(T-2)}}|Y]+E_0G^{\mathbf L^{(1):(T-2)}}\right)^2\\
&\leq&\frac{(4(T-1)C)^2}{4\alpha^2}\\
&\leq&M-1
\end{eqnarray*}
for any $\mathbf L^{(1):(T-2)}\in{\mathcal L^c(M)}^{T-2}$, and we have $K(\mathbf L^{(1):(T-2)})\in\mathcal L^c(M)$. Iterating through $K(\tilde L^{(1)},\tilde L^{(2)},\ldots,\tilde L^{(t)},L^{(t+1)},\ldots,L^{(T-2)})$ for $t=1,\ldots,T-3$ using \eqref{iteration}, we conclude that $\mathcal K$ is closed in ${\mathcal L^c(M)}^{T-2}$.

Now we show that $\mathcal K$ is a contraction. Consider, for any $\mathbf L^{(1):(T-2)}=(L^{(1)},\ldots,L^{(T-2)})\in{\mathcal L^c(M)}^{T-2}$ and $\mathbf L^{(1):(T-2)'}=(L^{(1)'},\ldots,L^{(T-2)'})\in{\mathcal L^c(M)}^{T-2}$,
\begin{eqnarray}
&&E_0|K(\mathbf L^{(1):(T-2)})-K(\mathbf L^{(1):(T-2)'})|^2\notag\\
&\leq&E_0|G^{\mathbf L^{(1):(T-2)}}-G^{\mathbf L^{(1):(T-2)'}}|^2\label{interim9}
\end{eqnarray}
since $R^{\mathbf L^{(1):(T-2)}}(x,y)$ is the orthogonal projection of $G^{\mathbf L^{(1):(T-2)}}(x,y)$ onto the space $\{V(X,Y)\in\mathcal L_2:E_0[V|X]=0,E_0[V|Y]=0\text{\ a.s.}\}$, and hence is a contraction. We have \eqref{interim9} less than or equal to
\begin{equation}
\frac{C}{4((T-2)!)^2\alpha^2}E_0[U(X,Y)^2]\label{interim10}
\end{equation}
for some constant $C>0$, where $X$, $Y$ are i.i.d. under the marginal distribution of $P_0$, $U(x,y)=\sum_{t=2}^T\sum_{\mathbf r^t\in\mathcal S_{1:(T-2)}}E_0[|\underline L^{(\mathbf r^t)}-\underline L^{(\mathbf r^t)'}||X_{t-1}=x,X_t=y]$ and
$$\underline L^{(\mathbf r^t)}=\prod_{\substack{s=2,\ldots,T\\s\neq t}}L^{(r_s^t)}(X_{s-1},X_s),\ \ \underline L^{(\mathbf r^t)'}=\prod_{\substack{s=2,\ldots,T\\s\neq t}}L^{(r_s^t)'}(X_{s-1},X_s).$$
By Jensen's inequality, \eqref{interim10} is further bounded above by
\begin{equation}
\frac{(T-1)^2C}{4\alpha^2}\sum_{t=2}^T\sum_{\mathbf r^t\in\mathcal S_{1:(T-2)}}E_0\left(E_0[|\underline L^{(\mathbf r^t)}-\underline L^{(\mathbf r^t)'}||X_{t-1},X_t]\right)^2.\label{interim4}
\end{equation}
%
%

Let us focus on $E_0\left(E_0[|\underline L^{(\mathbf r^t)}-\underline L^{(\mathbf r^t)'}||X_{t-1},X_t]\right)^2$. Note that, by telescoping, we have
\begin{eqnarray*}
&&\underline L^{(\mathbf r^t)}-\underline L^{(\mathbf r^t)'}\\
&=&\left(L^{(r_2^t)}(X_1,X_2)-L^{(r_2^t)'}(X_1,X_2)\right)L^{(r_3^t)}(X_2,X_3)\cdots L^{(r_{t-1}^t)}(X_{t-2},X_{t-1})L^{(r_{t+1}^t)}(X_t,X_{t+1}){}\\
&&{}L^{(r_{t+2}^t)}(X_{t+1},X_{t+2})\cdots L^{(r_T^t)}(X_{T-1},X_T){}\\
&&{}+L^{(r_2^t)'}(X_1,X_2)\bigg(L^{(r_3^t)}(X_2,X_3)\cdots L^{(r_{t-1}^t)}(X_{t-2},X_{t-1})L^{(r_{t+1}^t)}(X_t,X_{t+1})L^{(r_{t+2}^t)}(X_{t+1},X_{t+2}){}\\
&&{}\cdots L^{(r_T^t)}(X_{T-1},X_T){}\\
&&{}-L^{(r_3^t)'}(X_2,X_3)\cdots L^{(r_{t-1}^t)'}(X_{t-2},X_{t-1})L^{(r_{t+1}^t)'}(X_t,X_{t+1})L^{(r_{t+2}^t)'}(X_{t+1},X_{t+2})\cdots L^{(r_T^t)'}(X_{T-1},X_T)\bigg)\\
&\vdots&\\
&=&\sum_{\substack{s=2,\ldots,T\\s\neq t}}\prod_{\substack{k=2\\k\neq t}}^{s-1}L^{(r_k^t)'}(X_{k-1},X_k)\left(L^{(r_s^t)}(X_{s-1},X_s)-L^{(r_s^t)'}(X_{s-1},X_s)\right)\prod_{\substack{k=s+1\\k\neq t}}^{T}L^{(r_k^t)}(X_{k-1},X_k).
\end{eqnarray*}
Hence
\begin{eqnarray*}
&&E_0[|\underline L^{\mathbf r^t}-{\underline L'}^{\mathbf r^t}||X_{t-1},X_t]\\
&\leq&\sum_{\substack{s=2,\ldots,T\\s\neq t}}E_0\Bigg[\prod_{\substack{k=2\\k\neq t}}^{s-1}L^{(r_k^t)'}(X_{k-1},X_k)\left|L^{(s-t)}(X_{s-1},X_s)-L^{(s-t)'}(X_{s-1},X_s)\right|\prod_{\substack{k=s+1\\k\neq t}}^{T}L^{(k-t)}(X_{k-1},X_k){}\\
&&{}\Bigg|X_{t-1},X_t\Bigg]
\end{eqnarray*}
and so
\begin{eqnarray}
&&E_0\left(E_0[|\underline L^{\mathbf r^t}-{\underline L'}^{\mathbf r^t}||X_{t-1},X_t]\right)^2\notag\\
&\leq&(T-2)^2\sum_{\substack{s=2,\ldots,T\\s\neq t}}E_0\Bigg(E_0\Bigg[\prod_{\substack{k=2\\k\neq t}}^{s-1}L^{(r_k^t)'}(X_{k-1},X_k)\left|L^{(r_s^t)}(X_{s-1},X_s)-L^{(r_s^t)'}(X_{s-1},X_s)\right|{}\notag\\
&&{}\prod_{\substack{k=s+1\\k\neq t}}^{T}L^{(r_k^t)}(X_{k-1},X_k)\Bigg|X_{t-1},X_t\Bigg]\Bigg)^2.\label{interim3}
\end{eqnarray}
Let us now consider each summand above. Without loss of generality, consider the case $s<t$. We have
\begin{eqnarray}
&&E_0\left[\prod_{k=2}^{s-1}L^{(r_k^t)'}(X_{k-1},X_k)\left|L^{(r_s^t)}(X_{s-1},X_s)-L^{(r_s^t)'}(X_{s-1},X_s)\right|\prod_{\substack{k=s+1\\k\neq t}}^{T}L^{(r_k^t)}(X_{k-1},X_k)\Bigg|X_{t-1},X_t\right]\notag\\
&=&E_0\left[\prod_{k=2}^{s-1}L^{(r_k^t)'}(X_{k-1},X_k)\left|L^{(r_s^t)}(X_{s-1},X_s)-L^{(r_s^t)'}(X_{s-1},X_s)\right|\prod_{k=s+1}^{t-1}L^{(r_k^t)}(X_{k-1},X_k)\Bigg|X_{t-1}\right]{}\notag\\
&&{}E_0\left[\prod_{k=t}^{T}L^{(r_k^t)}(X_{k-1},X_k)\Bigg|X_t\right]\notag\\
&&{}\mbox{\ \ where we denote $\prod_{k=s+1}^{t-1}L^{(r_k^t)}(X_{k-1},X_k)=1$ if $s=t-1$}\notag\\
&=&E_0\left[\prod_{k=2}^{s-1}L^{(r_k^t)'}(X_{k-1},X_k)\left|L^{(r_s^t)}(X_{s-1},X_s)-L^{(s-t)'}(X_{s-1},X_s)\right|\prod_{k=s+1}^{t-1}L^{(r_k^t)}(X_{k-1},X_k)\Bigg|X_{t-1}\right]\notag\\
&&{}\mbox{\ \ since $\prod_{k=t}^{T}L^{(r_k^t)}(X_{k-1},X_k)$ is a forward martingale by Lemma \ref{lemma:properties};}\notag\\
&=&E_0\Bigg[E_0\Bigg[\prod_{k=2}^{s-1}L^{(r_k^t)'}(X_{k-1},X_k)\left|L^{(r_s^t)}(X_{s-1},X_s)-L^{(r_s^t)'}(X_{s-1},X_s)\right|\prod_{k=s+1}^{t-1}L^{(r_k^t)}(X_{k-1},X_k){}\notag\\
&&{}\Bigg|\mathcal F_{s:(t-1)}\Bigg]\Bigg|X_{t-1}\Bigg]\notag\\
&=&E_0\Bigg[E_0\left[\prod_{k=2}^{s-1}L^{(r_k^t)'}(X_{k-1},X_k)\left|L^{(r_s^t)}(X_{s-1},X_s)-L^{(r_s^t)'}(X_{s-1},X_s)\right|\Bigg|X_s\right]{}\notag\\
&&{}\prod_{k=s+1}^{t-1}L^{(r_k^t)}(X_{k-1},X_k)\Bigg|X_{t-1}\Bigg]\notag\\
&=&E_{\tilde f}\left[E_0\left[\prod_{k=2}^{s-1}L^{(r_k^t)'}(X_{k-1},X_k)\left|L^{(r_s^t)}(X_{s-1},X_s)-L^{(r_s^t)'}(X_{s-1},X_s)\right|\Bigg|X_s\right]\Bigg|X_{t-1}\right]\label{interim1}
\end{eqnarray}
where $E_{\tilde f}$ is under the change of measure generated by $\prod_{k=s+1}^{t-1}L^{(r_k^t)}(X_{k-1},X_k)$. On the other hand, we have
\begin{eqnarray}
&&E_0\left[\prod_{k=2}^{s-1}L^{(r_k^t)'}(X_{k-1},X_k)\left|L^{(r_s^t)}(X_{s-1},X_s)-L^{(r_s^t)'}(X_{s-1},X_s)\right|\Bigg|X_s\right]\notag\\
&=&E_0\left[E_0\left[\prod_{k=2}^{s-1}L^{(r_k^t)'}(X_{k-1},X_k)\left|L^{(r_s^t)}(X_{s-1},X_s)-L^{(r_s^t)'}(X_{s-1},X_s)\right|\Bigg|X_{s-1},X_s\right]\Bigg|X_s\right]\notag\\
&=&E_0\left[E_0\left[\prod_{k=2}^{s-1}L^{(r_k^t)'}(X_{k-1},X_k)\Bigg|X_{s-1}\right]\left|L^{(r_s^t)}(X_{s-1},X_s)-L^{(r_s^t)'}(X_{s-1},X_s)\right|\Bigg|X_s\right]\notag\\
&=&E_0\left[\left|L^{(r_s^t)}(X_{s-1},X_s)-L^{(r_s^t)'}(X_{s-1},X_s)\right|\Bigg|X_s\right]\label{interim2}
\end{eqnarray}
by the backward martingale property of $\prod_{k=2}^{s-1}L^{(r_k^t)'}(X_{k-1},X_k)$. So, by \eqref{interim1}, \eqref{interim2} and Jensen's inequality, we have
\begin{eqnarray}
&&E_0\Bigg(E_0\Bigg[\prod_{k=2}^{s-1}L^{(r_k^t)'}(X_{k-1},X_k)\left|L^{(r_s^t)}(X_{s-1},X_s)-L^{(r_s^t)'}(X_{s-1},X_s)\right|{}\notag\\
&&{}\prod_{k=s+1}^{T}L^{(r_k^t)}(X_{k-1},X_k)\Bigg|X_{t-1},X_t\Bigg]\Bigg)^2\notag\\
&=&E_0\left(E_{\tilde f}\left[E_0\left[\left|L^{(r_s^t)}(X_{s-1},X_s)-L^{(r_s^t)'}(X_{s-1},X_s)\right|\Bigg|X_s\right]\Bigg|X_{t-1}\right]\right)^2\notag\\
&=&E_0\left(E_{\tilde{\tilde f}}\left[\left|L^{(r_s^t)}(X_{s-1},X_s)-L^{(r_s^t)'}(X_{s-1},X_s)\right|\Bigg|X_{t-1}\right]\right)^2\notag\\
&\leq&E_0\left[|L^{(r_s^t)}(X_{s-1},X_s)-L^{(r_s^t)'}(X_{s-1},X_s)|^2\right]\notag\\
\end{eqnarray}
where $E_{\tilde{\tilde f}}$ denotes the expectation under the change of measure induced by $\prod_{k=s+1}^{t-1}L^{(r_k^t)}(X_{k-1},X_k)$ from step $t-1$ backward to $s$ and then following the benchmark $P_0$ before $s$. The last inequality follows from Jensen's inequality, and by the construction that $X_s$ is marginally distributed as $P_0$ and that the transition from $X_s$ and $X_{s-1}$ is under the baseline $P_0$.

Therefore, \eqref{interim3} is bounded by
$$(T-2)^2\sum_{\substack{s=2,\ldots,T\\s\neq t}}E_0\left[\left|L^{(r_s^t)}(X_{s-1},X_s)-L^{(r_s^t)'}(X_{s-1},X_s)\right|^2\right],$$
and hence \eqref{interim4} is bounded by
\begin{eqnarray*}
&&\frac{(T-1)^2(T-2)^2C}{4\alpha^2}\sum_{t=2}^T\sum_{\mathbf r^t\in\mathcal S_{1:(T-2)}}\sum_{\substack{s=2,\ldots,T\\s\neq t}}E_0[|L^{(r_s^t)}(X_{s-1},X_s)-L^{(r_s^t)'}(X_{s-1},X_s)|^2]\\
&=&\frac{(T-1)^3(T-2)^2(T-3)!C}{4\alpha^2}\sum_{t=1}^{T-2}E_0[|L^{(t)}-L^{(t)'}|^2]\\
&\leq&\frac{(T-1)^3(T-2)^2(T-3)!C}{4\alpha^2}d(\underline L^{(1):(T-2)},\underline L^{(1):(T-2)'})^2.
\end{eqnarray*}
When $\alpha$ is large enough, this gives $E_0|K(\mathbf L^{(1):(T-2)})-K(\mathbf L^{(1):(T-2)'})|^2\leq d(\mathbf L^{(1):(T-2)},\mathbf L^{(1):(T-2)'})^2$, and the computation above can be iterated over $(\tilde L^{(1)},\tilde L^{(2)},\ldots,\tilde L^{(t)},L^{(t+1)},\ldots,L^{(T-2)})$ for $t=1,\ldots,T-3$ in \eqref{iteration}. Then we get that
$$d(\mathcal K(\mathbf L^{(1):(T-2)}),\mathcal K(\mathbf L^{(1):(T-2)'}))^2\leq\frac{(T-1)^3(T-2)^2(T-3)!C}{4\alpha^2}d(\mathbf L^{(1):(T-2)},\mathbf L^{(1):(T-2)'})^2$$
which gives
\begin{align*}
d(\mathcal K(\mathbf L^{(1):(T-2)}),\mathcal K(\mathbf L^{(1):(T-2)'}))&\leq\frac{(T-1)^{3/2}(T-2)((T-3)!)^{1/2}C^{1/2}}{2\alpha}d(\mathbf L^{(1):(T-2)},\mathbf L^{(1):(T-2)'})\\
&\leq cd(\mathbf L^{(1):(T-2)},\mathbf L^{(1):(T-2)'})
\end{align*}
for some constant $0<c<1$, when $\alpha$ is large enough. Hence we conclude that $\mathcal K$ is a contraction on ${\mathcal L^c(M)}^{T-2}$.

By the Banach fixed point theorem, $\mathcal K$ has a fixed point $\underline L^*$. We are left to show that all components of $\underline L^*$ are identical. To this end, let $\underline L^*=(L^{(1)*},L^{(2)*},\ldots,L^{(T-2)*})$ and note that by the definition of fixed point and the iteration in \eqref{iteration},
\begin{align*}
\tilde L^{(1)*}&=K(L^{(1)*},L^{(2)*},\ldots,L^{(T-2)*})=L^{(1)*}\\
\tilde L^{(2)*}&=K(\tilde L^{(1)*},L^{(2)*},\ldots,L^{(T-2)})=K(L^{(1)*},L^{(2)*},\ldots,L^{(T-2)*})=L^{(2)*}\\
\tilde L^{(3)*}&=K(\tilde L^{(1)*},\tilde L^{(2)*},L^{(3)*}\ldots,L^{(T-2)})=K(L^{(1)*},L^{(2)*},L^{(3)*}\ldots,L^{(T-2)*})=L^{(3)*}\\
&\vdots\\
\tilde L^{(T-2)*}&=K(\tilde L^{(1)*},\tilde L^{(2)*},\ldots\tilde L^{(T-3)*},L^{(T-2)})=K(L^{(1)*},L^{(2)*},\ldots L^{(T-3)*},L^{(T-2)*})=L^{(T-2)*}
\end{align*}
and hence
$$L^{(1)*}=L^{(2)*}=\cdots=L^{(T-2)*}=K(L^{(1)*},L^{(2)*},\ldots,L^{(T-2)*}).$$
Denote all these $L^{(t)*}$'s as $L^*$. By the definition of $K$, it is clear that $L^*=K(L^*,L^*,\ldots,L^*)$ implies that $L^*$ satisfies \eqref{optimal solution many}. We conclude the lemma.
\endproof

To prepare for the next step, we introduce a new quantity $\bar G^{\mathbf L^{(1):(T-1)}}\in\mathbb R$, given any $\mathbf L^{(1):(T-1)}=(L^{(1)},\ldots,L^{(T-1)})\in\mathcal L^c(M)^{T-1}$, as
\begin{equation}
\bar G^{\mathbf L^{(1):(T-1)}}=\sum_{\mathbf r\in\mathcal S_{1:(T-1)}}E_0\left[h\prod_{s=2}^TL^{(r_s)}(X_{s-1},X_s)\right]\label{HL}
\end{equation}
where $\mathbf r=(r_2,\ldots,r_T)$ is a vector of length $T-1$ and $\mathcal S_{1:(T-1)}$ is the symmetric group of all permutations of $(1,\ldots,T-1)$.
Note that $\bar G^{\mathbf L^{(1):(T-1)}}$ is invariant to any ordering of the indices in $L^{(1)},\ldots,L^{(T-1)}$.

At this point it is more convenient to consider a $(T-1)!$-scaling of the Lagrangian in \eqref{inner maximization}, i.e.
\begin{equation}
(T-1)!\times\left(E_0\left[h\prod_{t=2}^TL(X_{t-1},X_t)\right]-\alpha E_0(L-1)^2\right)=\bar G^{\mathbf L}-\alpha(T-1)!E_0(L-1)^2\label{scaled objective}
\end{equation}
where $\mathbf L=(L,\ldots,L)\in\mathcal L^c(M)^{T-1}$. 
Now we shall consider a generalized version of the objective in \eqref{scaled objective}:
$$\bar G^{\mathbf L^{(1):(T-1)}}-\alpha(T-2)!\sum_{t=1}^{T-1}E_0(L^{(t)}-1)^2$$
as a function of $\mathbf L^{(1):(T-1)}$. We have the following monotonicity property of the mapping $K$ on this generalized objective:
\begin{lemma}
Starting from any $L^{(1)},\ldots,L^{(T-2)}\in\mathcal L^c(M)$, consider the sequence $L^{(k)}=K(L^{(k-T+2)},L^{(k-T+3)},\ldots,L^{(k-1)})$ for $k=T-1,T,\ldots$, where $K$ is defined in \eqref{K}. The quantity
$$\bar G^{\mathbf L^{(k):(k+T-2)}}-\alpha(T-2)!\sum_{t=1}^{T-1}E_0(L^{(k+t-1)}-1)^2$$
is non-decreasing in $k\geq1$. \label{lemma:monotonicity}
\end{lemma}

\proof{Proof of Lemma \ref{lemma:monotonicity}.}
Consider
\begin{eqnarray*}
&&\bar G^{\mathbf L^{(k):(k+T-2)}}-\alpha(T-2)!\sum_{t=1}^{T-1}E_0(L^{(k+t-1)}-1)^2\\
&=&E_0[G^{\mathbf L^{(k+1):(k+T-2)}}(X,Y)L^{(k)}(X,Y)]-\alpha(T-2)!E_0(L^{(k)}-1)^2-\alpha(T-2)!\sum_{t=2}^{T-1}E_0(L^{(k+t-1)}-1)^2\\
&&\mbox{\ \ \ \ by the symmetric construction of $\bar G^{\mathbf L^{(k):(k+T-2)}}$}\\
&\leq&E_0[G^{\mathbf L^{(k+1):(k+T-2)}}(X,Y)L^{(k+T-1)}(X,Y)]-\alpha(T-2)!E_0(L^{(k+T-1)}-1)^2-\alpha(T-2)!\sum_{t=2}^{T-1}E_0(L^{(k+t-1)}-1)^2\\
&&\mbox{\ \ \ \ by using Proposition \ref{prop:characterization bivariate}, treating the cost function as $G^{\underline L^{(k+1):(k+T-2)}}(x,y)/(T-2)!$,}\\
&&\mbox{\ \ \ \ and recalling the definition of $K$ in \eqref{K}}\\
&=&\bar G^{\mathbf L^{(k+1):(k+T-1)}}-\alpha(T-2)!\sum_{t=1}^{T-1}E_0(L^{(k+t)}-1)^2\\
&&\mbox{\ \ \ \ by the symmetric construction of $\bar G^{\mathbf L^{(k+1):(k+T-1)}}$.}
\end{eqnarray*}
This concludes the ascendency of $\bar G^{\mathbf L^{(k):(k+T-2)}}-\alpha(T-2)!\sum_{t=1}^{T-1}E_0(L^{(k+t-1)}-1)^2$.
\endproof

%

The final step is to conclude the convergence of the scaled objective value in \eqref{scaled objective} along the dynamic sequence defined by the iteration of $K$ to that evaluated at the solution of \eqref{optimal solution many}:
\begin{lemma}
Define the same sequence $L^{(k)}$ as in Lemma \ref{lemma:monotonicity}. We have
\begin{equation}
\bar G^{\mathbf L^{(k):(k+T-2)}}-\alpha(T-2)!\sum_{t=1}^{T-1}E_0(L^{(k+t-1)}-1)^2\to\bar G^{\mathbf L^*}-\alpha(T-1)!E_0(L^*-1)^2\label{convergence}
\end{equation}
as $k\to\infty$, where $L^*$ satisfies \eqref{optimal solution many}.
\label{lemma:convergence}
\end{lemma}

\proof{Proof of Lemma \ref{lemma:convergence}.}
First, convergence to the fixed point associated with the operator $\mathcal K$ under the $d$-metric implies componentwise convergence. Therefore, by Lemma \ref{lemma:contraction}, the sequence $L^{(k)}=K(L^{(k-T+2)},L^{(k-T+3)},\ldots,L^{(k-1)})$ for $k=T-1,T,\ldots$ defined in Lemma \ref{lemma:monotonicity} converges to $L^*$, the identical component of the fixed point $\mathbf L^*$ of $\mathcal K$, which has been shown to satisfy \eqref{optimal solution many} in Lemma \ref{lemma:contraction}.

Now consider each term in \eqref{convergence}. For the first term,
\begin{equation}
\bar G^{\mathbf L^{(k):(k+T-2)}}-\bar G^{\mathbf L^*}\leq C\sum_{\mathbf r\in\mathcal S_{1:(T-1)}}E_0|\underline L^{(\mathbf r)}-\underline L^*| \label{interim5}
\end{equation}
for some constant $C>0$, where $\mathbf r=(r_2,\ldots,r_T)$ is a vector of length $T-1$, $\mathcal S_{1:(T-1)}$ is the symmetric group of all permutations of $(1,\ldots,T-1)$, and
$$\underline L^{(\mathbf r)}=\prod_{t=2}^{T}L^{(r_s)}(X_{t-1},X_t),\ \ \underline L^*=\prod_{t=2}^TL^*(X_{t-1},X_t).$$
Then, by the same technique used in the proof of contraction in Lemma \ref{lemma:contraction}, we have \eqref{interim5} bounded by
$$C(T-2)(T-2)!\sum_{t=2}^TE_0|L^{(k+t-2)}(X,Y)-L^*(X,Y)|\to0$$
as $k\to\infty$, where $X$, $Y$ are i.i.d. each under the marginal distribution of $P_0$. 
For the second term in \eqref{convergence}, since $\|L^{(k)}-L^*\|_2\to0$, we have immediately that $\|L^{(k)}\|_2\to\|L^*\|_2$ and so
$$(T-2)!\sum_{t=2}^TE_0(L^{(k+t-2)}-1)^2-(T-1)!E_0(L^*-1)^2\to0$$
as $k\to\infty$. Hence we have proved the lemma.
\endproof

With these lemmas in hand, the proof of Proposition \ref{prop:characterization} is immediate:
\proof{Proof of Proposition \ref{prop:characterization}.}
For any $L\in\mathcal L^c(M)$, denote $L^{(1)}=L^{(2)}=\cdots=L^{(T-2)}=L$ and define the sequence $L^{(k)}$ for $k\geq T-1$ as in Lemma \ref{lemma:monotonicity}. By Lemmas \ref{lemma:monotonicity} and \ref{lemma:convergence} we conclude Proposition \ref{prop:characterization}.
\endproof

We are now ready to prove Theorem \ref{thm:expansion}:
\proof{Proof of Theorem \ref{thm:expansion}.}
By Theorem \ref{thm:duality}, for any small enough $\eta$, we need to find $\alpha^*$ and $L^*\in\mathcal L^c(M)$ such that \eqref{optimality} holds and $E_0(L^*-1)^2=\eta$. Then $L^*$ will be the optimal solution to \eqref{max P many}. To start with, for any large enough $\alpha$ in $E_0[h\underline L]-\alpha E_0(L-1)^2$, we have the optimal solution $L^*(x,y)=1+\frac{R^{L^*}(x,y)}{2\alpha}$ by Proposition \ref{prop:characterization}. We have
\begin{eqnarray}
&&H^{L^*}(x,y)\notag\\
&=&\sum_{t=2}^TE_0\left[h\prod_{\substack{s=2,\ldots,T\\s\neq t}}\left(1+\frac{R^{L^*}(X_{s-1},X_s)}{2\alpha}\right)\Bigg|X_{t-1}=x,X_t=y\right]\notag\\
&=&\sum_{t=2}^TE_0[h|X_{t-1}=x,X_t=y]+\frac{1}{2\alpha}\sum_{\substack{s,t=1,\ldots,T\\s\neq t}}E_0[hR^{L^*}(X_{s-1},X_s)|X_{t-1}=x,X_t=y]+\bar O\left(\frac{1}{\alpha^2}\right)\notag\\
&&\mbox{\ \ where $\bar O(1/\alpha^q)$ for $q>0$ satisfies $|\bar O(1/\alpha^q)\alpha^q|\leq C$ for some $C$, for any $\alpha>0$,}\notag\\
&&\mbox{\ \ uniformly over $x$ and $y$; the equality follows from the boundedness of $R^{L^*}(x,y)$}\notag\\
&&\mbox{\ \ inherited from $h$}\notag\\
&=&H(x,y)+\frac{1}{2\alpha}W^{L^*}(x,y)+\bar O\left(\frac{1}{\alpha^2}\right) \label{HL2}
\end{eqnarray}
where $H(x,y)=\sum_{t=1}^TE_0[h|X_{t-1}=x,X_t=y]$, and
$$W^{L^*}(x,y)=\sum_{\substack{s,t=1,\ldots,T\\s\neq t}}E_0[hR^{L^*}(X_{s-1},X_s)|X_{t-1}=x,X_t=y].$$
Now, \eqref{HL2} implies
\begin{equation}
R^{L^*}(x,y)=R(x,y)+\frac{S^{L^*}(x,y)}{2\alpha}+\bar O\left(\frac{1}{\alpha^2}\right)\label{interim R}
\end{equation}
where $R^L$ and $R$ are defined in \eqref{optimal residual} and \eqref{residual symmetrization}, and
$$S^{L^*}(x,y)=W^{L^*}(x,y)-E_0[W^{L^*}|x]-E_0[W^{L^*}|y]+E_0W^{L^*}.$$
Therefore we can write
\begin{align}
E_0(L^*-1)^2&=E_0\left(\frac{R^{L^*}(X,Y)}{2\alpha}\right)^2\notag\\
&=\frac{1}{4\alpha^2}E_0\left(R(X,Y)+\frac{S^{L^*}(X,Y)}{2\alpha}+\bar O\left(\frac{1}{\alpha^2}\right)\right)^2\notag\\
&=\frac{1}{4\alpha^2}\left(E_0R(X,Y)^2+\frac{1}{\alpha}E_0[R(X,Y)S^{L^*}(X,Y)]+O\left(\frac{1}{\alpha^2}\right)\right).\label{interim new1}
\end{align}
Note that the term $O(1/\alpha^2)$ in \eqref{interim new1} is continuous in $1/\alpha$ in a neighborhood of 0; in fact, it is a polynomial in $1/\alpha$ by tracing its definition. Hence \eqref{interim new1} reveals that, given any small enough $\eta>0$, we can find a large enough $\alpha^*$, and $L^*$ satisfying \eqref{optimality}, such that $E_0(L^*-1)^2=\eta$. More precisely, setting $E_0(L^*-1)^2=\eta$ and using \eqref{interim new1} gives 
\begin{eqnarray}
\frac{1}{2\alpha}&=&\sqrt{\frac{\eta}{Var_0(R(X,Y))}}\left(1+\frac{1}{\alpha}\frac{E_0[R(X,Y)S^{L^*}(X,Y)]}{Var_0(R(X,Y))}+O\left(\frac{1}{\alpha^2}\right)\right)^{-1/2}\mbox{\ \ since $E_0R(X,Y)=0$}\notag\\
&=&\sqrt{\frac{\eta}{Var_0(R(X,Y))}}\left(1-\frac{1}{2\alpha}\frac{E_0[R(X,Y)S^{L^*}(X,Y)]}{Var_0(R(X,Y))}+O\left(\frac{1}{\alpha^2}\right)\right).\label{interim new}
\end{eqnarray}
Now \eqref{interim new} gives
$$\frac{1}{2\alpha}=\sqrt{\frac{\eta}{Var_0(R(X,Y))}}(1+o(1))$$
as $\eta\to0$. Substituting it into the right hand side of \eqref{interim new}, we have
$$\frac{1}{2\alpha}=\sqrt{\frac{\eta}{Var_0(R(X,Y))}}-\frac{E_0[R(X,Y)S^{L^*}(X,Y)]\eta}{Var_0(R(X,Y))^2}+o(\eta)$$
which upon substituting into the right hand side of \eqref{interim new} once more gives
\begin{equation}
\frac{1}{2\alpha}=\sqrt{\frac{\eta}{Var_0(R(X,Y))}}-\frac{E_0[R(X,Y)S^{L^*}(X,Y)]\eta}{Var_0(R(X,Y))^2}+O(\eta^{3/2}).\label{complementary slackness}
\end{equation}
Note that this asymptotic is valid since we assume $Var_0(R(X,Y))>0$. 

Now consider the objective value of \eqref{max many}:
\begin{eqnarray}
&&E_0[h\underline L^*]=E_0\left[h\prod_{t=2}^T\left(1+\frac{R^{L^*}(X_{t-1},X_t)}{2\alpha}\right)\right]\notag\\
&=&E_0h+\frac{1}{2\alpha}\sum_{t=2}^TE_0[hR^{L^*}(X_{t-1},X_t)]+\frac{1}{4\alpha^2}\sum_{\substack{s,t=2,\ldots,T\\s<t}}E_0[hR^{L^*}(X_{t-1},X_t)R^{L^*}(X_{s-1},X_s)]{}\notag\\
&&{}+O\left(\frac{1}{\alpha^3}\right).\label{interim6}
\end{eqnarray}
Using \eqref{interim R}, we can write \eqref{interim6} as
\begin{eqnarray}
&&E_0h+\frac{1}{2\alpha}\sum_{t=2}^TE_0[hR(X_{t-1},X_t)]+\frac{1}{4\alpha^2}\sum_{t=2}^TE_0[hS^{L^*}(X_{t-1},X_t)]{}\notag\\
&&{}+\frac{1}{4\alpha^2}\sum_{\substack{s,t=2,\ldots,T\\s<t}}E_0[hR^{L^*}(X_{t-1},X_t)R^{L^*}(X_{s-1},X_s)]+O\left(\frac{1}{\alpha^3}\right)\notag\\
&=&E_0h+\frac{1}{2\alpha}E_0[H_2(X,Y)R(X,Y)]+\frac{1}{4\alpha}\Bigg(E_0[H(X,Y)S^{L^*}]{}\notag\\
&&{}+\sum_{\substack{s,t=2,\ldots,T\\s<t}}E_0[hR(X_{s-1},X_s)R(X_{t-1},X_t)]\Bigg)+O\left(\frac{1}{\alpha^3}\right)\label{interim7}
\end{eqnarray}
by conditioning on $(X_{t-1},X_t)$ inside some of the expectations to get the equality. Note that $E_0[H(X,Y)R(X,Y)]=Var_0(R(X,Y))$, by the property that $R$ is an orthogonal projection of $H$ onto the space $\{V(X,Y)\in\mathcal L_2:E_0[V|X]=E_0[V|Y]=0\text{\ a.s.}\}$. Then, substituting \eqref{complementary slackness} into \eqref{interim7} gives
\begin{eqnarray}
&&E_0h+\sqrt{Var_0(R(X,Y))\eta}-\frac{E_0[R(X,Y)S^{L^*}(X,Y)]\eta}{Var_0(R(X,Y))}+\frac{\eta}{Var_0(R(X,Y))}{}\notag\\
&&{}\left(E_0[H(X,Y)S^{L^*}(X,Y)]+\sum_{\substack{s,t=2,\ldots,T\\s<t}}E_0[hR(X_{s-1},X_s)R(X_{t-1},X_t)]\right)+O(\eta^{3/2})\notag\\
&=&E_0h+\sqrt{Var_0(R(X,Y))\eta}+\frac{\eta}{Var_0(R(X,Y))}\Bigg(E_0[H(X,Y)S^{L^*}(X,Y)]-E_0[R(X,Y)S^{L^*}(X,Y)]{}\notag\\
&&{}+\sum_{\substack{s,t=2,\ldots,T\\s<t}}E_0[hR(X_{s-1},X_s)R(X_{t-1},X_t)]\Bigg)+O(\eta^{3/2}).\label{interim8}
\end{eqnarray}
Now note that $H-R$ is orthogonal to $S^{L^*}$ by construction. This concludes that \eqref{interim8} is equal to
$$E_0h+\sqrt{Var_0(R(X,Y))\eta}+\frac{\eta}{Var_0(R(X,Y))}\sum_{\substack{s,t=2,\ldots,T\\s<t}}E_0[hR(X_{s-1},X_s)R(X_{t-1},X_t)]+O(\eta^{3/2})$$
which gives the theorem.
\endproof


%
%

\section{Proofs Related to the Statistical Properties of Algorithm \ref{ANOVA procedure}}
\proof{Proof of Theorem \ref{unbiased}}
Using the representation from a two-way random effect model, each copy $Z_{ijk}$ can be written as
$$Z_{ijk}=\mu+\tau_{i\cdot}+\tau_{\cdot j}+\tau_{ij}+\epsilon_{ijk}$$
where $\mu=E_0[H(X,Y)]$, $\tau_{i\cdot}=E_0[H(X,Y)|X=x_i]-E_0[H(X,Y)]$, $\tau_{\cdot j}=E_0[H(X,Y)|Y=y_j]-E_0[H(X,Y)]$, $\tau_{ij}=R(x_i,y_j)=H(x_i,y_j)-E_0[H(X,Y)|X=x_i]-E_0[H(X,Y)|Y=y_j]+E_0[H(X,Y)]$, and $\epsilon_{ijk}=Z_{ijk}-R(x_i,y_j)$ is the residual error.

Note that we have the following properties: $\tau_{i\cdot},\tau_{\cdot j},\tau_{ij},\epsilon_{ij}$ all have mean 0 and are uncorrelated for any $i,j,k$. Also, given $x_i$, $\tau_{ij}$ has mean 0 for any $j$ and $\tau_{ij_1}$ and $\tau_{ij_2}$ are uncorrelated for any $j_1$ and $j_2$. Similarly, given $y_j$, $\tau_{ij}$ has mean 0 for any $i$ and $\tau_{i_1j}$ and $\tau_{i_2j}$ are uncorrelated for any $i_1$ and $i_2$. Given $x_i$ and $y_j$, $\epsilon_{ijk}$ have mean 0 for any $k$ and $\epsilon_{ijk_1}$ and $\epsilon_{ijk_2}$ are independent for any $k_1$ and $k_2$.

We denote $\sigma_X^2=E\tau_{i\cdot}^2$, $\sigma_Y^2=E\tau_{\cdot j}^2$, $\sigma_{XY}^2=E\tau_{ij}^2$, $\sigma_\epsilon^2=E\epsilon_{ijk}^2$. We let $\sigma^2=Var_0(Z_{ijk})=\sigma_X^2+\sigma_Y^2+\sigma_{XY}^2+\sigma_\epsilon^2$ be the total variance.

The estimator in Procedure \ref{ANOVA procedure} is
$$\frac{1}{n}(s_I^2-s_\epsilon^2)=\frac{1}{n}\left(\frac{1}{(K-1)^2}\sum_{i,j=1}^Kn(\bar Z_{ij}-\bar Z_{i\cdot}-\bar Z_{\cdot j}+\bar Z)^2-\frac{1}{K^2(n-1)}\sum_{i,j=1}^K\sum_{l=1}^n(Z_{ijl}-\bar Z_{ij})^2\right).$$
Consider the terms one by one. First, by the sum-of-squares decomposition in two-way ANOVA, we can write the interaction sum-of-squares as
\begin{equation}
SS_I=\sum_{i,j=1}^Kn(\bar Z_{ij}-\bar Z_{i\cdot}-\bar Z_{\cdot j}+\bar Z)^2=SS_T-SS_X-SS_Y-SS_\epsilon\label{SSI}
\end{equation}
where
$$SS_T=\sum_{i=1}^K\sum_{j=1}^K\sum_{k=1}^n(Z_{ijk}-\bar Z)^2,\ \ SS_X=\sum_{i=1}^KKn(\bar Z_i-\bar Z)^2,\ \ SS_Y=\sum_{i=1}^KKn(\bar Z_j-\bar Z)^2,\ \ SS_\epsilon=\sum_{i=1}^K\sum_{j=1}^K\sum_{k=1}^n(Z_{ijk}-\bar Z_{ij})^2$$
are the total sum-of-squares, the treatment sum-of-squares for the first factor, the treatment sum-of-squares for the second factor, and the residual sum-of-squares respectively. We can further write
$$SS_T=\sum_{i,j,k}(Z_{ijk}-\mu)^2-K^2n(\bar Z-\mu)^2.$$
Note that
$$E_0\left[\sum_{i,j,k}(Z_{ijk}-\mu)^2\right]=K^2n\sigma^2$$
and
$$E_0[K^2n(\bar Z-\mu)^2]=K^2nVar_0(\bar Z)$$
so that
$$E_0[SS_T]=K^2n(\sigma^2-Var_0(\bar Z)).$$
On the other hand,
$$SS_X=Kn\left(\sum_i(\bar Z_{i\cdot}-\mu)^2-K(\bar Z-\mu)^2\right)$$
so that
$$E_0[SS_X]=K^2n(Var_0(\bar Z_{i\cdot})-Var_0(\bar Z)).$$
Similarly,
$$E_0[SS_Y]=K^2n(Var_0(\bar Z_{\cdot j})-Var_0(\bar Z)).$$
We also have
$$E_0[SS_\epsilon]=K^2(n-1)\sigma_\epsilon^2.$$

Therefore, we have
\begin{align}
E_0[SS_I]&=K^2n(\sigma^2-Var_0(\bar Z))-K^2n(Var_0(\bar Z_{i\cdot})-Var_0(\bar Z))-K^2n(Var_0(\bar Z_{\cdot j})-Var_0(\bar Z))-K^2(n-1)\sigma_\epsilon^2\notag\\
&=K^2n(\sigma_X^2+\sigma_Y^2+\sigma_{XY}^2)+K^2nVar_0(\bar Z)-K^2n(Var_0(\bar Z_{i\cdot})+Var_0(\bar Z_{\cdot j}))+K^2\sigma_\epsilon^2.\label{interim ANOVA}
\end{align}

Now, since we can write
$$\bar Z=\frac{1}{K^2n}\left(K^2n\mu+Kn\sum_i\tau_{i\cdot}+Kn\sum_j\tau_{\cdot j}+n\sum_{i,j}\tau_{ij}+\sum_{i,j,k}\epsilon_{ijk}\right)$$
we have, by the zero correlations among the terms,
\begin{align*}
Var_0(\bar Z)&=\frac{1}{(K^2n)^2}\left(K^2n^2K(\sigma_X^2+\sigma_Y^2)+n^2K^2\sigma_{XY}^2+K^2n\sigma_\epsilon^2\right)\\
&=\frac{1}{K}(\sigma_X^2+\sigma_Y^2)+\frac{1}{K^2}\sigma_{XY}^2+\frac{1}{K^2n}\sigma_\epsilon^2.
\end{align*}
Similarly, we can write
$$Var_0(\bar Z_{i\cdot})=\sigma_X^2+\frac{\sigma_Y^2}{K}+\frac{\sigma_{XY}^2}{K}+\frac{\sigma_\epsilon^2}{Kn}$$
and
$$Var_0(\bar Z_{\cdot j})=\sigma_Y^2+\frac{\sigma_X^2}{K}+\frac{\sigma_{XY}^2}{K}+\frac{\sigma_\epsilon^2}{Kn}.$$

Therefore, \eqref{interim ANOVA} becomes
\begin{eqnarray*}
&&K^2n(\sigma_X^2+\sigma_Y^2+\sigma_{XY}^2)+K^2n\left(\frac{1}{K}(\sigma_X^2+\sigma_Y^2)+\frac{1}{K^2}\sigma_{XY}^2+\frac{1}{K^2n}\sigma_\epsilon^2\right){}\\
&&{}-K^2n\left(\left(1+\frac{1}{K}\right)(\sigma_X^2+\sigma_Y^2)+\frac{2\sigma_{XY}^2}{K}+\frac{2\sigma_\epsilon^2}{Kn}\right)+K^2\sigma_\epsilon^2\\
&=&(K-1)^2n\sigma_{XY}^2+(K-1)^2\sigma_\epsilon^2.
\end{eqnarray*}
Together with $E_0[s_\epsilon^2]=\sigma_\epsilon^2$, we get
$$E_0\left[\frac{1}{n}(s_I^2-s_\epsilon^2)\right]=\sigma_{XY}^2.$$
\endproof

\proof{Proof of Theorem \ref{sampling error}}
We follow the notation in the proof of Theorem \ref{unbiased}. By Cauchy-Schwarz inequality, we have
\begin{eqnarray}
&&Var_0\left(\frac{1}{n}(s_I^2-s_\epsilon^2)\right)\\
&=&Var_0\left(\frac{1}{(K-1)^2}\sum_{i,j=1}^K(\bar Z_{ij}-\bar Z_{i\cdot}-\bar Z_{\cdot j}+\bar Z)^2-\frac{1}{K^2n(n-1)}\sum_{i,j=1}^K\sum_{l=1}^n(Z_{ijl}-\bar Z_{ij})^2\right)\notag\\
&\leq&Var_0\left(\frac{1}{(K-1)^2}\sum_{i,j=1}^K(\bar Z_{ij}-\bar Z_{i\cdot}-\bar Z_{\cdot j}+\bar Z)^2\right)+Var_0\left(\frac{1}{K^2n(n-1)}\sum_{i,j=1}^K\sum_{l=1}^n(Z_{ijl}-\bar Z_{ij})^2\right)\notag\\
&&{}+2\sqrt{Var_0\left(\frac{1}{(K-1)^2}\sum_{i,j=1}^K(\bar Z_{ij}-\bar Z_{i\cdot}-\bar Z_{\cdot j}+\bar Z)^2\right)Var_0\left(\frac{1}{K^2n(n-1)}\sum_{i,j=1}^K\sum_{l=1}^n(Z_{ijl}-\bar Z_{ij})^2\right)}.\label{interim ANOVA4}
\end{eqnarray}

Now, by \eqref{SSI}, we have
\begin{align*}
\sum_{i,j=1}^Kn(\bar Z_{ij}-\bar Z_{i\cdot}-\bar Z_{\cdot j}+\bar Z)^2&\leq SS_T-SS_\epsilon\\
&=\sum_{i,j,k}(Z_{ijk}-\bar Z)^2-\sum_{i,j,k}(Z_{ijk}-\bar Z_{ij})^2\\
&=\sum_{i,j,k}(\bar Z^2-2\bar ZZ_{ijk}+2\bar Z_{ij}Z_{ijk}-\bar Z_{ij}^2)\\
&=\sum_{i,j}(n\bar Z^2-2n\bar Z\bar Z_{ij}+2n\bar Z_{ij}^2-n\bar Z_{ij}^2)\\
&=\sum_{i,j}n(\bar Z_{ij}-\bar Z)^2\\
&=\sum_{i,j}n((\bar Z_{ij}-\mu)^2-K(\bar Z-\mu)^2)\\
&\leq\sum_{i,j}n(\bar Z_{ij}-\mu)^2
\end{align*}
and so
\begin{align}
Var_0\left(\frac{1}{(K-1)^2}\sum_{i,j=1}^K(\bar Z_{ij}-\bar Z_{i\cdot}-\bar Z_{\cdot j}+\bar Z)^2\right)
&\leq\frac{1}{(K-1)^4}E\left(\sum_{i,j=1}^K(\bar Z_{ij}-\mu)^2\right)^2\notag\\
&=\frac{1}{(K-1)^4}\left(Var_0\left(\sum_{i,j=1}^K(\bar Z_{ij}-\mu)^2\right)+K^2\left(\sigma_X^2+\sigma_Y^2+\sigma_{XY}^2+\frac{\sigma_\epsilon^2}{n}\right)^2\right).\label{interim ANOVA3}
\end{align}
Consider
\begin{equation}
Var_0\left(\sum_{i,j=1}^K(\bar Z_{ij}-\mu)^2\right)=Var_0\left(\sum_{i,j}(\tau_{i\cdot}+\tau_{\cdot j}+\tau_{ij}+\bar\epsilon_{ij})^2\right)\label{interim ANOVA1}
\end{equation}
where $\bar\epsilon_{ij}=(1/n)\sum_k\epsilon_{ijk}$.

Note that by construction, given $x_i$ and $y_j$, $\epsilon_{ijk}$ is the sum of $T$ independent copies of $h(\mathbf X_T)$. By writing out this summation and noting that the expectation of each summand is zero conditional on $x_i$ and $y_j$, it is routine to check that $E\epsilon_{ijk}^2=O(T)$, $E\epsilon_{ijk}^3=O(T)$ and $E\epsilon_{ijk}^4=O(T^2)$. To see this, note that $E\epsilon_{ijk}^2$ consists of $T$ terms of second moment of a summand conditional on $x_i$ and $y_j$, $E\epsilon_{ijk}^3$ consists of $T$ terms of the third moment of a summand, and $E\epsilon_{ijk}^4$ consists of $T$ terms of the fourth moment of a summand and also order $T^2$ terms of the product of two second moments of summands. Using these expressions, because $\epsilon_{ijk}$ are all independent conditional on $x_i$ and $y_j$, we use similar observations to get further that $E\bar\epsilon_{ij}^2=O(T/n)$, $E\bar\epsilon_{ij}^3=O(T/n^2)$, and $E\bar\epsilon_{ij}^4=O((1/n^4)(nT^2+n^2T))=O(T^2/n^3+T/n^2)$. Now, \eqref{interim ANOVA1} is equal to
\begin{eqnarray}
&&\sum_{i,j}Var_0\left((\tau_{i\cdot}+\tau_{\cdot j}+\tau_{ij}+\bar\epsilon_{ij})^2\right){}\notag\\
&&{}+2\sum_{(i_1,j_1),(i_2,j_2):(i_1,j_1)\neq(i_2,j_2)}Cov_0\left((\tau_{i_1\cdot}+\tau_{\cdot j_1}+\tau_{i_1j_1}+\bar\epsilon_{i_1j_1})^2,(\tau_{i_2\cdot}+\tau_{\cdot j_2}+\tau_{i_2j_2}+\bar\epsilon_{i_2j_2})^2\right)\notag\\
&\leq&\sum_{i,j}E(\tau_{i\cdot}+\tau_{\cdot j}+\tau_{ij}+\bar\epsilon_{ij})^4{}\notag\\
&&{}+2\sum_{(i_1,j_1),(i_2,j_2):(i_1,j_1)\neq(i_2,j_2)}Cov_0\left((\tau_{i_1\cdot}+\tau_{\cdot j_1}+\tau_{i_1j_1}+\bar\epsilon_{i_1j_1})^2,(\tau_{i_2\cdot}+\tau_{\cdot j_2}+\tau_{i_2j_2}+\bar\epsilon_{i_2j_2})^2\right).\label{interim ANOVA2}
\end{eqnarray}
By the assumption that $E[H(X,Y)^4]=M$, we have $E(\tau_{i\cdot}+\tau_{\cdot j}+\tau_{ij})^4=E(H(X,Y)-\mu)^4=O(M)$ and so
$$\sum_{i,j}E(\tau_{i\cdot}+\tau_{\cdot j}+\tau_{ij}+\bar\epsilon_{ij})^4=O\left(K^2\left(M+\frac{T^2}{n^3}+\frac{T}{n^2}\right)\right).$$
For the second term in \eqref{interim ANOVA2}, note that $(\tau_{i_1\cdot}+\tau_{\cdot j_1}+\tau_{i_1j_1}+\bar\epsilon_{i_1j_1})^2$ and $(\tau_{i_2\cdot}+\tau_{\cdot j_2}+\tau_{i_2j_2}+\bar\epsilon_{i_2j_2})^2$ are independent if $i_1\neq i_2$ and $j_1\neq j_2$. Hence the number of non-zero summands there is at most $O(K^2)$. Each non-zero summand is at most of order $O(M)$ or $O(T^2/n^3+T/n^2)$. So the whole second term in \eqref{interim ANOVA2} is $O(K^2(M+T^2/n^3+T/n^2))$. In overall, \eqref{interim ANOVA2} is $O(K^2(M+T^2/n^3+T/n^2))$, and \eqref{interim ANOVA3} is of order
\begin{equation}
O\left(\frac{1}{(K-1)^4}\left(K^2\left(M+\frac{T^2}{n^3}+\frac{T}{n^2}\right)+K^2\left(\sqrt M+\frac{T}{n}\right)^2\right)\right)=O\left(\frac{1}{K^2}\left(\frac{T^2}{n^2}+M\right)\right).\label{interim ANOVA6}
\end{equation}

For the second term in \eqref{interim ANOVA4}, we have
\begin{eqnarray}
&&Var_0\left(\frac{1}{K^2n(n-1)}\sum_{i,j,k}(Z_{ijk}-\bar Z_{ij})^2\right)\notag\\
&=&\frac{1}{K^4n^2(n-1)^2}Var_0\left(\sum_{i,j,k}\epsilon_{ijk}^2\right)\notag\\
&=&\frac{1}{K^4n^2(n-1)^2}\left(\sum_{i,j,k}Var_0(\epsilon_{ijk}^2)+2\sum_{(i_1,j_1,k_1),(i_2,j_2,k_2):(i_1,j_1,k_1)\neq(i_2,j_2,k_2)}Cov_0(\epsilon_{i_1j_1k_1}^2,\epsilon_{i_2j_2k_2}^2)\right).\label{interim ANOVA5}
\end{eqnarray}
Now, $Var_0(\epsilon_{ijk}^2)\leq E\epsilon_{ijk}^4=O(T^2)$. Also, in the second sum of \eqref{interim ANOVA5}, all summands with $i_1\neq i_2$ and $j_1\neq j_2$ are zero. So the number of non-zero summands is at most $O(K^2n^2)$, and each of term is of order at most $O(T^2)$. Therefore, \eqref{interim ANOVA5} is of order
\begin{equation}
O\left(\frac{1}{K^4n^2(n-1)^2}(K^2nT^2+K^2n^2T^2)\right)=O\left(\frac{T^2}{K^2n^2}\right).\label{interim ANOVA7}
\end{equation}

Since the third term in \eqref{interim ANOVA4} is of order less than either the first or the second term, \eqref{interim ANOVA6} and \eqref{interim ANOVA7} imply that \eqref{interim ANOVA4} is of order $O((1/K^2)(M+T^2/n^2))$, and we conclude the theorem.
\endproof

\section{Proofs Related to Higher Order Dependency Assessment} 
\subsection{Detailed Derivation for Example \ref{Gaussian example}}
For $(X_{t-2},X_{t-1},X_t)$ that is multivariate normal with mean 0 and covariance matrix $\Sigma$ under $P_f$, where
$$\Sigma=\left[\begin{array}{ccc}1&\rho_1&\rho_2\\\rho_1&1&\rho_1\\\rho_2&\rho_1&1\end{array}\right]$$
we have, by definition,
$$P_{\tilde f_1}(X_{t-2},X_{t-1},X_t)=P_{\tilde f_1}(X_{t-1}|X_{t-2})P_{\tilde f_1}(X_t|X_{t-1})=P_f(X_{t-1}|X_{t-2})P_f(X_t|X_{t-1}).$$
Now from the property of multivariate normal distribution, $(X_{t-1}|X_{t-2}=x_{t-2})\sim N(\rho_1x_{t-2},1-\rho_1^2)$ and $(X_t|X_{t-1}=x_{t-1})\sim N(\rho_1x_{t-1},1-\rho_1^2)$ under $P_f$ and hence $P_{\tilde f_1}$. Moreover, $(X_{t-2},X_{t-1},X_t)$ are still multivariate normal under $P_{\tilde f_1}$. The correlations between $X_{t-1}$ and $X_t$, and between $X_{t-2}$ and $X_{t-1}$ are still $\rho_1$. The characteristic function of $X_t|X_{t-2}$ under $P_{\tilde f_1}$ is given by
\begin{align*}
E_{\tilde f_1}[e^{i\theta X_t}|X_{t-2}]&=E_{\tilde f_1}[E_{\tilde f_1}[e^{i\theta X_t}|X_{t-1}]|X_{t-2}=x_{t-2}]\\
&=E_{\tilde f_1}[e^{i\rho_1X_{t-1}\theta-(\theta^2/2)(1-\rho_1^2)}|X_{t-2}=x_{t-2}]\\
&=e^{i\rho_1^2x_{t-2}\theta-((\rho_1\theta)^2/2)(1-\rho_1^2)-(\theta^2/2)(1-\rho_1^2)}\\
&=e^{i\rho_1^2x_{t-2}\theta-(\theta^2/2)(1-\rho_1^4)}
\end{align*}
which is equal to the characteristic function of $N(\rho_1^2X_{t-2},1-\rho_1^4)$. This shows that $(X_t|X_{t-2}=x_{t-2})\sim N(\rho_1^2x_{t-2},1-\rho_1^4)$, which implies that $X_{t-2}$ and $X_t$ has correlation $\rho_1^2$.

\subsection{Proof of Theorem \ref{thm:2-lag}}\label{proof:2-lag}
We focus on the maximization formulation. 
First, let $Z^*$ be the optimal value of \eqref{max multi-lag}. Introduce the formulation
\begin{equation}
\renewcommand\arraystretch{1.4}
\begin{array}{ll}
\max&E_f[h(\mathbf X_T)]\\
\mbox{subject to}&E_0\left(\frac{dP_f(X_{t-1},X_t)}{dP_0(X_{t-1},X_t)}-1\right)^2\leq\eta_1\\
&\{X_t:t=1,\ldots,T\}\mbox{\ is a 1-dependent stationary process under $P_f$}\\
&P_f(x_t)=P_0(x_t)\\
&P_f\in\mathcal P_0,
\end{array} \label{multi-lag2}
\end{equation}
i.e.~the formulation \eqref{max P many} with $\eta$ replaced by $\eta_1$. Note that \eqref{multi-lag2} is a subproblem to \eqref{max multi-lag}, in the sense that any feasible solution of \eqref{multi-lag2} is feasible in \eqref{max multi-lag}. This is because by definition $\phi_2^2(P_f(x_{t-2},x_{t-1},x_t))=0$ for any 1-dependent measure $P_f$. Now let $Z'$ be the optimal value of \eqref{multi-lag2}, and consider the decomposition
\begin{equation}
Z^*-E_0h=(Z^*-Z')+(Z'-E_0h). \label{triangle}
\end{equation}
We know that $Z'-E_0h\leq\sqrt{Var_0(R(X_{t-1},X_t))\eta_1}+O(\eta_1)$ by Theorem \ref{thm:expansion}. On the other hand, let $P_{f^*}$ be an $\epsilon$-nearly optimal solution for \eqref{max multi-lag}, i.e. $E_{f^*}h\geq Z^*-\epsilon$ for some small $\epsilon>0$. We argue that $Z^*-Z'\leq Z^*-E_{\tilde f^*_1}h$ where $P_{\tilde f^*_1}$ is the measure corresponding to the 1-dependent counterpart of $P_{f^*}$. Note that $P_{\tilde f^*_1}$ certainly satisfies all constraints in \eqref{multi-lag2}, inherited from the properties of $P_{f^*}$.
We then must have $Z'\geq E_{\tilde f^*_1}h$ by the definition of $Z'$ and the feasibility of $P_{\tilde f^*_1}$ in \eqref{multi-lag2}, and so $Z^*-Z'\leq Z^*-E_{\tilde f^*_1}h$.

Next, note that the optimal value of
\begin{equation}
\renewcommand\arraystretch{1.4}
\begin{array}{ll}
\max&E_f[h(\mathbf X_T)]\\
\mbox{subject to}&E_{\tilde f^*_1}\left(\frac{dP_f(X_{t-2},X_{t-1},X_t)}{dP_{\tilde f^*_1}(X_{t-2},X_{t-1},X_t)}-1\right)^2\leq\eta_2\\
&\{X_t:t=1,\ldots,T\}\mbox{\ is a 2-dependent stationary process under $P_f$}\\
&P_f(x_{t-1},x_t)=P_{\tilde f^*_1}(x_{t-1},x_t)\\
&P_f\in\mathcal P_0
\end{array} \label{interim formulation}
\end{equation}
which we denote $Z''$, is at least $Z^*-\epsilon$, because $P_{f^*}$ is a feasible solution of \eqref{interim formulation} and its objective value is at least $Z^*-\epsilon$. Therefore we have $Z^*-Z'\leq Z''-E_{\tilde f^*_1}h+\epsilon$.

Next, we note that \eqref{interim formulation} is equivalent to
\begin{equation}
\renewcommand\arraystretch{1.4}
\begin{array}{ll}
\max&E_f[h(\mathbf X_T)]\\
\mbox{subject to}&E_{\tilde f^*_1}\left(\frac{dP_f((X_{t-1},X_t)|(X_{t-2},X_{t-1}))}{dP_{\tilde f^*_1}((X_{t-1},X_t)|(X_{t-2},X_{t-1}))}-1\right)^2\leq\eta_2\\
&\{(X_{t-1},X_t):t=1,\ldots,T\}\mbox{\ is a 1-dependent stationary process under $P_f$}\\
&P_f(x_{t-1},x_t)=P_{\tilde f^*_1}(x_{t-1},x_t)\\
&P_f\in\mathcal P_0
\end{array} \label{multi-lag3}
\end{equation}
by using an augmented state space representation. Under this representation, formulation \eqref{multi-lag3} reduces into \eqref{max P many} with state defined as $(X_{t-1},X_t)$, since $dP_f((X_{t-1},X_t)|(X_{t-2},X_{t-1}))/dP_{\tilde f^*_1}((X_{t-1},X_t)|(X_{t-2},X_{t-1}))=dP_f((X_{t-1},X_t),(X_{t-2},X_{t-1}))/dP_{\tilde f^*_1}((X_{t-1},X_t),(X_{t-2},X_{t-1}))$. Then Theorem \ref{thm:expansion} (the non-i.i.d.~baseline version discussed in Section \ref{sec:non-iid}) implies an optimal value of $E_{\tilde f^*_1}h+\sqrt{Var_{\tilde f^*_1}(S_{\tilde f^*_1}(X_{t-2},X_{t-1},X_t))\eta_2}+O(\eta_2)$, where $S_{\tilde f^*_1}(X_{t-2},X_{t-1},X_t)$ is the projection of 
\begin{align*}
\tilde H_{\tilde f^*_1}(x,y,z)&:=\sum_{t=3}^TE_{\tilde f^*_1}[h|(X_{t-2},X_{t-1})=(x,y),(X_{t-1},X_t)=(y,z)]\\
&=\sum_{t=3}^TE_{\tilde f^*_1}[h|X_{t-2}=x,X_{t-1}=y,X_t=z]
\end{align*}
onto
$$\tilde{\mathcal M}_{\tilde f^*_1}:=\left\{V(X,Y,Z)\in\mathcal L_2(P_{\tilde f^*_1}):E_{\tilde f^*_1}[V|X,Y]=E_{\tilde f^*_1}[V|Y,Z]=0\mbox{\ a.s.}\right\}$$
where $(X,Y)$ and $(Y,Z)$ are two consecutive states in the augmented state space representation of $P_{\tilde f^*_1}$. Therefore, $Z^*-Z'\leq\sqrt{Var_{\tilde f^*_1}(S_{\tilde f^*_1}(X_{t-2},X_{t-1},X_t))\eta_2}+O(\eta_2)+\epsilon$ (where $O(\eta_2)$ can be shown to hold uniformly over all $\tilde f^*_1$, or $\eta_1$, by the boundedness of $h$). Then from \eqref{triangle} we have $Z^*-E_0h\leq\sqrt{Var_{\tilde f^*_1}(S_{\tilde f^*_1}(X_{t-2},X_{t-1},X_t))\eta_2}+O(\eta_2)+\epsilon+\sqrt{Var_0(R(X,Y))\eta_1}+O(\eta_1)$. Finally, with $\epsilon$ being arbitrary, we use Lemmas \ref{lemma:projection} and \ref{lemma:continuity} depicted below to conclude Theorem \ref{thm:2-lag}. 
\endproof

\begin{lemma}
The object $S(x,y,z)$ defined in \eqref{H2} is precisely the projection of
\begin{align*}
\tilde H_{0}(x,y,z)&:=\sum_{t=3}^TE_0[h|(X_{t-2},X_{t-1})=(x,y),(X_{t-1},X_t)=(y,z)]\\
&=\sum_{t=3}^TE_0[h|X_{t-2}=x,X_{t-1}=y,X_t=z]
\end{align*}
onto
$$\tilde{\mathcal M}_0:=\left\{V(X,Y,Z)\in\mathcal L_2(P_0):E_0[V|X,Y]=E_0[V|Y,Z]=0\mbox{\ a.s.}\right\}.$$
\label{lemma:projection}
\end{lemma}

\begin{lemma}
$Var_{\tilde f^*_1}(S_{\tilde f^*_1}(X,Y,Z))\to Var_0(S(X,Y,Z))$ as $\eta_1\to0$. 
\label{lemma:continuity}
\end{lemma}



\proof{Proof of Lemma \ref{lemma:projection}.}
We first check that $S(X,Y,Z)\in\tilde M_0$. Note that from \eqref{H2} we can write
$$S(x,y,z)=\tilde H_0(x,y,z)-E_0[\tilde H_0|x,y]-E_0[\tilde H_0|y,z]+E_0[\tilde H_0|y].$$
So
$$E_0[S(X,Y,Z)|X,Y]=E_0[\tilde H_0|X,Y]-E_0[\tilde H_0|X,Y]-E_0[\tilde H_0|Y]+E_0[\tilde H_0|Y]=0$$
since $X$ and $Z$ are independent. Similarly, $E_0[S(X,Y,Z)|Y,Z]=0$, and so $S(X,Y,Z)\in\tilde{\mathcal M}_0$.

Next we show that $S$ and $\tilde H_0-S$ are orthogonal. Consider
\begin{eqnarray*}
&&E_0[S(\tilde H_0-S)]\\
&=&E_0(\tilde H_0(X,Y,Z)-E_0[\tilde H_0|X,Y]-E_0[\tilde H_0|Y,Z]+E_0[\tilde H_0|Y]){}\\
&&{}(E_0[\tilde H_0|X,Y]+E_0[\tilde H_0|Y,Z]-E_0[\tilde H_0|Y])\\
&=&E_0(E_0[\tilde H_0|X,Y])^2+E_0(E_0[\tilde H_0|Y,Z])^2-E_0(E_0[\tilde H_0|Y])^2{}\\
&&{}-E_0(E_0[\tilde H_0|X,Y]+E_0[\tilde H_0|Y,Z]-E_0[\tilde H_0|Y])^2\\
&=&-2E_0[E_0[\tilde H_0|X,Y]E_0[\tilde H_0|Y,Z]]+2E_0[E_0[\tilde H_0|X,Y]E_0[\tilde H_0|Y]]{}\\
&&{}+2E_0[E_0[\tilde H_0|Y,Z]E_0[\tilde H_0|Y]]-2E_0(E_0[\tilde H_0|Y])^2\\
&=&-2E_0(E_0[\tilde H_0|Y])^2+2E_0(E_0[\tilde H_0|Y])^2+2E_0(E_0[\tilde H_0|Y])^2-2E_0(E_0[\tilde H_0|Y])^2\\
&=&0
\end{eqnarray*}
where the second-to-last step is via conditioning on $Y$, and by using the fact that given $Y$, $E_0[\tilde H_0|X,Y]$ is independent of $E_0[\tilde H_0|Y,Z]$. Therefore $S(X,Y,Z)$ is the projection of $\tilde H_0$ onto $\tilde{\mathcal M}_0$.

\proof{Proof of Lemma \ref{lemma:continuity}.}
To facilitate the proof, denote $\mathcal P^{\tilde f^*_1}$ as the projection operator onto $\tilde{\mathcal M}_{\tilde f^*_1}$ in the space $\mathcal L_2(P_{\tilde f^*_1})$, and similarly denote $\mathcal P^0$ as the projection onto $\tilde{\mathcal M}_0$ in the space $\mathcal L_2(P_0)$. Note that since $E_{\tilde f^*_1}S_{\tilde f^*_1}=0$, $\sqrt{Var_{\tilde f^*_1}(S_{\tilde f^*_1})}$ is exactly $\|S_{\tilde f^*_1}\|_{\tilde f^*_1}$, where we denote $\|\cdot\|_{\tilde f^*_1}$ as the $\mathcal L_2(P_{\tilde f^*_1})$-norm. In a similar fashion, $\sqrt{Var_0(S)}=\|S\|_0$, where we denote $\|\cdot\|_0$ as the $\mathcal L_2(P_0)$-norm. With these notations, we are set to prove that $\|\mathcal P^{\tilde f^*_1}\tilde H_{\tilde f^*_1}\|_{\tilde f^*_1}\to\|\mathcal P^0\tilde H_0\|_0$ as $\eta_1\to0$.

Next, we observe that by the boundedness of $h$ and that convergence in $\chi^2$-distance implies convergence in distribution (since $\chi^2$-distance dominates Kullback-Leibler (\cite{tsybakov2008introduction}), whose convergence implies weak convergence), we have $\tilde H_{\tilde f^*_1}(x,y,z)\to\tilde H_0(x,y,z)$ pointwise a.s. as $\eta_1\to0$. Hence by dominated convergence $\|\tilde H_{\tilde f^*_1}-\tilde H_0\|_0\to0$, where $\tilde H_{\tilde f^*_1}$ and $\tilde H_0$ are coupled in the natural way under the measure $P_0$. Moreover, we have $\|V\|_{\tilde f^*_1}\to\|V\|_0$ for any bounded $V=V(X,Y)$.

Now consider
\begin{equation}
\|\mathcal P^{\tilde f^*_1}\tilde H_{\tilde f^*_1}\|_{\tilde f^*_1}-\|\mathcal P^0\tilde H_0\|_0=(\|\mathcal P^{\tilde f^*_1}\tilde H_{\tilde f^*_1}\|_{\tilde f^*_1}-\|\mathcal P^{\tilde f^*_1}\tilde H_{\tilde f^*_1}\|_0)+(\|\mathcal P^{\tilde f^*_1}\tilde H_{\tilde f^*_1}\|_0-\|\mathcal P^0\tilde H_0\|_0).\label{interim higher1}
\end{equation}
The first term in \eqref{interim higher1} goes to 0 as $\eta_1\to0$ by our observations above. The second term is bounded from above by $\|\mathcal P^{\tilde f^*_1}\tilde H_{\tilde f^*_1}-\mathcal P^0\tilde H_0\|_0$, which in turn is bounded by
\begin{equation}
\|\mathcal P^{\tilde f^*_1}\tilde H_{\tilde f^*_1}-\mathcal P^0\tilde H_{\tilde f^*_1}\|_0+\|\mathcal P^0\tilde H_{\tilde f^*_1}-\mathcal P^0\tilde H_0\|_0.\label{interim higher2}
\end{equation}
The second term in \eqref{interim higher2} is dominated by $\|\tilde H_{\tilde f^*_1}-\tilde H_0\|_0$ by the contraction property of the projection $\mathcal P^0$, which converges to 0 as $\eta_1\to0$ by the prior observations. We are left to show that the first term in \eqref{interim higher2} also goes to 0.

To this end, write
\begin{equation}
\|\mathcal P^{\tilde f^*_1}\tilde H_{\tilde f^*_1}-\mathcal P^0\tilde H_{\tilde f^*_1}\|_0=\|\mathcal P^{\tilde f^*_1}\tilde H_{\tilde f^*_1}-\mathcal P^0\mathcal P^{\tilde f^*_1}\tilde H_{\tilde f^*_1}\|_0+\|\mathcal P^0\mathcal P^{\tilde f^*_1}\tilde H_{\tilde f^*_1}-\mathcal P^0\tilde H_{\tilde f^*_1}\|_0.\label{interim higher}
\end{equation}
We tackle the first and the second terms in \eqref{interim higher} one-by-one. For the first term, using the explicit expression for $\mathcal P^0$, we can write
\begin{eqnarray*}
&&\mathcal P^{\tilde f^*_1}\tilde H_{\tilde f^*_1}-\mathcal P^0\mathcal P^{\tilde f^*_1}\tilde H_{\tilde f^*_1}\\
&=&E_0[\mathcal P^{\tilde f^*_1}\tilde H_{\tilde f^*_1}|X,Y]+E_0[\mathcal P^{\tilde f^*_1}\tilde H_{\tilde f^*_1}|Y,Z]-E_0[\mathcal P^{\tilde f^*_1}\tilde H_{\tilde f^*_1}|Y]\\
&=&E_{\tilde f^*_1}[\mathcal P^{\tilde f^*_1}\tilde H_{\tilde f^*_1}|X,Y]+E_{\tilde f^*_1}[\mathcal P^{\tilde f^*_1}\tilde H_{\tilde f^*_1}|Y,Z]-E_{\tilde f^*_1}[\mathcal P^{\tilde f^*_1}\tilde H_{\tilde f^*_1}|Y]+o(1)
\end{eqnarray*}
a.s., by the weak convergence from $P_{\tilde f^*_1}$ to $P_0$. But since $\mathcal P^{\tilde f^*_1}\tilde H_{\tilde f^*_1}\in\tilde{\mathcal M}_{\tilde f^*_1}$, $E_{\tilde f^*_1}[\mathcal P^{\tilde f^*_1}\tilde H_{\tilde f^*_1}|X,Y]$, $E_{\tilde f^*_1}[\mathcal P^{\tilde f^*_1}\tilde H_{\tilde f^*_1}|Y,Z]$ and hence $E_{\tilde f^*_1}[\mathcal P^{\tilde f^*_1}\tilde H_{\tilde f^*_1}|Y]$ are all 0. By dominated convergence we have $\|\mathcal P^{\tilde f^*_1}\tilde H_{\tilde f^*_1}-\mathcal P^0\mathcal P^{\tilde f^*_1}\tilde H_{\tilde f^*_1}\|_0\to0$.

For the second term, we use the following observation. The conditions $E_{\tilde f^*_1}[V|X,Y]=0$ and $E_{\tilde f^*_1}[V|Y,Z]=0$ in the definition of the closed subspace $\tilde{\mathcal M}_{\tilde f^*_1}$ is equivalent to the conditions $\langle V,\nu(X,Y)\rangle_{\tilde f^*_1}=0$ and $\langle V,\rho(Y,Z)\rangle_{\tilde f^*_1}=0$ for any measurable functions $\nu$ and $\rho$. Consequently, any elements in $\tilde{\mathcal M}_{\tilde f^*_1}^\perp$ can be expressed in the form $\nu(X,Y)+\rho(Y,Z)$. In particular, $\tilde H_{\tilde f^*_1}-\mathcal P^{\tilde f^*_1}\tilde H_{\tilde f^*_1}$ is in this form. But then $\mathcal P^0(\tilde H_{\tilde f^*_1}-\mathcal P^{\tilde f^*_1}\tilde H_{\tilde f^*_1})=0$. So the second term in \eqref{interim higher} is 0. We thus conclude that $\|\mathcal P^{\tilde f^*_1}\tilde H_{\tilde f^*_1}-\mathcal P^0\tilde H_{\tilde f^*_1}\|_0\to0$, and from \eqref{interim higher1}, \eqref{interim higher2} and \eqref{interim higher}, we obtain that $\|\mathcal P^{\tilde f^*_1}\tilde H_{\tilde f^*_1}\|_{\tilde f^*_1}-\|\mathcal P^0\tilde H_0\|_0\to0$. This proves the lemma.
\endproof

\subsection{Proofs Related to the Statistical Properties of Algorithm \ref{ANOVA procedure 2-dependence}}
\proof{Proof of Theorem \ref{unbiased 2-dependence}.}
Note that $Var_0(S(X,Y,Z))=Var_0(E_0[S(X,Y,Z)|Y])+E_0(Var_0[S(X,Y,Z)|Y])=E_0[Var_0(S(X,Y,Z)|Y)]$ since $E_0[S(X,Y,Z)|Y]=0$ by the definition of $S(X,Y,Z)$. Now, given $Y$, Algorithm \ref{ANOVA procedure 2-dependence} is unbiased for $Var_0(S(X,Y,Z)|Y)$. This can be seen easily by observing that Algorithms \ref{ANOVA procedure} and \ref{ANOVA procedure 2-dependence} are identical except the conditioning on $y$ and the definition of the inner samples, and that we have already shown unbiasedness for Algorithm \ref{ANOVA procedure} in Theorem \ref{unbiased}.
\endproof

\proof{Proof of Theorem \ref{sampling error 2-dependence}.}
For convenience, let $\xi$ be an output of Algorithm \ref{ANOVA procedure 2-dependence} given the sample $Y$. We have $Var_0(\xi)=Var_0(E_0[\xi|Y])+E_0[Var_0(\xi|Y)]=Var_0(Var_0(S(X,Y,Z)|Y))+E_0[Var_0(\xi|Y)]$, the last equality being a consequence of the unbiasedness of $\xi$ for $Var_0(S(X,Y,Z)|Y)$ as shown in the proof of Theorem \ref{unbiased 2-dependence}. Now, under the condition that $E[G(X,Y,Z)^4|Y]\leq M_1$, $Var_0(\xi|Y)$ can be analyzed similarly as in the proof of Theorem \ref{sampling error}, and shown to have order $O((1/K^2)(T^2/n^2+M_1))$. Therefore, together with the condition that $Var_0(Var_0(S(X,Y,Z)|Y))\leq M_2$, we have $Var_0(\xi)=O(M_2+(1/K^2)(T^2/n^2+M_1))$.
\endproof

\section{Duality Theorem}\label{sec:duality}
\begin{theorem}[Adapted from \cite{luenberger69}, Chapter 8 Theorem 1]
Suppose one can find $\alpha^*\geq0$ and $L^*\in\mathcal C$ such that
\begin{equation}
\xi(L)-\alpha^*E_0(L-1)^2\leq\xi(L^*)-\alpha^*E_0(L^*-1)^2 \label{optimality}
\end{equation}
for any $L\in\mathcal C$. Then $L^*$ solves
$$\begin{array}{ll}
\max&\xi(L)\\
\text{subject to}&E_0(L-1)^2\leq E_0(L^*-1)^2\\
&L\in\mathcal C.
\end{array}
$$
Here $L$ is taken as $L(X,Y)$, $\mathcal C$ as $\mathcal L$ and $\xi(L)$ as $E_0[h(X,Y)L(X,Y)]$, for use in the proof of Proposition \ref{prop:max solution}, and we have $L=L(X_{t-1},X_t)$ for any $t$, $\mathcal C$ as $\mathcal L^c(M)$, and $\xi(L)$ as $E_0[h(\mathbf X_T)\underline L]$ for the proof of Theorem \ref{thm:expansion}.
\label{thm:duality}
\end{theorem}

\end{APPENDICES}

\end{document}